\documentclass[preprint2]{aastex}

\shorttitle{Guide Star Catalog 2.3}
\shortauthors{CASB/OATo}
\usepackage{natbib}

\begin{document}

\title{THE SECOND-GENERATION GUIDE STAR CATALOG: \\
  DESCRIPTION AND PROPERTIES }


\author{
Barry M. Lasker           \altaffilmark{1,10},
Mario G. Lattanzi        \altaffilmark{2},
Brian J. McLean           \altaffilmark{1},\\
Beatrice Bucciarelli     \altaffilmark{2}, Ronald Drimmel
\altaffilmark{2}, Jorge Garcia                \altaffilmark{5},
Gretchen Greene         \altaffilmark{1}, Fabrizia Guglielmetti
\altaffilmark{6}, Christopher Hanley    \altaffilmark{1}, George
Hawkins         \altaffilmark{1}, Victoria G. Laidler
\altaffilmark{1,3}, Charles Loomis          \altaffilmark{1,3},
Michael Meakes          \altaffilmark{1}, Roberto Mignani
\altaffilmark{7}, Roberto Morbidelli      \altaffilmark{2}, Jane
Morrison             \altaffilmark{4}, Renato Pannunzio
\altaffilmark{2}, Amy Rosenberg           \altaffilmark{1}, Maria
Sarasso         \altaffilmark{2}, Richard L. Smart
\altaffilmark{2}, Alessandro Spagna   \altaffilmark{2}, Conrad R.
Sturch \altaffilmark{1,3,9}, Antonio Volpicelli \altaffilmark{2},
Richard L. White        \altaffilmark{1}, David Wolfe
\altaffilmark{1}, and Andrea Zacchei \altaffilmark{8}}

\altaffiltext{1} {Space Telescope Science Institute, 3700 San
Martin Drive,  Baltimore, MD 21218, USA} \altaffiltext{2}
{INAF--Osservatorio Astronomico di Torino, Strada Osservatorio 20,
10025 Pino Torinese, TO, Italy} \altaffiltext{3} {Computer
Sciences Corporation, Space Telescope Science Institute, 3700 San
Martin Drive, Baltimore, MD 21218, USA} \altaffiltext{4} {Steward
Observatory, 9425 N.Weather Hill Dr., Tucson, AZ 5743-5484, USA}
\altaffiltext{5} {Gemini Observatory Southern Operations Center,
c/o AURA, Casilla 603, La Serena, Chile} \altaffiltext{6}
{Max-Planck-Institut f�r Plasmaphysik, Ber:MF Geb:L4  Zi:338,
Boltzmannstrasse 2, 85748 Garching bei Muenchen, Germany}
\altaffiltext{7} {Mullard Space Science Laboratory, University
College London, Holmbury St. Mary, Dorking, Surrey, RH5 6NT, UK}
\altaffiltext{8} {Osservatorio Astronomico di Trieste, Via Tiepolo
11, 34131, Trieste, Italy} \altaffiltext{9}{Retired}
\altaffiltext{10}{Deceased}

\begin{abstract}
The Guide Star Catalog II (GSC-II) is an all-sky database of
objects derived from the uncompressed Digitized Sky Surveys
that the Space Telescope Science Institute has created from the
Palomar and UK Schmidt survey plates and made available to the
community.

Like its predecessor (GSC-I), the GSC-II was primarily created to
provide guide star information and observation planning support
for {\it Hubble Space Telescope}. This version, however, is already employed at some of the
ground-based new-technology telescopes such as GEMINI, VLT, and
TNG, and will also be used to provide support for the James Webb Space Telescope
(JWST) and
Gaia space missions as well as the Large Sky Area Multi-Object
Fiber Spectroscopic Telescope, one of the major ongoing
scientific projects in China.

 Two catalogs have already been extracted from the GSC-II database
and released to the astronomical community. A magnitude-limited
($R_F$ = 18.0) version, GSC2.2, was distributed soon after its
production in 2001, while the GSC2.3 release has been available
for general access since 2007.

 The GSC2.3 catalog described in this paper contains
astrometry, photometry, and classification for 945,592,683 objects
down to the magnitude limit of the plates. Positions are tied to
the International Celestial Reference System;
for stellar sources, the all-sky average absolute error
per coordinate ranges from $0.''2$ to $0.''28$ depending on
magnitude. When dealing with extended objects, astrometric errors
are ~20\% worse in the case of galaxies and approximately a factor
of 2 worse for blended images.

Stellar photometry is determined to 0.13-0.22 mag as a function of
magnitude and photographic passbands ($R_F$, $B_J$, $I_N$).
Outside of the galactic plane, stellar classification is reliable
to at least 90\% confidence for magnitudes brighter than
$R_F=19.5$, and the catalog is complete to ~$R_F$=20.

\end{abstract}



\section{Introduction}
The Guide Star Catalog II (GSC-II) is an astronomical database
constructed from the scanned images of 9541 Palomar and UK Schmidt
photographic sky survey plates digitized at Space Telescope Science Institute (STScI).
 These same
plate images are also known as the Digitized Sky Survey (DSS), and
are accessible directly from the STScI eb site as well as from a
number of major astronomical data centers around the world. A
subset of the original images, based on the second-epoch Palomar
surveys only (also known as DPOSS), is available separately from
Caltech \citep{2003BASBr..23Q.197D}. Whilst all the images
represent over 8 terabytes of data that are archived at STScI, the
use of the H-transform compression technique
\citep{1992doss.conf..167W,1994SPIE.2199..703W} has enabled a
reduction in data volume to $\approx 1$ terabyte for distribution
to data centers.

Each individual uncompressed image was processed to detect the
objects, and calibrations were obtained for each plate by
polynomial modeling against photometric and astrometric reference
catalogs. The average accuracy in position for stellar objects of
intermediate magnitude ($V\lesssim 18.5$) is $\sim 0.3''$, whilst
the corresponding astrometric precision, on a scale of
approximately 0.5 degrees, is close to $0.2''$. Photometric
calibrations, carried out in the natural plate passband, show
typical errors of 0.1-0.2 mag and systematic offsets lower than
0.1 mag. Stars are correctly classified as such with at least 90\%
confidence for magnitudes brighter than $R_F=19.5$.

The plate catalogs were loaded into a database system known as
Catalog of Objects and Measured Parameters from All Sky
Surveys (COMPASS), \citep[see][]{1998asal.confE...3L}. This database was
built using "Objectivity", a commercial Object-Oriented Database
Management System; it constitutes a repository for  $\approx 5$
billion measurements, and is $\approx2.5$ terabytes in size. Once
the individual observations have been cross-matched to link
different measurements of unique astronomical objects, a catalog
can be exported from the database as a collection of FITS binary
tables. Then, the export catalog is made available to the
community via Web services and also provided to a number of data
centers. Over the last few years there have been a number of
interim releases from the GSC-II, primarily to consortium members
for telescope operations. The latest public release, the
GSC~2.3.2, contains 945,592,683 unique objects and is $\approx
170$ GB in size. Details about the release content and
access methods are available from the ST ScI and Osservatorio Astronomico
di TOrino (OATo) Web sites.

In this paper we describe the construction, calibration, and overall quality
of GSC~2.3.2 , the current catalog release (for simplicity, GSC~2.3 in the rest of this paper),
which was investigated through internal tests and a number of external comparisons to suitable literature
data.

\subsection{Background}
In the age of new-technology telescopes and space-based missions,
it is useful to recall that
an important part of observational astronomy has historically been the
creation of catalogs containing the reference and (often) target
objects required to support observing programs. From a
historical viewpoint it is fascinating that astrometry---the oldest
branch of astronomy---is so vital to the success of modern high-tech observatories.
As the technological complexity (and cost) of building and operating
telescopes has increased enormously over the last 20 years, so has the
effort to provide the best scientific return for these investments.
One of the important issues is the optimization of observing efficiency, which
depends on the use of proper pointing and input catalogs as well as
on the access to digitized versions of the optical sky surveys.

The need for a deeper, all-sky catalog was highlighted early by the
creation of the first Guide Star Catalog (GSC-I) to satisfy the
pointing requirements of the {\it Hubble Space Telescope} ({\it HST}). An in-depth
description of the GSC-I catalog may be found in a set of three papers
\citep{1990AJ.....99.2019L,1990AJ.....99.2059R,1990AJ.....99.2082J}.
The GSC-I was used for {\it HST} observation planning as well as target
acquisition and tracking and has proved to be very reliable for its
intended purpose. In addition, since its publication on CD-ROM, this
catalog had become widely used by many ground-based telescopes
to speed up the process of finding guide stars.

Although the GSC-I has been used with great success operationally
and scientifically, it became clear (even during its construction)
that it was possible to improve its usefulness by addressing the
known systematic calibration errors. Indeed, it was clear that an
increase in scope to include multicolor, multi-epoch data would
eventually lead to the requirements for a new much improved
archive, the GSC-II. In addition to the risk of damaging the
UV-sensitive MAMA detectors in the second generation of {\it HST}
instruments by O stars as faint as 19th magnitude, it was
generally acknowledged that the proper motions of the guide stars
during the proposed lifetime of {\it HST} would pose a potential problem
for the accurate pointing of the telescope during the latter years
of its operation. It was also realized that a fainter archive with
multicolors such as the GSC-II would address a number of other
astronomical needs such as: supporting adaptive optics on the next
generation of large telescopes, remote or queue scheduling
capabilities, and improved detector instruments requiring color
information of the target objects.

In order to be prepared for this, STScI negotiated access to the
original plates of the Palomar Observatory Sky Survey (POSS-II) and,
in partnership with the
Anglo-Australian Observatory (AAO), undertook the Second Epoch
Southern (SES) red survey with the UK Schmidt Telescope Unit (UKSTU).  These
new surveys, when combined with the plate material used in the GSC-I
and the earlier POSS-I and UKSTU surveys, provide the material to
generate a GSC-II based on, at least, two epochs and three passbands.

\begin{figure*}[t]
\centering
\includegraphics[scale=.70,angle=0]{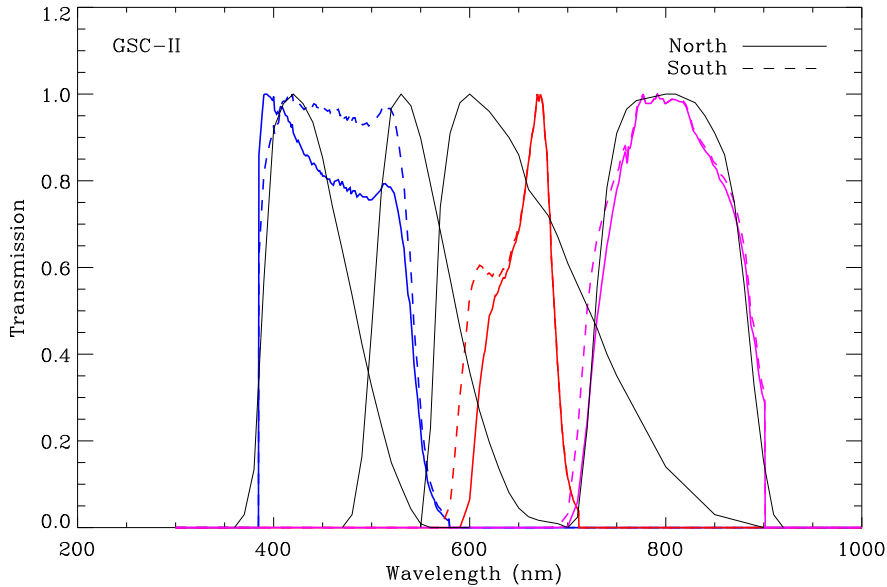}
\caption{Transmission curves of the photographic passbands $B_J, R_F,$ and $I_N$
for the Palomar (solid lines) and AAO (dashed line) Schmidt surveys, compared to the Johnson--Kron--Cousins  BVRI$_{\rm c}$ filters.
\label{fig:photometry:filters}}
\end{figure*}

\subsection{The GSC-II Project}
STScI and OATo first began a collaboration to start
developing GSC-II in 1989. STScI was primarily motivated by telescope operations,
and OATo was interested in the scientific applications of this catalog for
galactic structure.

Once the project was underway, additional resources were contributed by
ESO and GEMINI (who wished to use GSC-II for VLT and GEMINI telescope operations respectively), ST-ECF (as part of the NASA-ESA {\it HST} funding agreement), and the
Astrophysics Division of ESA (for science projects). A significant
fraction ($\sim 30\%$) of the plate processing was performed at ST-ECF using
the same pipeline software system described later.

\section{Astronomical data}

\subsection{Photographic Plates}
All of the survey plates scanned in this project were originally taken
by the Palomar Schmidt telescope in California, USA, and the UK Schmidt
telescope at Coonabarabran, Australia. Table \ref{GSCI-IIplates} summarizes the main
characteristics of the various plate material and a complete
description of the photographic surveys can be found in \citet{1995ASPC...84..137M} and references therein.

In general, there is a strong similarity between the northern and
southern surveys since they used the same photographic emulsions;
however, there are small differences in some of the filters used, which result
in slight differences in the transmission curves as shown in Figure \ref{fig:photometry:filters}.

\subsection{Astrometric Reference Catalogs}
Following the recommendation of the IAU XXIII General Assembly
(resolution B2),  the International Celestial Reference System
(ICRS)was adopted to be the reference system for the GSC-II. At
optical wavelengths, the ICRS is defined by the {\it Hipparcos} catalog;
however, the bright {\it Hipparcos} stars are heavily saturated on
Schmidt plates. On the other hand, the fainter stars  ($V>10$) of
the {\it Tycho} \citep{1997hity.book.....P} and {\it Tycho}-2
\citep{2000A&A...355L..27H} catalogs, do have measurable images on
the Schmidt plates, so that each survey plate is directly tied to
the ICRS through {\it Tycho}'s faint end. Because the {\it Tycho-2} catalog
was not available until after early 2000, a subset of the GSC~2.3
astrometric calibrations are based on the ACT catalog
\citep{1998AJ....115.2161U}. Slight deviations of the ACT and
{\it Tycho-2} catalogs from the ICRS system are well below our measuring
error.

\subsection{Photometric Reference Catalogs}
The photometric calibrators used for the GSC-II are the GSPC-II,
GSPC-I and {\it Tycho} catalogs, which sample respectively the faint, intermediate
and bright range of the photographic magnitudes.

The GSPC-I \citep[Guide Star Photometric Catalog][]{1988ApJS...68....1L}, which
was built to support the construction of GSC-I,  contains 1477
photoelectric sequences with $B$ and $V$ standard photometry of 9-15th mag
stars at the 0.1-0.15 mag level of accuracy.

\begin{figure}[t]
\plotone{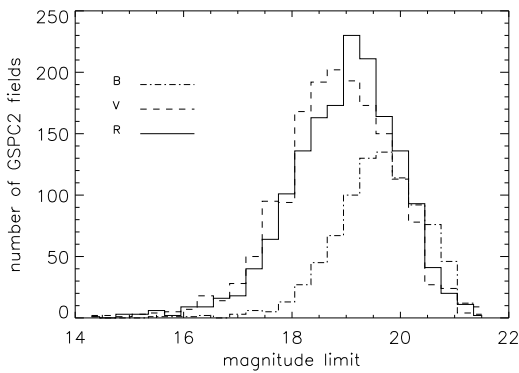}
\caption{Histogram of GSPC-II sequences magnitude limit.
\label{faint_limit}}
\end{figure}

\begin{figure*}
\epsscale{2.0}
\plottwo{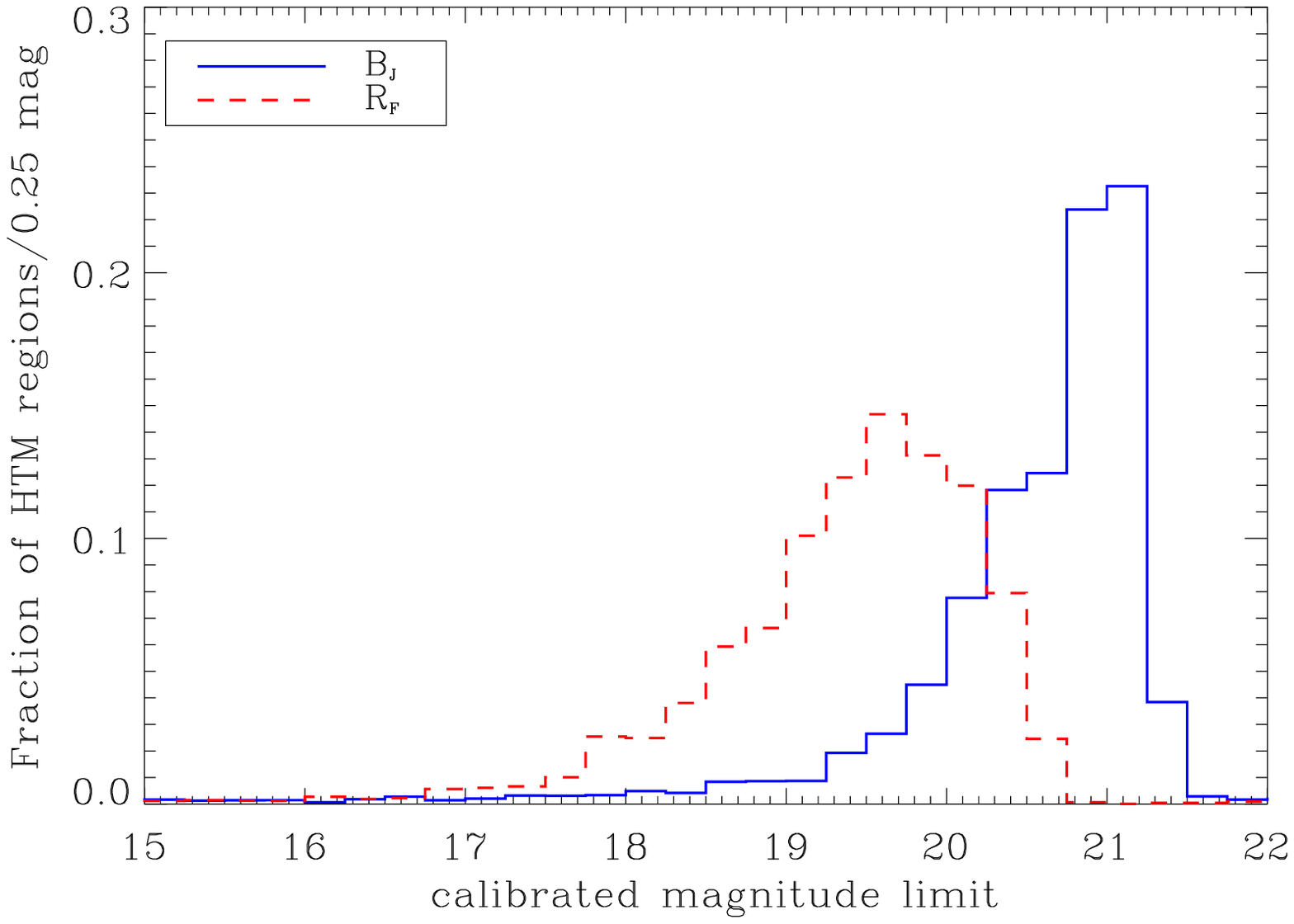}{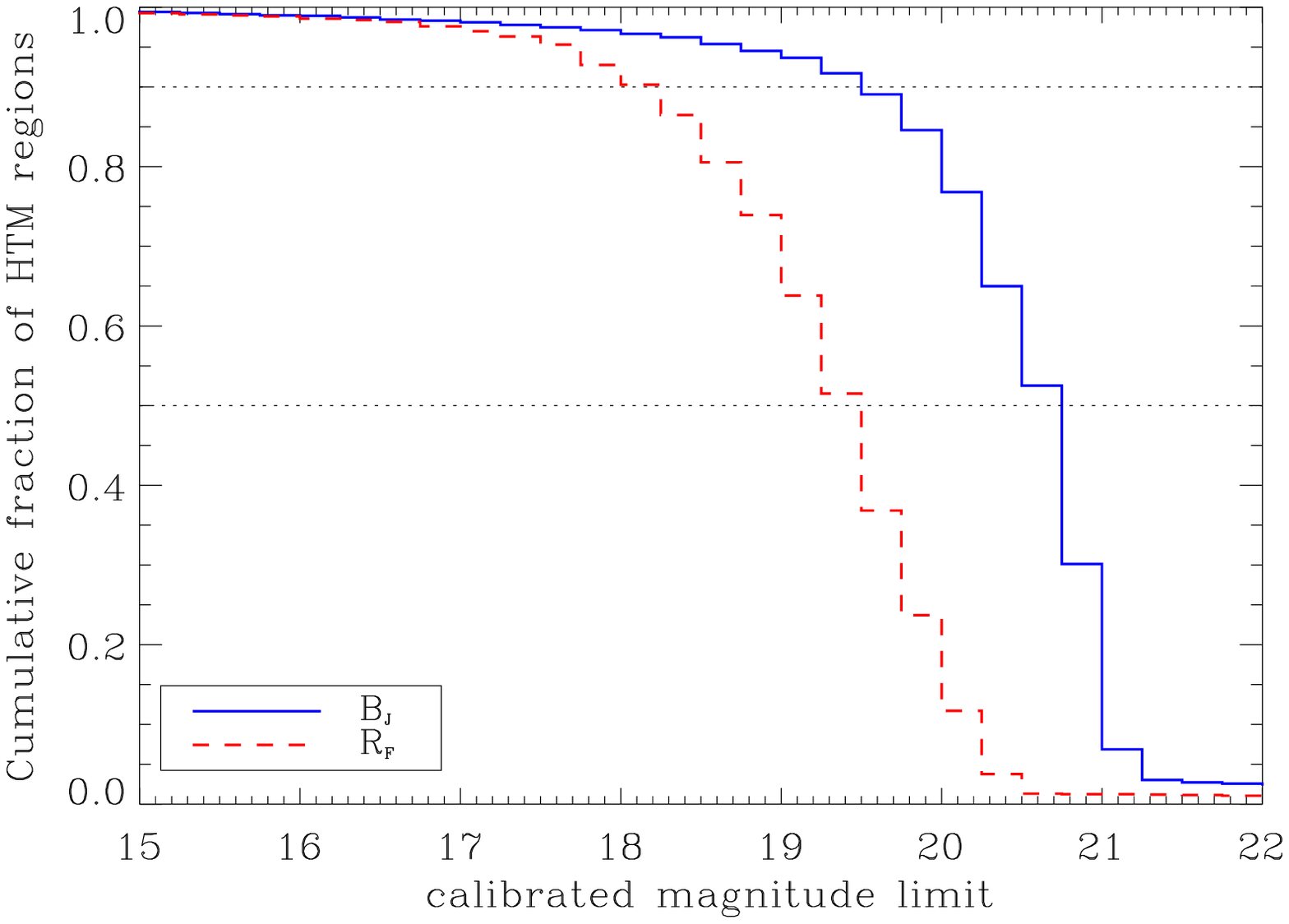}
\caption{GSC~2.3 calibrated magnitude limit. {\it Left panel}: distribution
of the faintest GSPC2 reference star available in the  HTM regions
and which result from the photometric calibration of the blue
$B_J$ and red $R_F$ plates.  {\it Right panel}:  cumulative
distribution of the HTM regions has a function of the faintest
GPSC2 reference stars.  Note that $\sim$90\% of the sky has been
calibrated with photometric sequences down to $R_F\ga 18$ and
$B_J\ga 19.5$, while for $\sim$50\% of the sky the photometric
sequences attain $R_F\ga 19.5$ and $B_J\ga 20.5$.
\label{gsc2:magmax}}
\end{figure*}

Because of the nonlinear response of the photographic plate, the need for fainter
calibrators was a requirement for the construction of GSC-II. A lack of
such calibrators in the astronomical literature motivated the start of
a decade-long series of observations, only now coming to
completion, which led to the construction of GSPC-II
\citep{2001A&A...368..335B,2006yCat.2272....0B}.
This is an all-sky catalog of CCD photometric sequences centered in each
survey plate field. Each sequence
provides $B$, $V$ and $R$ standard photometry at the 0.05-0.1 mag
precision level down to approximately V=19.5. The number of stars per
sequence varies from a dozen to several hundred depending on the
field of view of the CCD and the galactic latitude.
At the time of GSC~2.3 calibrations, 0.5\% of the plates still did not have an
extended photometric sequence. In these cases, the photometric calibration
relied on existing standards and deriving artificial points, as explained in
Section 3.3.2.
Figure \ref{faint_limit} shows the magnitude limit distribution of
all GSPC-II calibration sequences.
The {\it Tycho} catalog \citep{1997hity.book.....P}
was used in the GSC-II photometric calibration to stabilize the bright end of the
density-to-intensity curve.
The calibration limits of GSC~2.3 are given in Fig. \ref{gsc2:magmax} which shows the
distribution of the faintest GSCP-II reference stars available in the Hierarchical
Triangulated Mesh (HTM) regions and
used for the photometric calibration of the blue $B_J$ and red $R_F$ plates.

\section{Plate processing pipeline}

\subsection{Plate digitization}
The processing of photographic plates begins with the use of
microdensitometers, machines capable of measuring photographic
transmission $T$. This is the fraction of light
measured as a current, transmitted by the portion of the
developed emulsion as illuminated. The PDS machines used for the
digitization of GSC-II plates measure the semispecular density which is
defined as the light transmitted and diffused within an angle of $< 24^\circ$.
 This is then converted into photographic density $\rho$ through the relation
$$ \rho = -\log_{10}T.$$
As $T$ can assume all values from zero (complete opaque emulsion) to
1 (perfectly transparent media), $\rho$ can vary from zero to infinity;
in practice, $\rho$ usually lies in the range $0-5.0$.
The Perkin--Elmer 2020G PDSs originally selected for digitizing the STScI
plate collection were first modified to cope with the accuracy requirements
imposed by the GSC-I project \citep[see][]{1990AJ.....99.2019L}.
The increase in the number of plates (5$\times$) to be digitized for the GSC-II
and DSS projects  also required a significant increase in scanning throughput.
For this reason, the two scanning machines available at STScI
were refurbished with entirely new critical subsystems to
increase their speed without affecting their spatial and photometric
capability; the improved twin digitizing devices were named GAMMA 1 and
GAMMA 2 \citep{1994AAS...184.2701L}.

The Guide Star Automatic Measuring MAchine (GAMMA) is a
laser-illuminated multichannel scanning microdensitometer, modularly
built upon the substrate of the modified PDS used in earlier STScI
work.

A rebuild of the $x$- and $y$-servos is based on an HP 5507 laser
transducer system with a custom card to implement the PDS "1/$N$"
functionality. The light source is a spatially filtered 2~mW HeNe
laser beam, expanded to 1 mm diameter (1/$e^2$ of total energy), moved
with a TeO(2)
acousto-optic deflector (AOD), and finally imaged on the plate as a 42
$\mu$m Gaussian spot. This size gives low aliasing with the adopted 15
$\mu$m sampling. The table moves in the $x$-direction at 100 mm~s$^{-1}$, while
the AOD steps the laser beam through a small number of channels (typically 5)
in the $y$-direction. The AOD is controlled by a frequency synthesizer
operated from a digital signal processor (DSP) which cycles through a
channel-to-frequency table, synchronized by pulses from the servo 1/$N$
logic.  The DSP also removes y-position errors due to residual servo
auto-lock disequilibrium by applying a small frequency correction to
the AOD, based on a linear function of the error signal from the $y$-servo.

The light-collecting optics consists of a standard microscope,
an exit slit, an integrating sphere to make the response insensitive
to the channel number, and a photomultiplier. Pixel integration times
are about 25-50 $\mu$sec, and the data conversion is done with a 15-bit
floating point analog-to-digital-converter (ADC), which supplies the
photometric parameter $D_{ij} = a +
b\rho_{i,j}$ linearly related to the optical density measured at pixel
$i,j$.

A separate photodiode to monitor the laser flicker is also
provided for photometric normalization. A dedicated VAXStation with
IEEE-488 and CAMAC interfaces controls each GAMMA machine. Separate programs
operate each subsystem (servo, data collection, laser monitoring, AOD,
console) and communicate with each other via shared memory and common
event flags. This matching of the software and hardware architectures
results in easily maintainable code.

Nominal performance of the scanning machines was verified on a regular
basis using several numerical methods. A sample of three areas from each
scan was tested using a scan-line correlation and constrained
minimization to test for scan-line "shear" as a
function of channel. Scans where the shear exceeded 0.1 pixel were
rejected and further tests and mechanical maintenance were then
performed.  In general the intrapass channels showed a very high
degree of correlation and shear rarely exceeded 0.05 pixel.  Periodic
engineering scans were also performed on a routine basis of Max Levy 2
slanted and vertical rulings in order to tune the AOD system for
optimal performance.

\subsection{Image processing}
After digitization of the Schmidt plate material on the GAMMA
machines, each plate scan was processed through an image processing
pipeline, as detailed in the following sections.

\subsubsection{Sky Background Determination}
A single value of the sky background standard deviation is calculated for the
whole plate by subdividing it into a regular grid of 0.5 mm cells for the
estimation of local density histograms and pixels
statistics. Then, the median values of each cell are computed, and a sequence
of median filters is applied at different scales (8,4,2,1 cells) to produce
an average sky image. The $\sigma$ calculation for each cell is found by
taking the slope of the cumulative histogram of pixel values in the cell at
the 50\% level, to minimize the disturbance of stellar contribution. Pixels at
this level are nearly all sky, and if the sky noise behaves as a Gaussian, the
slope of the cumulative pixel histogram is $1 /(\sqrt{2\pi}\sigma)$ which can
be used to find $\sigma$ for the individual cells.
The final $\sigma$ for the plate is then found from the mode of the histogram
of the individual cell $\sigma$'s. This is a good approximation for $\sigma$
over the whole plate from the unsmoothed sky using all pixels in the plate.
Finally, the edges of the average sky image  are examined in an attempt
to avoid processing the "bad" edges of the plate as well as the label and
sensitometer areas: areas in which the mean values are more than 5 $\sigma$
away from the mean sky are set to a negative value, which is a flag for the
object detection program to avoid processing.

\subsubsection{Object Detection}
\begin{figure*}
\epsscale{2.0} \plottwo{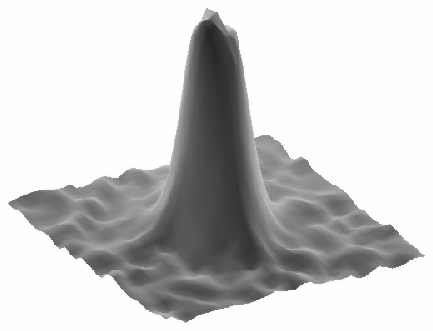}{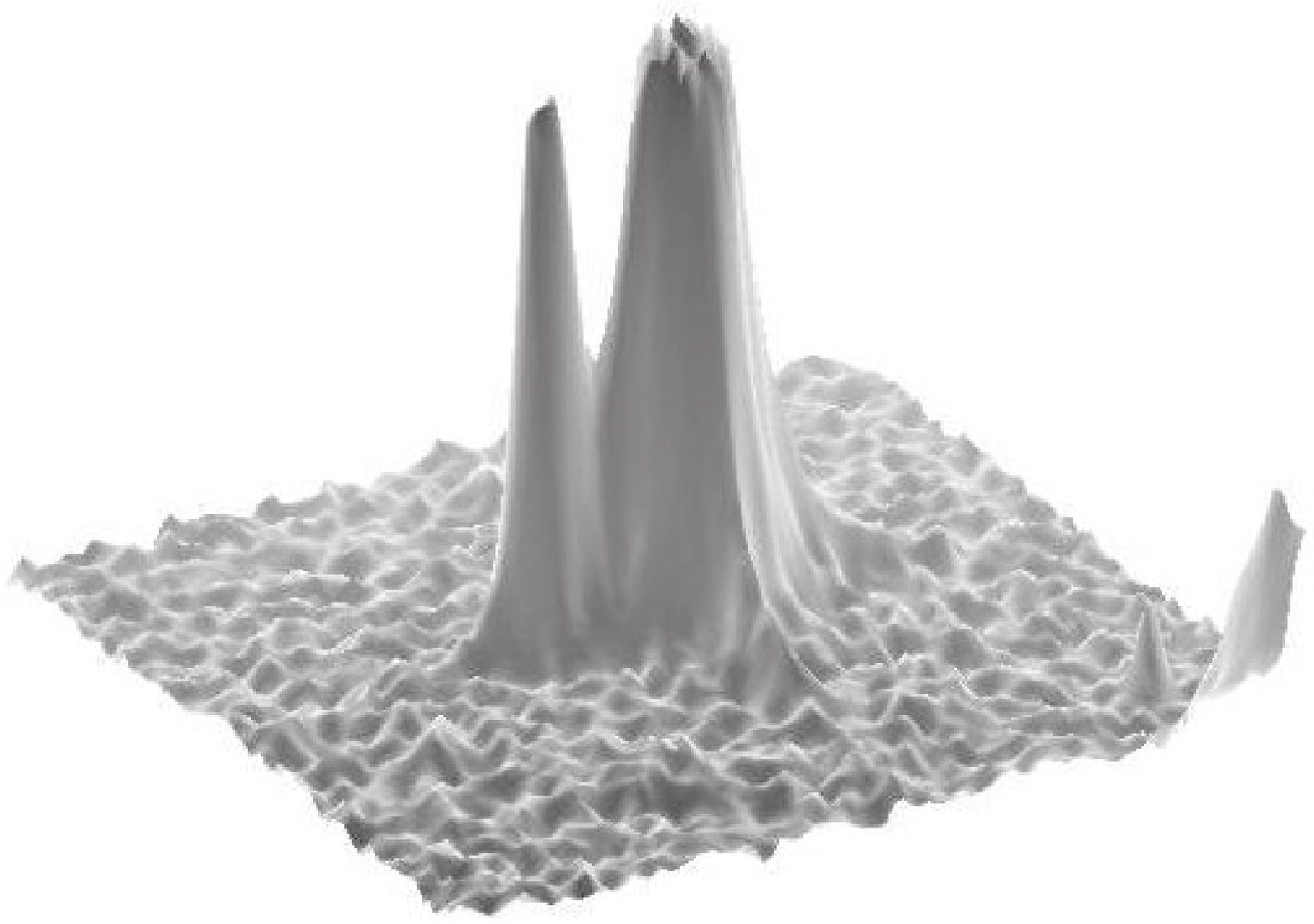}
 \caption{Cutout
of a bright star (left panel) and of a resolved binary star (right
panel). The irregular truncation of the peak of the brightest
object is due to the saturation limit of the photographic
emulsion. \label{cutouts}}
\end{figure*}

This task used the same implementation of the ROE COSMOS image processing
software used for the GSC-I \citep[see][]{1990AJ.....99.2019L,1980ComputJ.....23.262,1984VA.....27..433M}
which connects a minimum number of pixels that are brighter than a
threshold level above the sky.
The threshold level for object detection was nominally set at 3$\sigma$,
though it was increased to 4$\sigma$ in cases of plates with large
grain noise or in crowded fields.
A single value of sigma was used for the entire plate, as described
above.  Fifteen additional thresholds, equally spaced (in density
space) between the detection threshold and the saturation level, were
also defined, and minimum areas were determined at each of these
thresholds as well. A small image "cutout" was then extracted and saved
around each detected object, with ample margins to include the wings
of extended objects. An example of image cutouts is given in Figure \ref{cutouts}.

\begin{figure*}
\epsscale{2.0} \plottwo{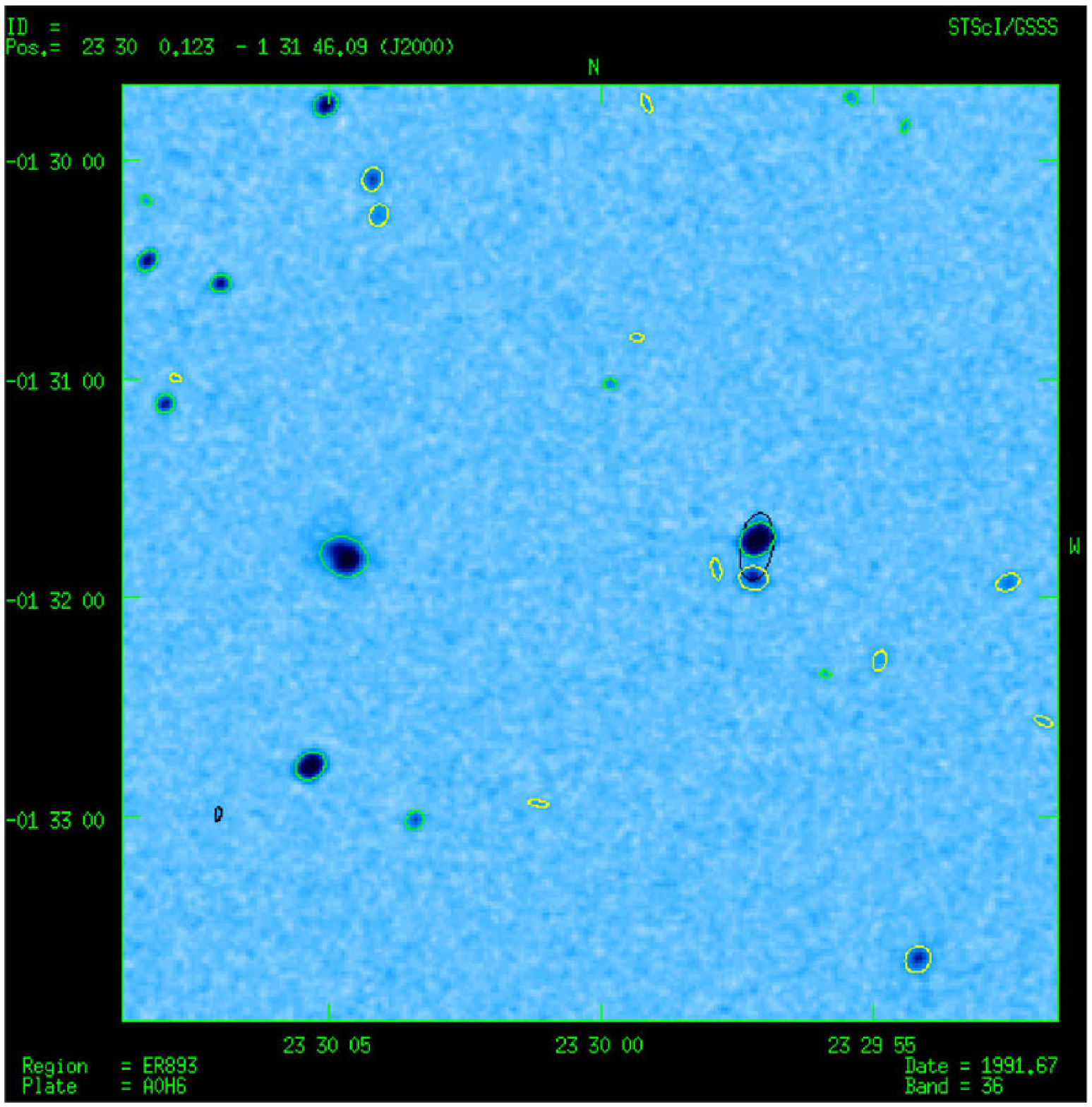}{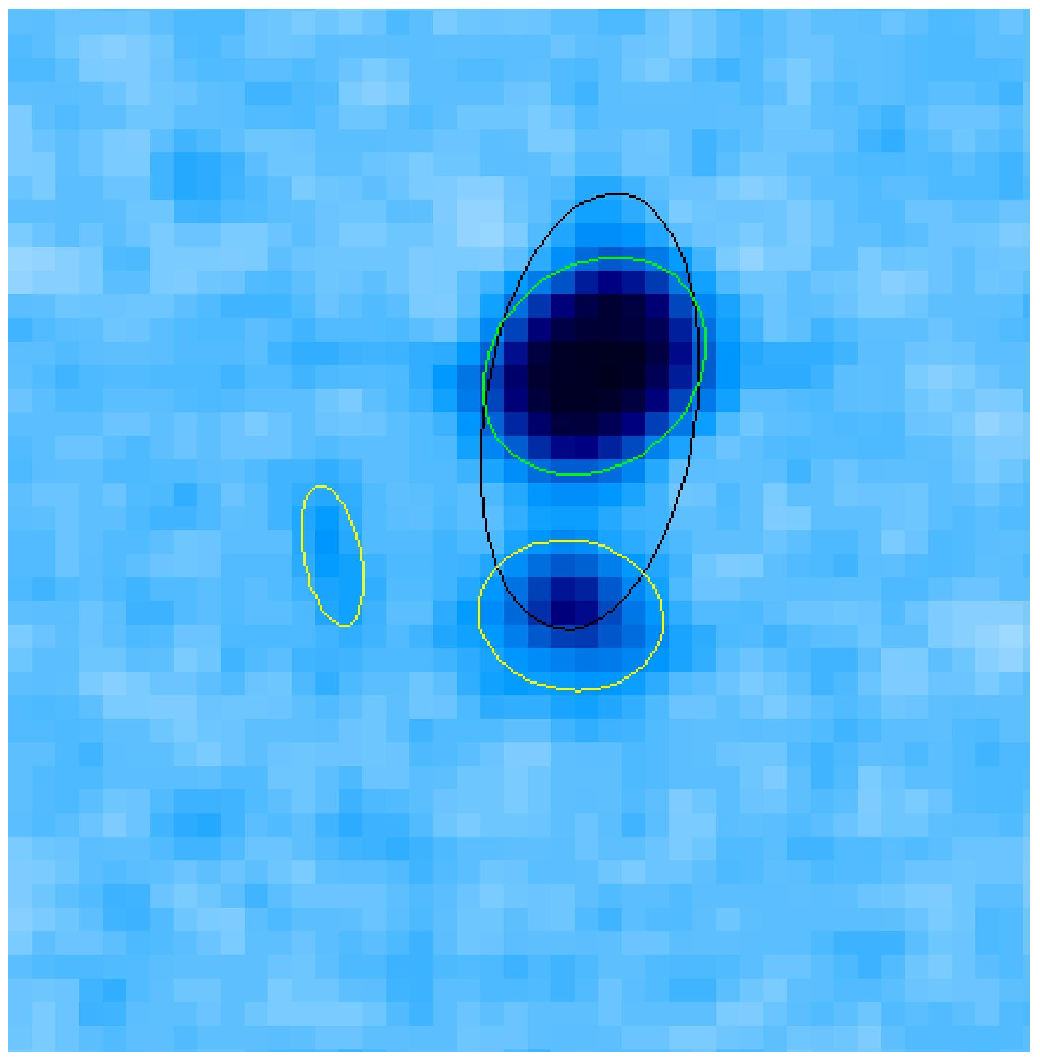}
\caption{Typical high latitude field of $4'\times 4'$ (left panel)
showing various stellar and extended objects, plus a deblended
binary system (zoomed in the right panel). The identified objects
are encircled by ellipses whose size and orientation are defined
by the moments of their density distribution; the line color
indicates the object classification: green for "stars", yellow for
``nonstars", and black for either ``defects" or ``blends".
\label{deblender}}
\end{figure*}

From the size, integrated density, and moment values of each object,
parameters such as object's shape and ($x,y$) of its centroider were derived.
These and any further image processing were performed on the cutout images
rather than the full scan image. The maximum
cutout extracted was $4.25' \times 4.25'$, so, if a detected object was larger
than this, it received no further processing (deblending, centroiding,
etc). This results in a small number of "holes" in the GSC-II catalog
around the brightest stars, galaxies, and the occasional very bright
satellite trails.

\subsubsection{Blend resolution}
This is based upon the published \citep{1990MNRAS.247..311B}
ROE COSMOS deblender algorithm,
which works well also in relatively crowded regions. Each new object
is re-thresholded, and at each new threshold the
connectivity algorithm is applied (\cite{Thanisch}) to check if the object
has split into two or more objects.
Faint outlying pixels around blended objects are fractionally allocated
according to the faint pixel allocation technique as described in these references.
The deblender uses the same thresholds as the object detection
task, but applies a single local sky value, i.e., the one recorded
by the object detection.
Once an object has been identified as single, its size, integrated density
and moments are computed and parameters such as $x,y$ and shape finally
calculated. An example of deblended objects formed by the two companions of
a parent binary is shown in Figure \ref{deblender}.

\subsubsection{Centroiding}
The centroider algorithm performs a fit to the cutout image with an elliptical
Gaussian to find the object's center.  After the image is normalized,
the centroid is calculated iteratively, with the barycentric position
as starting point, and assuming a flat sky. The centroider returns the
amplitude, $x$, $y$, and their covariance matrix, as well as the $x$ slope, $y$ slope,
and level of the fitted background. Deblended images occurring in the same
cutout require special handling. First, an attempt is made to perform a
simultaneous fit to all objects in the deblended family. If this
fails to converge, each deblended component is run individually
through the single centroider after applying the fractional pixel
allocation determined by the deblender. If the centroider failed to
converge on an object after a reasonable number of iterations, or if
the centroider failed for some other reason, only the barycentric
position determined from the object detection or deblending was used in
the calibration. If the object had been classified as a plate artifact,
which has a very low chance of convergence, the iterative procedure was not
attempted.  For objects that occurred in very large cutouts (for
example, faint stars in the vicinity of a bright star), the centroider
was also omitted, due to resource constraints.

\subsubsection{Classification Parameters}
In order to provide additional information for the image classifier (see Table \ref{features}),
two additional types of numerical features known as spike parameters and texture parameters are computed.
Bright point sources on Schmidt plates display
diffraction spikes. These were used to good effect in GSC-I for
star/nonstar discrimination in the brighter magnitude ranges. Spike
parameters result from the comparison of the image gradient along
the spike, which should be weak, with the gradient along the thin
dimension of the spike, which should be strong. Texture parameters are
based on the co-occurrence matrix of the object cutout (with
fractional pixel allocation applied in the case of deblended
objects),see \citet{1983sma..conf...69M} for further details.

\subsubsection{Summary of Known Limitations}
The sky processor is less efficient in very crowded fields or
around bright stars and halos, such as filter ghosts. As already
mentioned, the object detection procedure fails in the presence of
large objects, such as bright stars or galaxies, or connected
objects that will require a cutout greater than 4.25'x4.25'. Also,
the use of a unique sky standard deviation for the whole plate
becomes problematic at some low galactic regions in the presence
of high extinction gradients; here, local detection thresholds
should be determined for optimal object detection.

GSC-II has both long- and short-exposure plates along the galactic plane.
In these regions, the plate processing pipeline was often run with customized
parameters.

Faint objects in the halos of bright ones were often missed due to both
the sky problem described above and a failure of the deblender
task. Objects brighter than 13th magnitude were sometimes broken up
into smaller objects. The halo or spikes of a bright star were usually
split up and identified as artifacts; if these artifacts were matched
between successive plates or if they were classified as
star/nonstar objects, then they were included in the export catalog.
In some cases, faint objects close to a bright stars have been erroneously
detected twice: firstly, as a secondary component in the wide cutout of the
bright star, and secondly as a direct detection.

\subsection{Plate calibrations}
For each plate used in the GSC-II, the results of the image processing
pipeline discussed in the previous sections provide a list of identified
objects with position measurements (and errors), integrated density measurements,
and a set of shape-related parameters. These are the raw data used
by the calibration procedures to derive celestial coordinates, natural
magnitudes, and classification.

\subsubsection{Astrometry}
The astrometric calibrations of the GSC~2.3 are based on reference
catalogs that provide local representations of a fundamental
coordinate system. The chosen astrometric reference catalogs are
the ACT and {\it Tycho-2}, as described in Section 2.2. Since, on
average, there are a few hundred reference stars per plate, the
plate positions are tied directly to the reference catalog and
hence to the ICRS. The brighter stars appearing in the reference
catalogs, i.e., $V < 8.5$, are heavily overexposed on the Schmidt
plates and thus not useful in calibrating the plates.

In order to compensate for differential refraction across the
plate, the measured $x$ and $y$ pixel coordinates are pre-corrected
for refraction using the method described by \cite{kon}.
Determining the relationship between $x,y$ and their associated
equatorial coordinates was done by traditional global plate
modeling, after an equidistant projection was applied to transform
equatorial into tangential coordinates. We found that a quadratic
model was suitable to represent the relationship between the $x$ and
$y$ plate measures (precorrected for refraction) and the
equidistant tangential coordinates ($\xi,\eta$). Only the objects
flagged as astrometric reference stars were used in a least-squares
reduction to determine the coefficients of the model.
These coefficients were then used to map all the $x,y$ measurements
to the associated ($\xi,\eta$) values. Having started calibrating
the plates in 1998, when {\it Tycho-2} was still not available, the
polynomial plate model was determined from the ACT for $\approx 73
\%$ of the plates and from  {\it Tycho-2} for the remaining.

\begin{figure}[ht]
\epsscale{1.0} \plotone{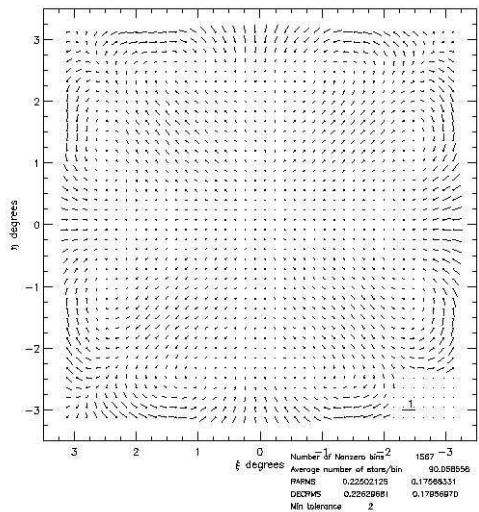}
\caption{XJ astrometric mask 40x40 bins. The scale of 1 arcsec is
plotted in the lower right-hand corner. \label{masks_xj}}
\end{figure}

Once a significant number of plates from the same survey had been reduced
(approximately 100), the estimated tangential coordinates were compared to
those calculated from the reference star's coordinates, and used to
correct for residual systematics according to the precepts of \cite{1990ApJ...358..359T}.
Specifically, for each plate the reference stars' residuals were accumulated in
4.4 mm $\times$ 4.4 mm bins, then added and averaged over all the plates to build a plate-based
{\it astrometric mask}, which yields the systematic pattern the global plate
model failed to remove.
A separate astrometric mask for each survey was determined using the
ACT catalog. Figure \ref{masks_xj} shows the mask applied to the northern blue
survey plates (XJ).

\subsubsection{Photometry}
GSC-II magnitudes were derived via modeling of the nonlinear
density-to-intensity response of plate-based photometric
calibrators, in the natural system defined by each individual
plate (Table \ref{GSCI-IIplates}). The calibrators are essentially based on
Johnson--Kron--Cousins B, V, R standards from the Guide Star
Photometric Catalog (GSPC) I \citep{1988ApJS...68....1L} and II \citep{2001A&A...368..335B};
additionally, $B_T$, $V_T$ photometry of {\it Tycho} stars was included in the calibration to constrain the
bright range of the response curve.

\begin{figure*}
\centering
\includegraphics[scale=0.7]{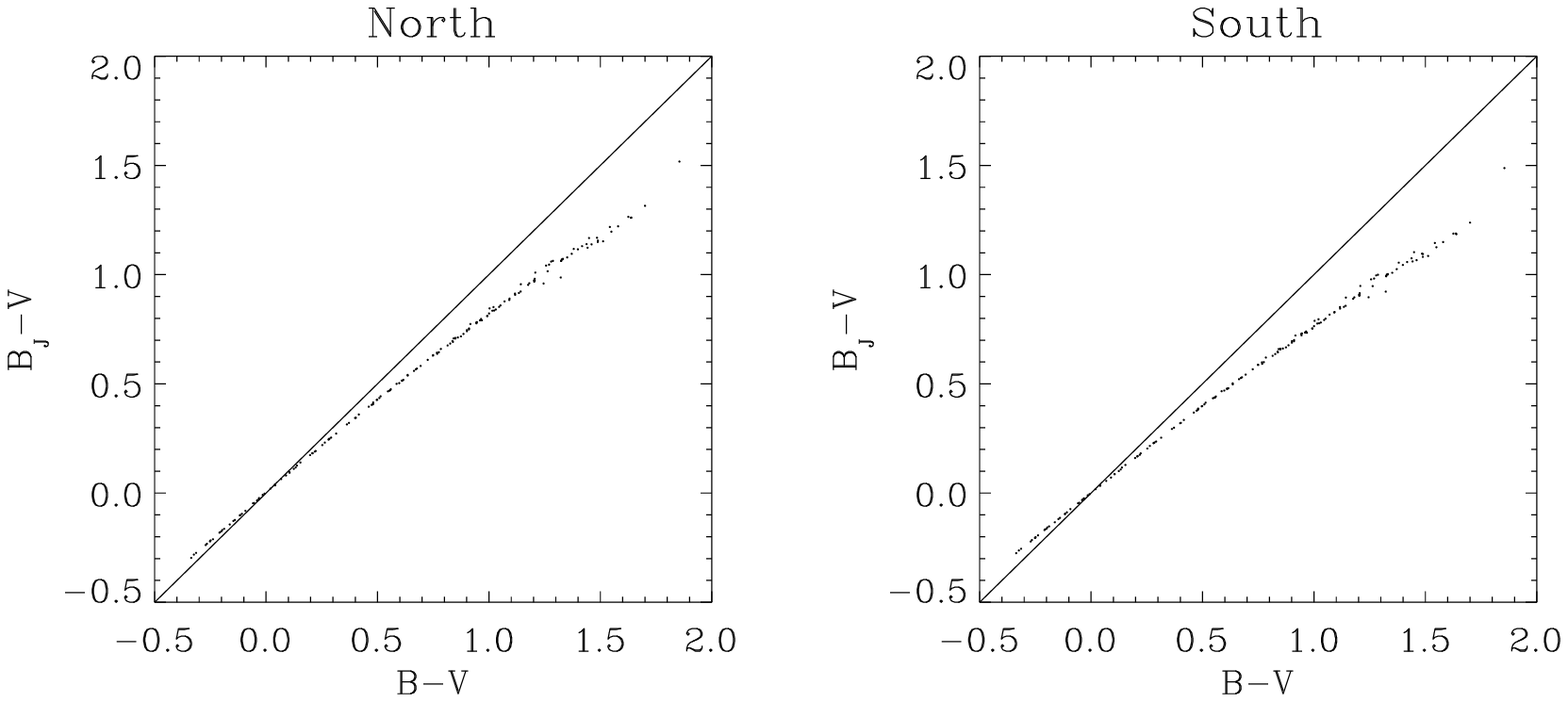}
\includegraphics[scale=0.7]{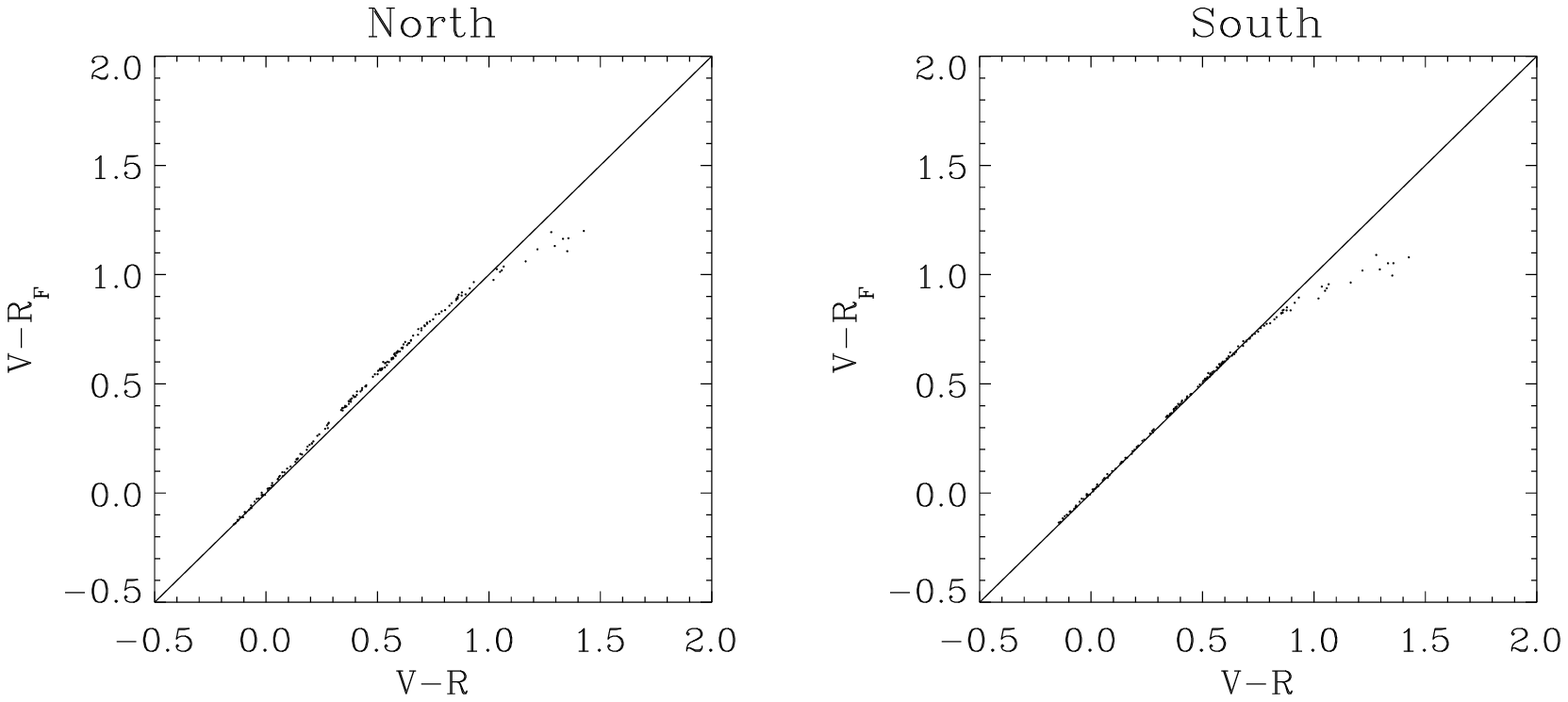}
\includegraphics[scale=0.7]{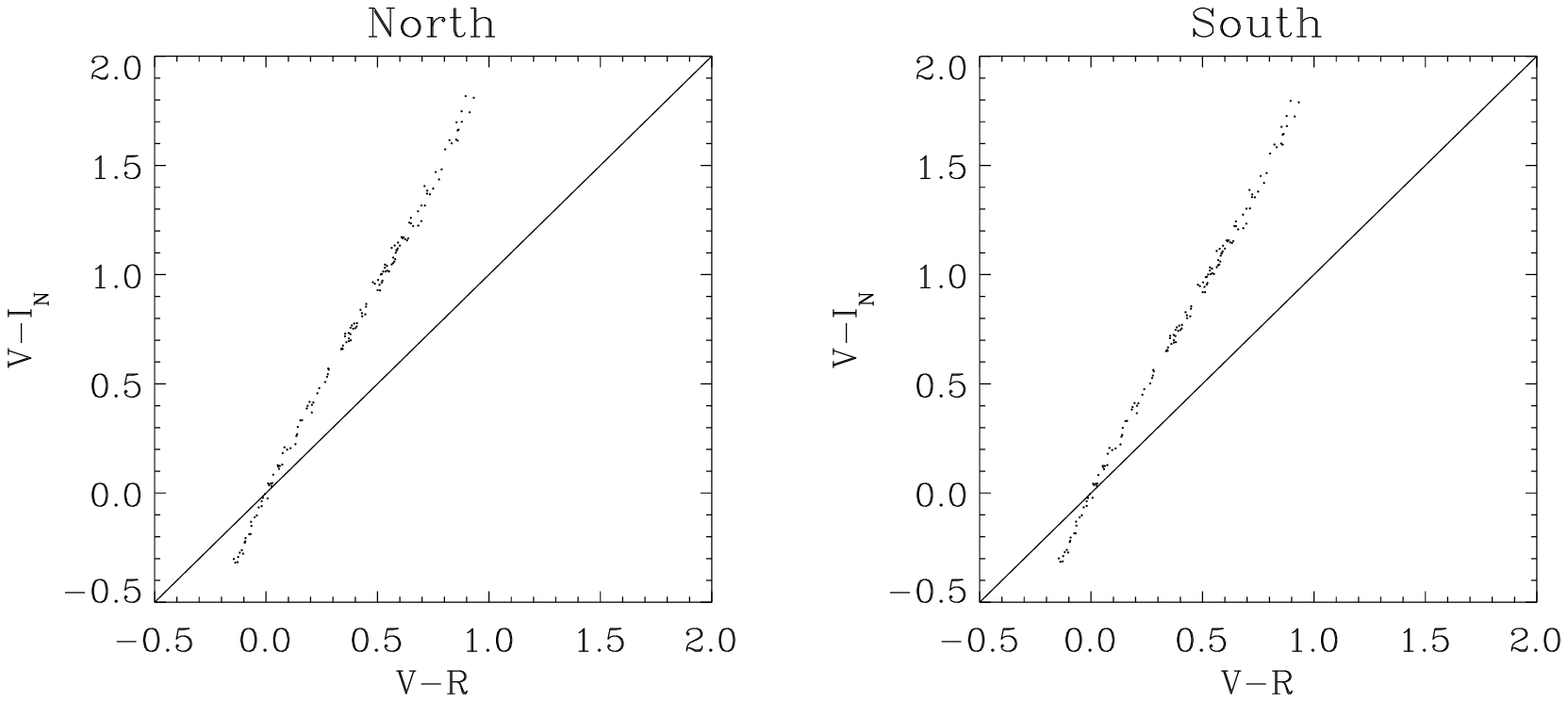}
\caption{Color--color plots used to transform GSPC-I/II calibrators from the
BVR$_{\rm c}$ photometric system to photographic passbands $B_J$, $R_F,$ and $I_N$ ($x=y$ lines
are displayed for reference).
\label{fig:photometry:transformations}}
\end{figure*}

Initially, the magnitudes of photometric calibrators have been transformed from the original Johnson--Kron--Cousins and {\it Tycho}
photometric systems to the natural
system of the plates by means of linear interpolation of the
color-color point diagrams shown in Figure
\ref{fig:photometry:transformations}. These were generated by
synthetic photometry \citep[IRAF SYNPHOT package][]{1994ASPC...61..339B} of stellar
spectra from the Bruzual--Persson--Gunn--Stryker Spectrophotometry
Atlas \citep{1983ApJS...52..121G}.
Note that in the usual case of photometric standards having both
B--V and V--R available, the transformation based on the color
closer to the requested photographic passband was utilized (e.g.\
$B_J-V_{\rm pg}$ and $V_{\rm pg}-R_F$ as a function of $B-V$ and $V-R$,
respectively).

\begin{figure*}
\centering
\includegraphics [scale=.40,angle=-90] {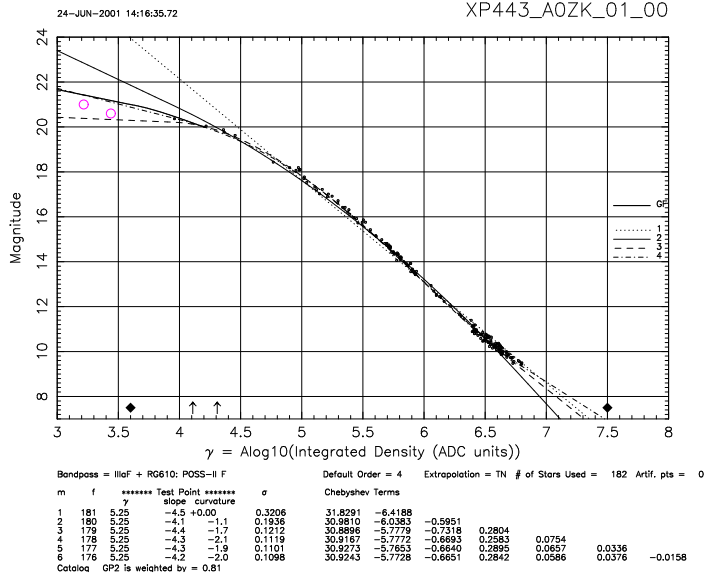}
\caption{Photometric calibration of plate XP443.  Small circles represent GSPC-I/II and Tycho stars used to fit a fourth-order polynomial (dot-dashed line) on $R_F$ vs. $P$. Artificial stars (large circles) were not applied in this case.
\label{fig:photometry:fit}}
\end{figure*}

\begin{figure*}
\centering
\includegraphics [scale=.40,angle=-90] {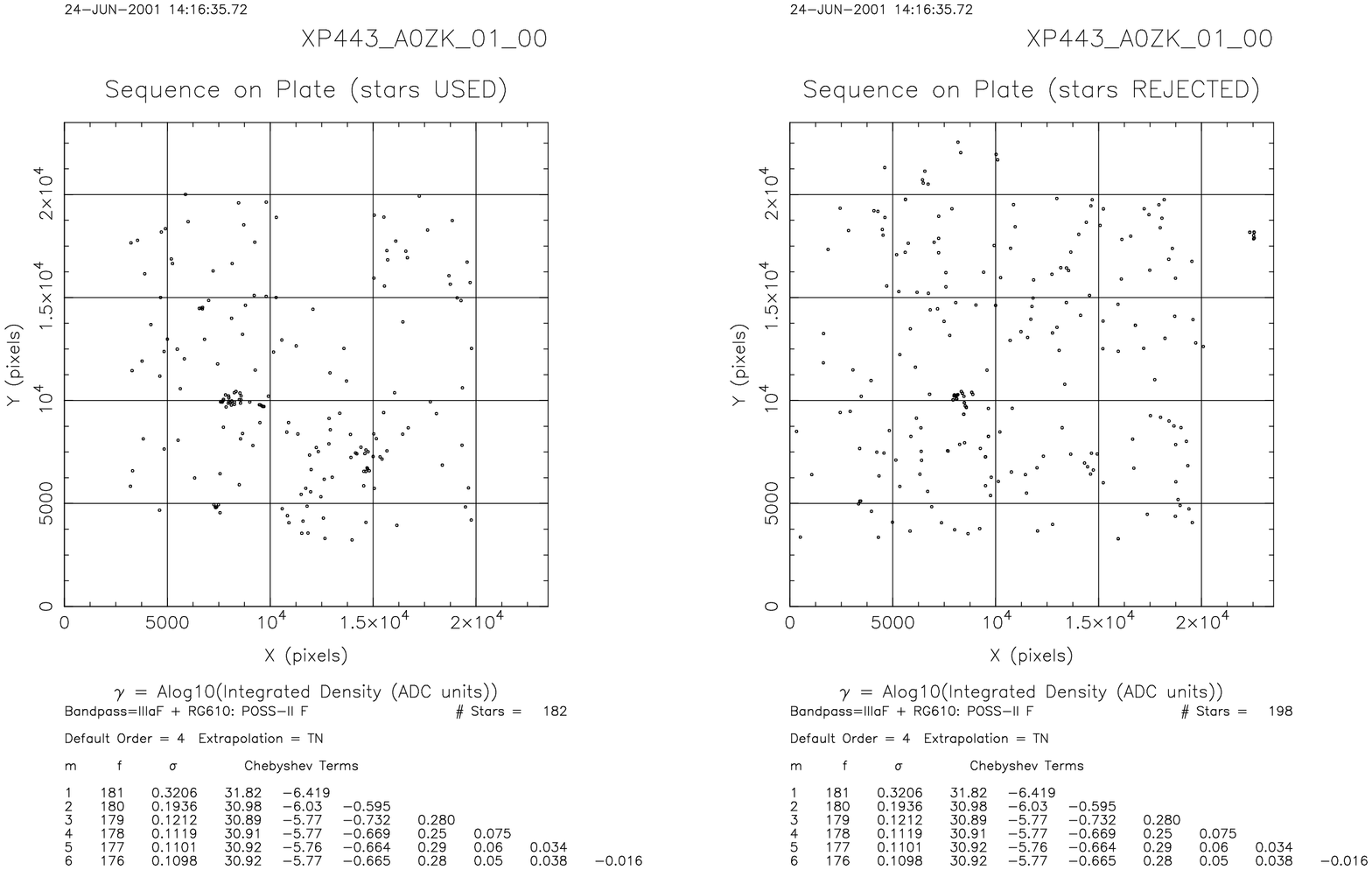}
\caption{Photometric calibration of plate XP443.  Plate distribution of the
photometric calibrators (left panel).  Note the faint GSPC-I/II
sequence located at $X\approx 8000$ and $Y\approx 10,000$, while
Tycho stars are spread over the plate. The right panel shows the objects
rejected for various reasons (blends, outliers, colors outside the range,
etc.), plus a GSPC-I/II sequence very close to the plate
border ($X\approx 22,000$,  $Y\approx 18,000$).
\label{fig:photometry:plate}}
\end{figure*}

The instrumental photometric parameter of choice was the {\it
integrated density}, defined as the logarithm of the total
density,

\begin{equation}
P=\log_{10}\sum_{i,j}\left( D_{i,j}-D_{\rm bck}\right)
\end{equation}

 calculated as the sum of the PDS-measured
density, $D_{i,j}$ (ADC units), above the local background,
$D_{\rm bck}$, for each  pixel pertaining to the selected object.
The fit to standard magnitudes was performed using Chebyshev
polynomials, namely:

$$ m = \sum_{i=0}^4 A_k ~~T_k (P') ,$$

where $T_0(x) = 0.5$ , $T_1(x) = x$, $T_2(x) = 2 x^2 -1$, $T_3(x)=
4 x^3- 3 x$, and $T_4(x) = 8 x^4 -8 x^2 +1$, $A_k$ are the
unknowns to be estimated, and $P'$ is the photometric parameter
linearly scaled to map the range [-1,1], where Chebyshev
polynomials can be suitably applied. A typical pipeline output,
as given in Figure \ref{fig:photometry:fit}, shows that these polynomial
bases fit nicely the characteristic curve over a large range of densities,
in particular at the faint end, where the large majority of
the objects lie. In this respect, extensive experiments have demonstrated that
the use of polynomials of order higher than fourth was not advisable
because of their high instability toward the plate magnitude
limit. The resulting coefficients were then used to convert the
measured integrated densities into magnitudes for all the objects
detected on the plate.

As a protection against gross errors in the calibration, reference
stars with large residuals were rejected from the fit, according
to the criterion
 $\Delta$ $>$ $\sigma\times \epsilon_1$, where $\Delta$ is the residual of the
star under test, $\sigma$ is the formal error of the fit, and
$\epsilon_1$ is an adjustable parameter, generally set equal to 2.
The procedure was stopped when no further stars were rejected, or
when the difference of formal error between successive iterations
was less than $\epsilon_2$, an adjustable parameter typically set
at 0.03. Occasionally, such iterations failed to achieve formally
acceptable solutions. In this case, manual intervention was
required, for example, to change the order of the Chebyshev
polynomial or to modify the acceptance criteria for the reference
stars.

The color transformations become poorly defined at extreme color indices;
 therefore, only standard stars in the ranges of $-0.34 \leq B - V \leq 1.85$
and $-0.14 \leq V-R \leq 1.42$ were accepted.
 Moreover, those stars for which the  integrated
density could not be adequately determined were discarded from the
initial set. This could depend on a variety of reasons, such as
image processing failures, defective images, and centroiding
biased by faint components of unresolved blends. Therefore, in
general, stars brighter than 8.5 mag as well as deblended objects
were eliminated.  Since {\it Tycho} stars are generally more numerous
than GSPC-II stars on any plate, thereby overconstraining the
bright side of the fit, a weighting scheme determined by the
magnitude range of the selected catalog divided by its number of
stars was applied. Figure \ref{fig:photometry:plate} shows an
example of the spatial distribution of the photometric
calibrators, with the GSPC-II sequence near the plate center and
the {\it Tycho} stars located all over the plate.

\begin{figure}
\epsscale {1.0}
\plotone{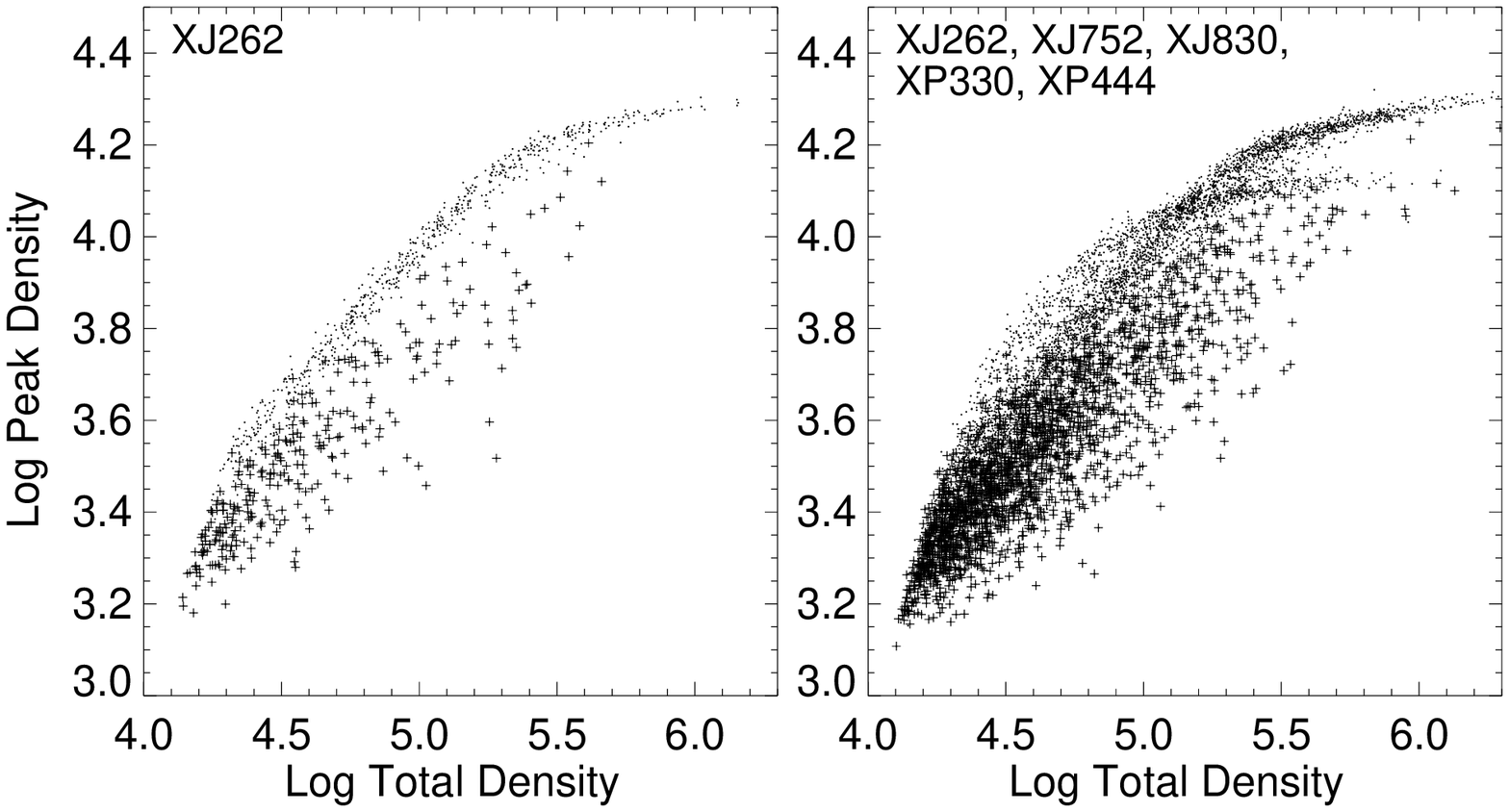}
\caption{Total density vs. peak density for stars (dots) and galaxies (plusses) measured on GSC-II photographic plates. (a) The distribution for a
single plate (XJ262) is well defined and could be used for classification. (b) The
distribution for five plates differs for different plates and so a classifier for all
plates could not be constructed from these parameters.
\label{unranked}}
\end{figure}

\begin{figure}
\epsscale{1.0}
\plotone{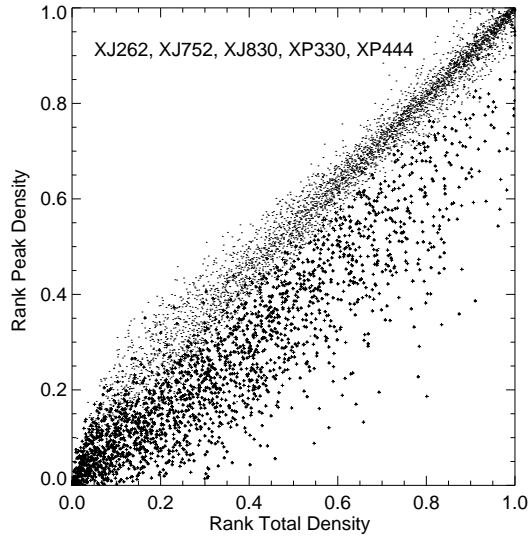}
\caption{Distribution of total density vs. peak density for objects from Figure \ref{unranked} after transformation using ranks. Stars and galaxies from different photographic plates separate very well. An accurate classifier can be constructed based on these ranked parameters.
\label{ranked}}
\end{figure}

Generally,  GSPC (I/II) stars range between 9th to 19th
magnitude (in $V$), while the GSC objects are both brighter and
fainter, such that extrapolations from the photometric reductions
are required. Because the images of bright stars (brighter than
9.0) are heavily exposed on Schmidt plates, their integrated
density is not well determined. Therefore, no extrapolation of the
photometric reduction for objects brighter than those used in the
calibration was performed, and the corresponding magnitude flag
set to $-99$. In the latest catalog version, all the bright
magnitude stars having a {\it Tycho-2} counterpart were replaced with
the {\it Tycho-2} magnitude value (see Section 4).

The method of extrapolation used for the faint stars was
tangential, i.e., a linear extrapolation with the slopes set to be
tangent to the end points (brightest and faintest standard stars)
of the fitted function. However, by using a fourth-order Chebyshev
polynomial, the extrapolated region was often unstable. In order
to stabilize this region, we introduced an {\it artificial}
reference star and assigned to it the plate magnitude limit  of
the survey---obtained from the literature and reported in
Table \ref{GSCI-IIplates}---which was associated to the integrated
density equal to that of
the smallest detectable image on the plate. In general, the use of
an artificial point improved the situation, especially on plates where
the faintest GSPC stars were in the range of 16-17 mag. On the
other hand, in cases where the standard sequence had a fainter
limit, the use of the artificial star was not needed and could
even cause GSPC stars to be rejected by the iterative process---an
undesirable effect---so the artificial point was not used on
these plates.

Variations in the photometric sensitivity across the Schmidt
plates are caused by sensitivity variations in the emulsion, telescope
vignetting, and point-spread function (PSF) changes across the field.
We have not corrected for such effects with this release.

It should also be emphasized that the photometric pipeline, which is tuned for
point-like objects, systematically overestimates the magnitude of bright
galaxies ($R_F <18$) which do not have a stellar PSF. This limitation was acceptable since the primary purpose
of GSC-II is to provide guide stars for telescope operations. It is possible that a future version may use a different
calibration technique to provide more accurate galaxy magnitudes.

\subsubsection{Image Classification}
The main motivation for classification in the GSC series is to reliably
identify stars, which can be used for guiding, from ``nonstars'', which cannot.
At the scale of the Schmidt
plates, such ``nonstars" include not only galaxies, but blends of
overlapping images which were not resolved into individual objects by
the deblending step of the plate processing pipeline. A third category,
``artifact", is used to label images that appear on the plates but do not
correspond to astronomical objects---for example, pieces of
diffraction spikes, halos, etc. In a subsequent release of the catalog,
we plan to replace the generic ``nonstar" category by ``galaxy" and
``blend" labels, at least for the brighter objects where such
discrimination is possible. In general, nonstars can be considered
to be primarily galaxies far from the galactic plane, and primarily blends close
to the plane.

\begin{figure*}
\centering
\includegraphics[scale=0.7]{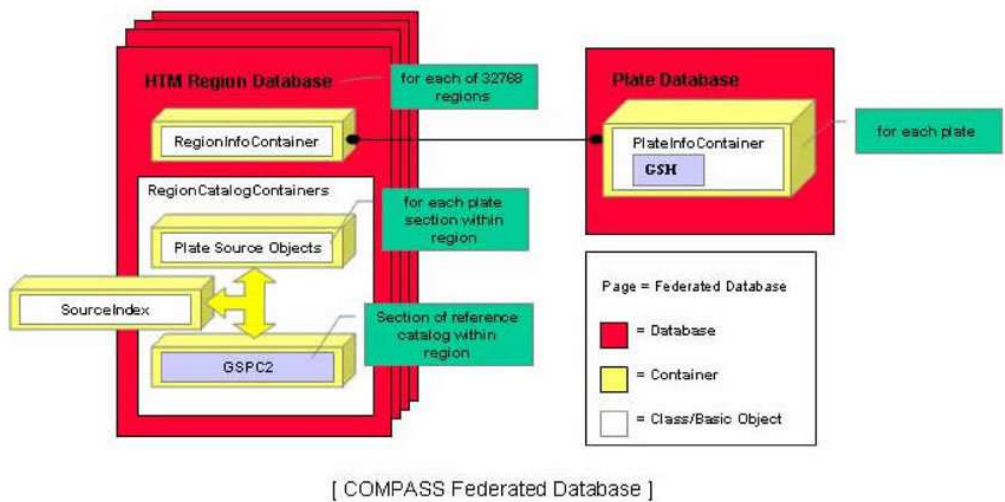}
\caption{Representation of the GSC-II database structure.
\label{DBDesignESS}}
\end{figure*}

A set of 30 features, describing the shape, uniformity, and brightness,
was calculated for each object, as detailed in Table \ref{features}.
This set was selected based on good star/galaxy separation in feature
plots.  The ``raw" features were then transformed by the statistical
process of "ranking", that is, ordering a population of objects by the
value of the feature, and mapping the full range of the feature into a
range of 0:1.  The ranked features were then used in the remainder of the
classification process \citep[see][]{1997scma.conf..135W}.
This ranking has the effect of significantly
reducing plate-to-plate variations in the features (see figures \ref{unranked} and \ref{ranked}), removing the
necessity of creating a specific training set for each plate. This can
be illustrated by considering the case of ellipticity, a common
classification feature: the roundest objects on any plate are likely to
be stars, even if one plate is poorly guided compared to the other.
The ranking was performed separately in different zones of
the plate, which was divided typically into a 7x7 grid, whereas some
especially crowded plates were gridded more finely due to memory
limitations. Objects lying near the grid boundaries were interpolated.

An oblique decision tree \citep[OC1][]{1994cs........8103M}
was used to perform
the classifications. Unlike classical, or axis-parallel, decision
trees that can split on only one variable at a time, OC1 can perform
splits on a linear combination of all the features. This permits OC1
to find and sensibly exploit relationships among the image
features. To lower the variance contribution
to the classification error, a set of five decision trees were created from
the training set of 5334 hand-classified objects based on stars and galaxies from
the deep CCD catalog of \citet{1996AJ....111..615P}, using different randomizations
in the tree-building process to produce five pruned trees from the same data.
During the plate classification
task, each object was classified by all five trees, and the results were
voted to produce a final, single plate, classification.

The training set and decision trees were tuned to produce the fewest
number of misclassifications over the entire population of objects on the
plate. Since the majority of the detected objects are faint objects, and since
bright and faint objects classify differently, this effectively meant
optimizing for faint objects at the expense of bright objects.

\section{The GSC-II Database}
The COMPASS database is the primary repository for all the data from the
plate image processing pipeline.  The GSC-II is exported from the
database as a subset of the multiband and multi-epoch surveys because
the same objects on the sky appear on different plates.  The database
establishes the associations between objects on the different surveys and provides a central
indexing scheme for which the export task can extract the calibrated
parameters for each object.

\subsection{Database Structure and Operations}
The database architecture is built on a hierarchical object-oriented
framework using Objectivity, a commercial Object-Oriented Database
Management System. We regarded as important elements for the
selection of this system its low development costs and its scalability
to multi-terabyte data sets, along with the capability of mapping the
object modeling of the raw data directly into the database schema. The
latter has allowed us to identify a unique correspondence between the
plate-based raw data and partitions of the celestial sphere (sky
regions) through the creation of a {\it federated database file system}
(see diagram in Figure \ref{DBDesignESS}), with each plate covering
multiple regions and each region containing several overlapping plate
sections.

The subdivision of the celestial sphere is based on the HTM
\citep[Hierarchical Triangulated Mesh][]{2001misk.conf..631K}, a
spherical quad-tree index library developed by the Sloan Digital Sky
Survey (SDSS) project at  Johns Hopkins University. The regions are
approximately equal area of the order of 1 square degree.  Several
large-scale archives have adopted HTM to facilitate interoperability for
data correlations in addition to providing a high-speed spatial index
into a data set.

Each of the HTM regions in the database contains a number of ``containers''
corresponding to the plate measurements, the reference catalog data (GSC-I, GSPC-II,
{\it Tycho-2}, ACT and SKY2000), and a master index which identifies each unique object
on the sky within that region and has links to the individual observations and catalog entries.

\subsection{Object Matching and Naming}
As each plate was processed and loaded into the database, a matching task
was run which identified objects on that plate with previously loaded objects.
Object matching was based  on position only and was run iteratively with search
tolerances increasing from 1 to 4 arcsec. The matching
code computed the distance with respect to all the previous observations
belonging to any source within the searching radius and assigned the new
object to the source containing the nearest neighbor. This procedure is repeated
with a larger radii on the remaining objects and unmatched objects are added as new sources to the master index.

A known problem with the matching is due to noise fluctuations in the
extended saturated cores of very bright stars resulting in the detection
of spurious objects. Typically, these bright stars are contained in the
SKY2000 and {\it Tycho-2} catalogs, so the matching of these spurious objects to
the same catalog entries is the cause of multiple entries in the GSC-II.
This is not a problem for {\it HST} operations since the {\it Tycho-2}/SKY2000 values take
precedence but we plan on fixing this issue in a later release.

In order to maintain backward compatibility for {\it HST} operations, the
master index was preloaded with the GSC-I objects so that the GSC-I
identifications were maintained.  As objects were added to the index,
an internal identification was assigned which was the HTM container
name (e.g., N012301) plus a running sequential number. This was adopted
to aid fast object lookups within the COMPASS database. It should be
noted that this name is converted to a 10-character base-36 number known
as the HSTid in GSC2.3 and later and that this is the ``official''
name. This was done because of the ten-character limit for the Guide
Star names throughout the {\it HST} ground systems and flight software once
{\it HST} pointing switched over to using the new catalog for pointing.

\subsection{The Export Catalog}

The GSC~2.3 export catalog provides positions, multicolor photometry, and morphological
classification for about one billion unique objects.
For each source, the GSC-II database may contain multiple measurements deriving from
the observations matched from the overlapping multi-epoch and multicolor plates.
In this section, the selection criteria adopted to filter the available data are
illustrated. It is important to recall in this context that the main driver for the
GSC-II project was to support and/or to improve on the efficiency of observation
planning and operations in space and ground observatories which
sponsored the project; the catalog extraction rules ultimately define the
properties of the catalog released.

The magnitude limits of the export catalog were set at
$B_J<$19.5  or $R_F<$18.5 for the first public version, GSC 2.2,
released on 2001 July, while for the GSC~2.3 almost all the
objects down to the magnitude limits of the plates (see Table \ref{GSCI-IIplates})
have been extracted from the database and included in the export
catalog.

The magnitude limits of the GSC 2.2 basically correspond
to the range which satisfies  our
{\it guide star} requirements for telescope operations in terms of positional
and photometric accuracy. In particular, the colors provided in the
export catalog are indispensable for both bright-object protection of the MAMA
detectors on {\it HST}
and adaptive optics operations of ground-based telescopes.
The classification between stellar sources and
nonstellar objects (extended objects
or blends), is critical for the selection of both {\it guide stars} and  adaptive
optics reference
stars and the GSC-II is sufficiently reliable  within the stated magnitude
range.

The deeper GSC~2.3 release was mainly motivated by our commitment
to distribute to the international astronomical community a large
and valuable data set for a wide variety of astrophysical studies.
Moreover, GSC-II data have been already used for various technical
analysis which required a complete deep all-sky catalog, such as
the guide star studies for the JWST
\citep{Spagna2001,2003AAS...202.0410S}, and the definition studies
of the Gaia mission \citep{2005ESASP.576..163D}.

\subsubsection{Source selection.}
All the database entries were screened, but only objects detected and measured at least on one
IIIaF or IIIaJ or IV-N plate were exported. Objects classed as
defect on only one plate were discarded. A magnitude threshold,
$m\le 25$, significantly fainter than the plate limits of all the
photographic surveys was applied in order to prune both faint
false positive detections and spurious objects.

In order to improve the completeness and accuracy
of the bright {\it guide stars}, which are strongly saturated in
the GSC-II long-exposure plates, the GSC~2.3 has been supplemented
with data from the {\it Tycho-2} and SKY2000 \citep{Myers2002} catalogs which contain stars down to
$V \approx 12$ and $8$, respectively. The data for any given object were taken
preferentially from the {\it Tycho-2} when available, then SKY2000 and finally from the Schmidt plate measurements.

The GSC~2.3 also contains  the names of the GSC-I
objects that have been matched with the GSC-II objects.

\subsubsection{Astrometry.}
Generally, GSC~2.3 provides positions,
($\alpha, \delta$), measured on the red IIIaF plate, and multiple
measurements are resolved by selecting the closest solution to the plate
center. In case the object was not detected on any IIIaF plate, then the
positions from the blue IIIaJ, visual IIaD, and infrared (IR) IV-N plates
have been used following this priority order. The position epoch is set
accordingly. The position errors are estimated by means of a simple
general model aimed to provide conservative uncertainties for telescope
operations. These figures do not correspond to formal
errors and cannot be used to define confidence limits in a statistical
sense.

If a source was selected from the {\it Tycho-2} catalog , the GSC~2.3 provides the positions,
the proper motions and the mean observing epoch in right ascension (R.A.),
which is slightly
different from the declination (decl.) epoch in the {\it Tycho-2} catalog.
For SKY2000 objects, coordinates and proper motions are provided at epoch J2000.0.
It should be noted that at this time we are {\it not publishing} computed motions due to a
small systematic error in the coordinates (see Section 5.3)

\subsubsection{Photometry.}
Photographic magnitudes $B_J$, $R_F$,
$V_{\rm pg}$, $I_N$,plus the blue POSS-I O are provided, if available, for
all the exported objects. In case of multiple measurements, which may
occur at the plate borders where different plates of the same band
overlap, we selected the observation closest to the plate center,
without applying any averaging procedure.  As a result of
the main source selection criterion, at least one $B_J$, $R_F$, or $I_N$
magnitude is always present, while additional blue $O$ and $V_{\rm pg}$
magnitudes are given for the objects also matched  on the POSS-I O,
Palomar Quick-V, and SERC Quick-V surveys. As for the positions,
magnitude errors are approximate and conservative estimates.

In the case of {\it Tycho-2} objects $B_T$ and $V_T$ magnitudes are provided, whilst
SKY2000 objects may provide additional Johnson UBV magnitudes.\\

\subsubsection{Source Classification.}

Classification voting was implemented on the set of observations
for an individual sky source.
For each source we have at least two classifications coming from
the plates in the different bands, but we may have up to 25 if the
object falls into a region of multiple plate overlap. The final
published classification of a matched object is the mode of the
individual classifications (ignoring artifact classification on the
assumption that any object matched on two plates is real). Unmatched
artifacts are excluded from the catalog. In the event of a tie,
e.g. as many stellar as nonstellar votes, we decided in favor of
non-stellar based on our experience that it is easier to misclassify
a galaxy as a star rather than vice versa. Individual classifications
based on older 25 $\mu$m data are excluded from the voting to ensure
consistency.

Each source was classed as star or nonstar and, in the
case of ties, nonstar classifications were selected to ensure
highest quality stellar sources. We used only the classification
of  the objects derived from the higher-resolution scans ($1.0''$
pixel);  classification of low resolutions scans ($1.7''$ pixel)
were utilized in the rare case of objects without any detection on
the high-resolution scans.  A quality flag which indicates an {\it
unanimity} voting was also added to the export catalog.
``Defects'' revealed and matched on more than one plate were classified as nonstars.

Since this catalog is primarily for the selection of {\it guide stars} and the GSC-I classification
was strongly biased toward this, then if an object  was cross-matched with the GSC-I that  classification was adopted and it superseded  the voting scheme described here.

Finally, the morphological parameters  (ellipticity and orientation angle) of
the image used to derive the positions were exported too.

\subsubsection{GSC~2.3 Export file format}
The export catalog is a distributed FITS file system based on the linear
quad-tree HTM sky partitioning scheme, with a level-6
triangulation implementation \citep[see][]{2000ASPC..216..141K}.
Each FITS file includes all calibrated celestial sources for a single
triangular section of sky (HTM level 6 leafnode), roughly corresponding
to one square degree of sky data. The file names are the
quad-tree representation of each triangular section, with the first
character being either an ``N'' or ``S'' representing north or south.

Each FITS file contains three sections, the required primary header,
an ASCII table extension (containing a lookup table that access
software can use to speed up reading data subsets), and a BINARY
table containing the actual catalog data.

Table \ref{tab:GSC23_fields} shows the fields officially released
with GSC~2.3, including object names, astrometric and photometric
parameters, object  classification, as well as some auxiliary
parameters and flags. The binary FITS tables also include several
empty fields  for future releases or containing data for internal testing;  these are
not part of GSC~2.3 and must not be used nor distributed.

As explained in the notes of Table \ref{tab:GSC23_fields}, all the
photometric fields may contain magnitudes for different, although
similar, passbands.  GSC-II users can identify the exact filter by
looking at the magnitude-code fields, for which we list in Table
\ref{tab:GSC23_pht_codes} the photometric codes used in GSC~2.3.

Further details on the processing of the GSC-II objects are  given
in the source status flag, whose structure is described in detail
in Table \ref{tab:status}.
{\it Tycho-2} stars are identified by a Quality Status Flag
of $999999nn$\footnote{Here $nn$ is the number of observations used to compute
Tycho 2 position and proper motion.} and
SKY2000 stars are identified by a Quality Status Flag of 88888800.

\subsection{GSC~2.3 distribution}
The export catalog, GSC~2.3, requires a storage capability  of
approximately 170 GB and it is currently maintained by
STScI and INAF-OATo. Copies of the catalog have been already
distributed to the project patrons (ESO, GEMINI, ESA) as well as the
main astronomical data centers and several research institutions.
Due to limited resources,
the GSC-II team does not plan to provide an extensive distribution
service of the complete catalog and requests will be reviewed on a
case-by-case basis.
However, public access to GSC-II is available through
both Web and VO-compliant (Virtual Observatory) interfaces at STScI, OATo and CDS.

\section{Catalog Properties}
The application of the "selection" criteria explained in the
previous section generated the export catalog, GSC~2.3, containing
945,592,683 objects in total, classified as stars (22\%) and
nonstars (78\%).

Table \ref{tab:GSC23_stats} presents the global statistics of
GSC~2.3, including the number of objects for which photographic
magnitudes, $B_J$, $V_{\rm pg}$, $R_F$, $I_N$, are provided, and the
number of objects related to GSC-I, {\it Tycho-2}, and SKY2000.

All-sky cumulative starcounts in the $B_J$, $R_F$, $I_N$ are provided
in Table \ref{tab:GSC23_counts}.

\begin{figure*}
\centering
\includegraphics[width=1\textwidth]{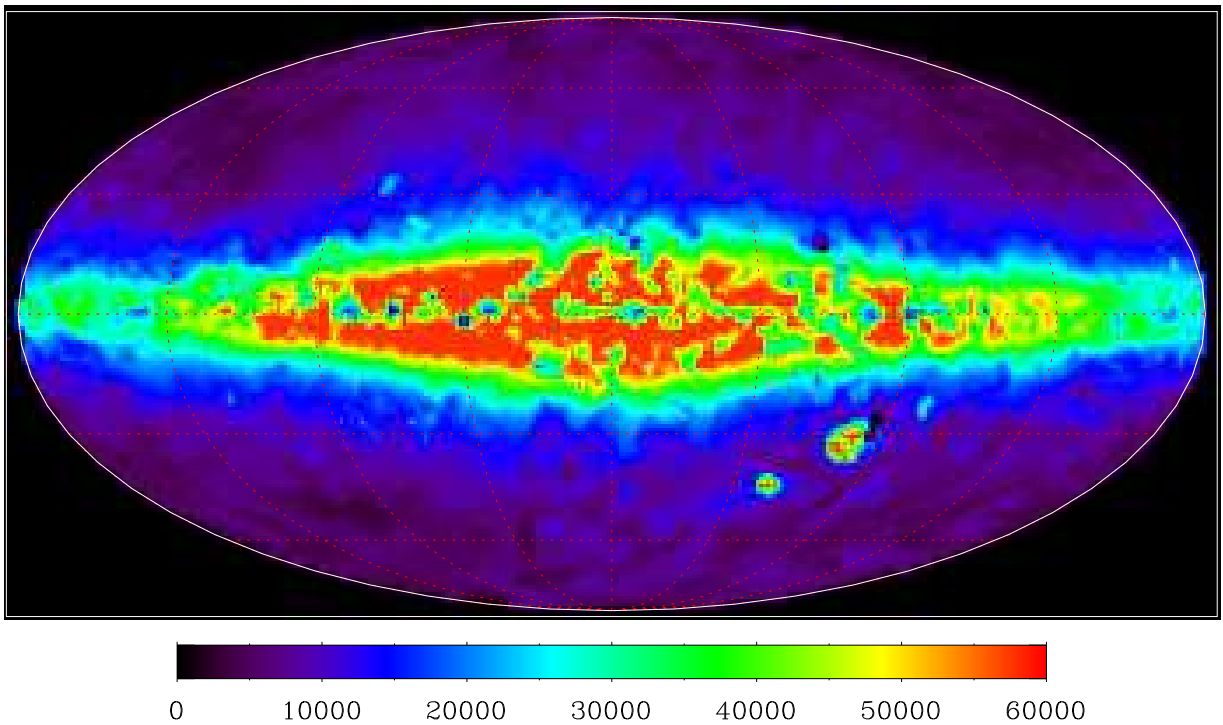}
\caption{GSC~2.3  all-sky map.  Cumulative counts in galactic
coordinates, including both stellar and extended objects.  The
color scale indicates the GSC~2.3 density ranging from 0 to 60,000
 objects per square degree. The image resolution is a
smooth version of the densities obtained from the HTM counts.
\label{fig:skymap}}
\end{figure*}

The GSC2.3 all-sky density map in galactic coordinates is
presented in Figure \ref{fig:skymap}, which includes both stellar
and nonstellar objects. The counts are dominated by stars close
to the galactic plane, and most of the objects classified as
nonstars are actually unresolved stellar blends belonging to
crowded fields toward low galactic latitudes. Also, an analysis
of the counts shows that at $R_F \le $20 about 4\% of the GSC2.3
sky has a density of above $10^5$ objects per square degree,
reaching in some cases up to $10^6$ before getting confusion
limited.

The reliability of the astrometry, photometry, and classification of GSC~2.3
can be investigated by comparisons with independent reference material as well
as by internal checks.
The latter take advantage of the significant plate overlap and can probe
the degree of catalog consistency down to the
very plate limit; nonetheless, they are not discussed here because, since
the plate-based nature of GSC-II is not preserved in the way the export
catalog was constructed, an error analysis based on the plate overlaps
would not give the right information about the GSC~2.3 errors.

\subsection {Comparison Catalogs}
\label{sec:comparison_catalogs} The UCAC~2
\citep{2004AJ....127.3043Z} contains almost 50 million objects
down to $R\approx 16$, covering the entire southern hemisphere and
a large fraction of the northern one (complete to
$\delta=40^\circ$, and up to $\delta=52^\circ$ in some areas). Its
astrometric precision is between 40 and 70 mas per coordinate
depending on magnitude, and systematic errors are within 10 mas;
these values are lower than the typical GSC-II errors, except for
the faintest UCAC~2 objects, so that this catalog appears ideal to
reveal the astrometric precision of GSC~2.3. The only drawback is
the magnitude limit being significantly brighter than that of
GSC~2.3. It is worth noting that the next release of this catalog
(UCAC~3) which is all-sky would be an excellent reference catalog
to recalibrate the plates since it is deeper than {\it Tycho-2} and
would reduce some of the systematic errors in  GSC-II that we
describe later.

In order to extend the astrometric error analysis to fainter
magnitudes we also matched GSC~2.3 with SDSS DR5
\citep{2007ApJS..172..634A}. This provides astrometrically and
photometrically calibrated data for an area of $\sim 8000$
 deg$^2$ around the northern Galactic cap; a total of about $6 \cdot 10^6$
objects including stars, galaxies and quasars are in common with GSC~2.3.
For point sources brighter than $r=20$, the astrometric accuracy of SDSS
is 75 mas per coordinate with an additional 20-30 mas systematic error
\citep[see][]{2003AJ....125.1559P}. At the survey limit ($r=22$), the astrometric
accuracy is limited by photon statistics to approximately 100 mas root mean square (rms).
The photometric calibration is accurate to roughly 0.02 mag in the $g$-, $r$-, and $i$-bands \citep{2004AN....325..583I}, so the SDSS can also be used to estimate
the GSC~2.3 photometric errors. A comparison against GSPC-II also provides a final
global validation of the photometry.

In order to compare the astrometry to an all-sky catalog of similar size, yet derived from independent
data, we also matched GSC~2.3 against the
Two Micron All Sky Survey (2MASS; \citet{2006AJ....131.1163S}) Point Source Catalog (PSC).
This lists 470,992,970 objects distributed over the entire sky,
providing near-IR photometry with signal-to-noise ratio (S/N)
better that 10 down to $J\simeq 14.2$, $H\simeq 15.1$, and
$K_s\simeq 14.2$.
 Note that the GSC~2.3 does include the bulk of 2MASS
objects, as these basically span the entire magnitude range of GSC-II,
thanks to the wide color spectrum covered by both catalogs
and the natural spread of stellar temperatures and galactic extinction.
The astrometric accuracy of 2MASS is 70--80 mas per coordinate over the magnitude range $9
< K_s \la  14$, while for brighter sources it is approximately 120 mas.
For fainter sources the error increases monotonically.
The systematic errors are of the order of 10 mas, on average, and up to
a maximum of 25 mas for the worse cases.

The USNO-B \citep{2003AJ....125..984M}
or the Southern SuperCosmos Survey \citep[SSS,][]{2001MNRAS.326.1279H} were not used as external checks since they
are derived from much of the same plate material as GSC-II, so do not provide independent measurements
of the quality. As a sanity check GSC~2.3 was compared to both USNO-B and SSS but the differences did not reveal anything
new that is considered significant.

In addition to the large-scale comparisons performed with the
above catalogs, we further checked object classification against
the selected-by-eye galaxy catalog of the NPM/SPM programs
\citep{1994gsso.conf...20K,2002yCat.1277....0P} and various NGP
surveys  \citep{1993ASPC...51..304I,2007MNRAS.378..198G}, the APM
\citep{1992IAUIn...2...31I,1996MNRAS.278.1025L} and APS
\citep{1997ASPC..127...55C} automated galaxy classification
catalogs.

\begin{figure*}[t]
\includegraphics[scale=0.6,angle=+90]{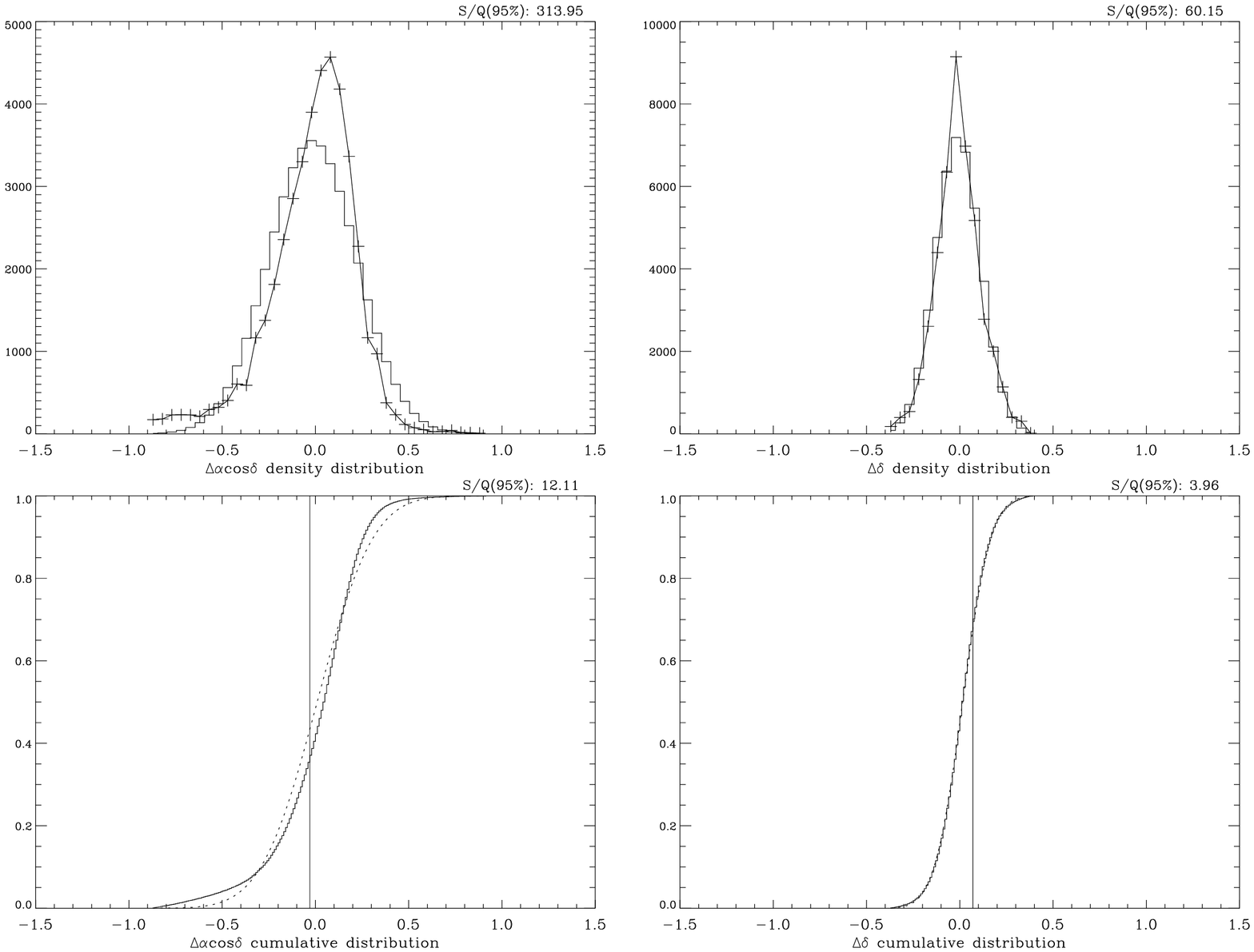}
\caption{Comparison to the UCAC~2 catalog for a sky region
centered at 0 h, spanning all declinations, and in the
magnitude range 14.5-15.0. The panels show the results of the
$\chi^2$ (top panels, the histogram is the theoretical $\chi^2$
distribution), and K-S test (bottom panels, the dotted line is
the theoretical distribution). The empirical distributions do not
satisfy the tests ($S/Q >1$, see the text). This is attributed to the
presence of residual systematic errors. \label{ucac2_all}}
\end{figure*}

\begin{figure*}[t]
\includegraphics[scale=0.6,angle=+90]{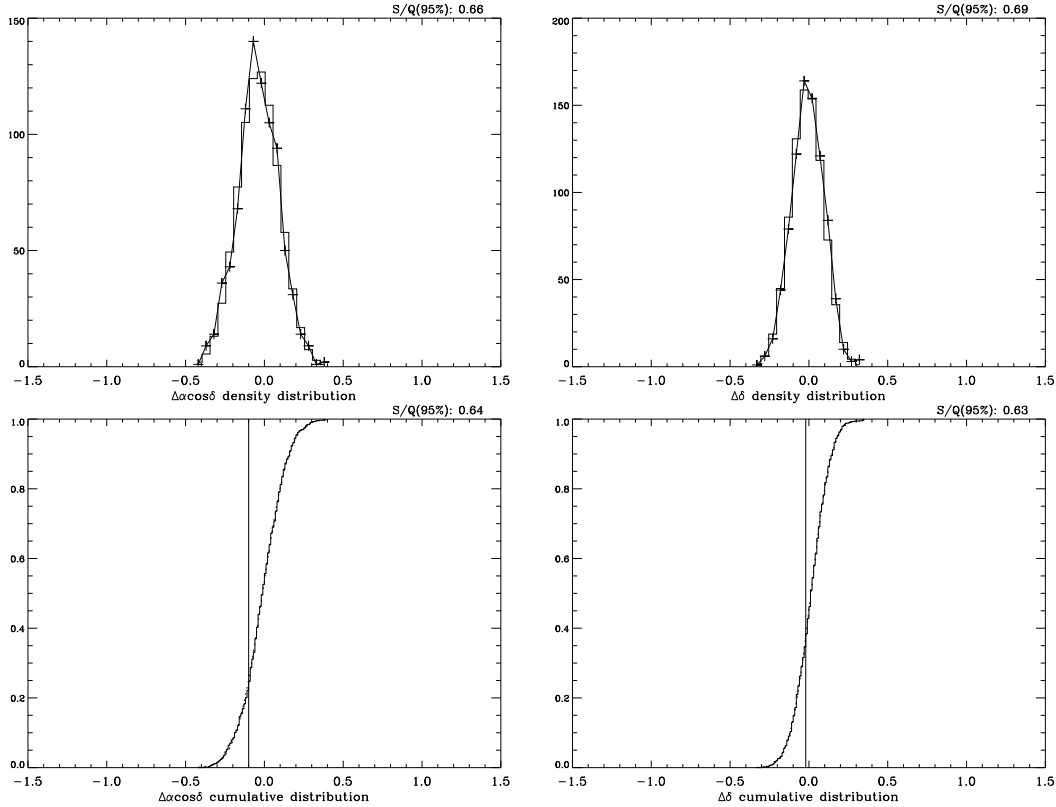}
\caption{As in Figure\ref{ucac2_all}, these panels show the results of the $\chi^2$ and
K-S tests for the GSC~2.3 versus UCAC~2 residuals, this time restricted to one plate and
to the magnitude range $R_F=14.5-15$. In this case, both significance tests are
satisfied ($S/Q < 1$).
\label{ucac2_plate}}
\end{figure*}

\subsection{Astrometric errors}
As discussed in Sect. 3.3 (astrometric calibrations), the GSC-II
is affected by systematic errors, which are function of both
position and magnitude. In the following, we investigate size and
variation of such errors by comparing the GSC2.3 astrometry to the
positions of selected external catalogs. These were calibrated
with {\it Tycho-2} stars, thus we are in a position to check at the same
time on the degree of global and local adherence of the GSC~2.3
positional system to that of the ICRF reference frame. We also
give an estimation of the positional random error for both
point-like and extended objects, which is the relevant quantity in
any application requiring knowledge of the {\it relative}
astrometric precision.

\subsubsection{Statistical tests}
A purely statistical approach to test the degree of randomness of
the astrometric errors is the method of hypothesis testing.
Specifically, we applied on our data two goodness-of-fit tests,
i.e., the $\chi^2$ and the Kolmogorov-Smirnov (K-S) tests, as
detailed in the following.

Let $\epsilon_{i1}$ and $\epsilon_{i2}$ be the actual coordinate
errors of star {\it i} from the GSC~2.3 and the comparison
catalogs, respectively. Assuming that $\epsilon_{i1}$ and
$\epsilon_{i2}$ are independent normally distributed random
variables with respective variance $\sigma^2_1$ and $\sigma^2_2$,
then their difference $d_i = \epsilon_{i1} - \epsilon_{i2}$ is
also normally distributed, with variance equal to
$\sigma^2_1+\sigma^2_2$.

This assumption on the distribution function of $d_i$ was the working
hypothesis to which the $\chi^2$ and the Kolmogorov-Smirnov tests were
applied.  Since each star has only one observation per plate, we had to
resort to an ``idealized'' experiment in order to have the required
redundancy.  To this end, we considered measurements of different stars
close in magnitude as belonging to the same parent population, i.e.,
each measure was treated as a single realization of the statistic under
test.  Such approach is correct as long as the measurement/model errors
are independent (or marginally dependent) on sky/plate position or other
effects not taken into account by the plate model; if any of these
conditions is not satisfied, the goodness-of-fit tests are doomed to
fail, whereas a successful test would confirm the existence of such
conditions.

Assuming a Gaussian probability density distribution (pdf) of the $d_i$, we estimated first the population
standard deviation by using the formula $\sigma=\sqrt{\frac{\pi}{2}}d$, where
$d=\frac{1}{N}\sum_i|d_i-\langle d \rangle|$ is the sample mean deviation, which is
less sensitive to outliers. Then, we applied a 3$\sigma$ rejection criteria
to the data and calculated the usual sample mean $\mu$ and standard deviation
$\sigma$ of the cleaned sample. Finally, by binning the residuals at 0.''05
steps, we built the test statistic

    \[\chi^2 = \sum_{i=1}^{k}\frac{(N_i-Np_{0i})^2}{Np_{0i}} \]

where $N$ is the total number of residuals, $N_i$ is the number of residuals in the i-th bin, and

 \[p_{0i} = 1/(\sqrt{2\pi}\sigma) \int_{x_i-\Delta x/2}^{x_i+\Delta x/2}
\exp(\frac{-(t-\mu)^2}{2\sigma^2}) dt \]

the probability of a residual falling into the $i$th of the $k$
bins, which can be evaluated using the {\it error function}. Note
that the $\sigma$ and $\mu$ parameters used in the integral above
are estimated from the sample itself.

The approximate distribution of $\chi^2$, valid for large samples, is that of the
{\it chi-square} probability function with $k-3$ degrees of freedom.

An alternative significance test we have applied to our sample is
the double-sided K-S  statistic, which is
exact in the sense that does not require data binning.

The K-S test statistic is defined as \[  D_k = max |S_k(x) - F(x)| \]
with

\[
S_k(x)  =\cases{ 0 & for $x < d_{1}$ \cr
  k/N & for $d_{k}\leq x < d_{k+1}$ \cr
  1 & for $d_{k}\leq x$ \cr }
\]

and

\[F(x) = 1/(\sqrt{2\pi}\sigma) \int_{-\infty}^{x}
\exp(\frac{-(s-\mu)^2}{2\sigma^2}) ds \]
\[= 0.5 + \frac{\mbox{erf}(x)}{2} \]

where erf is the error function.

\begin{figure*}[t]
\epsscale{2.0} \plottwo{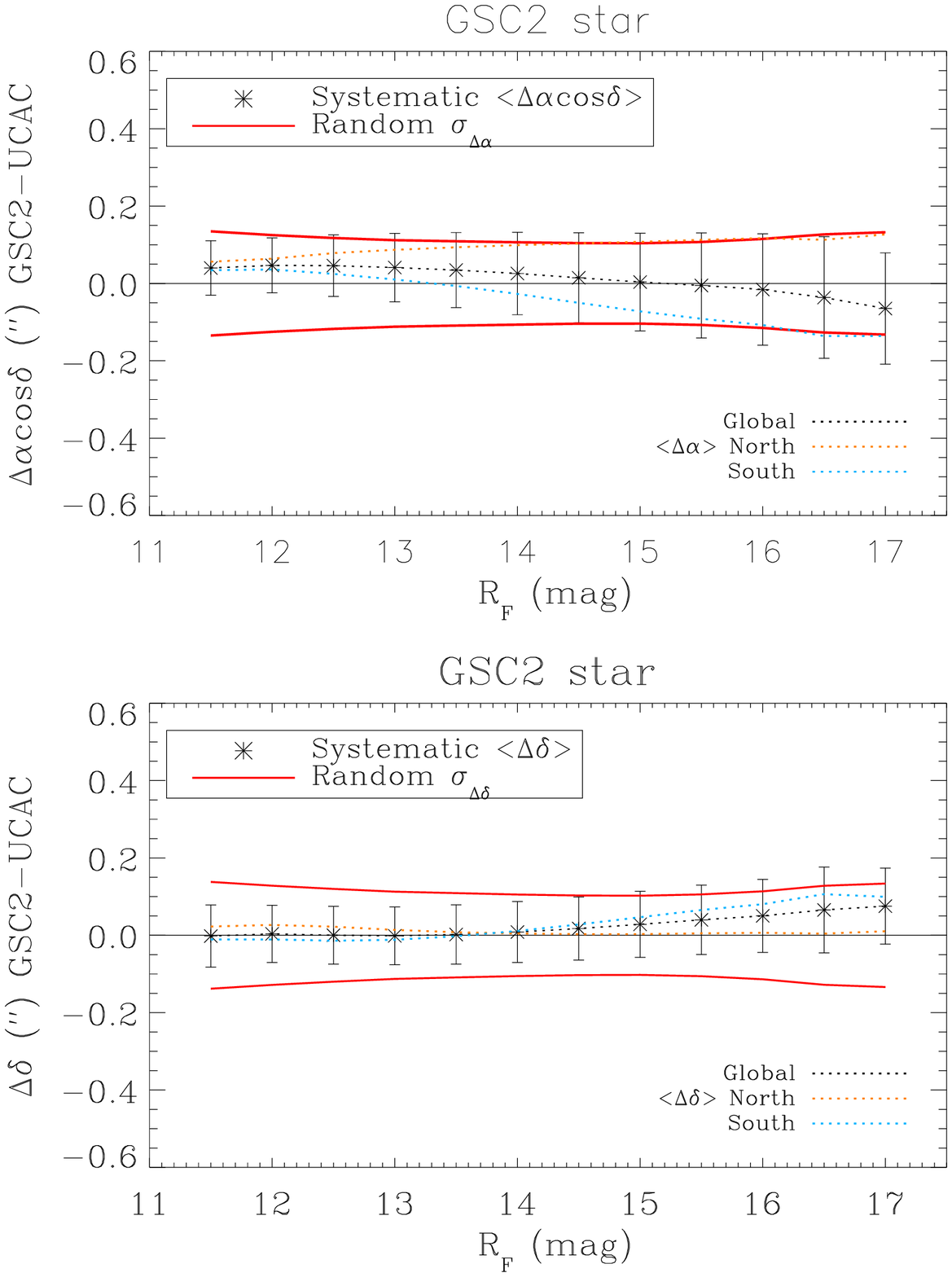}{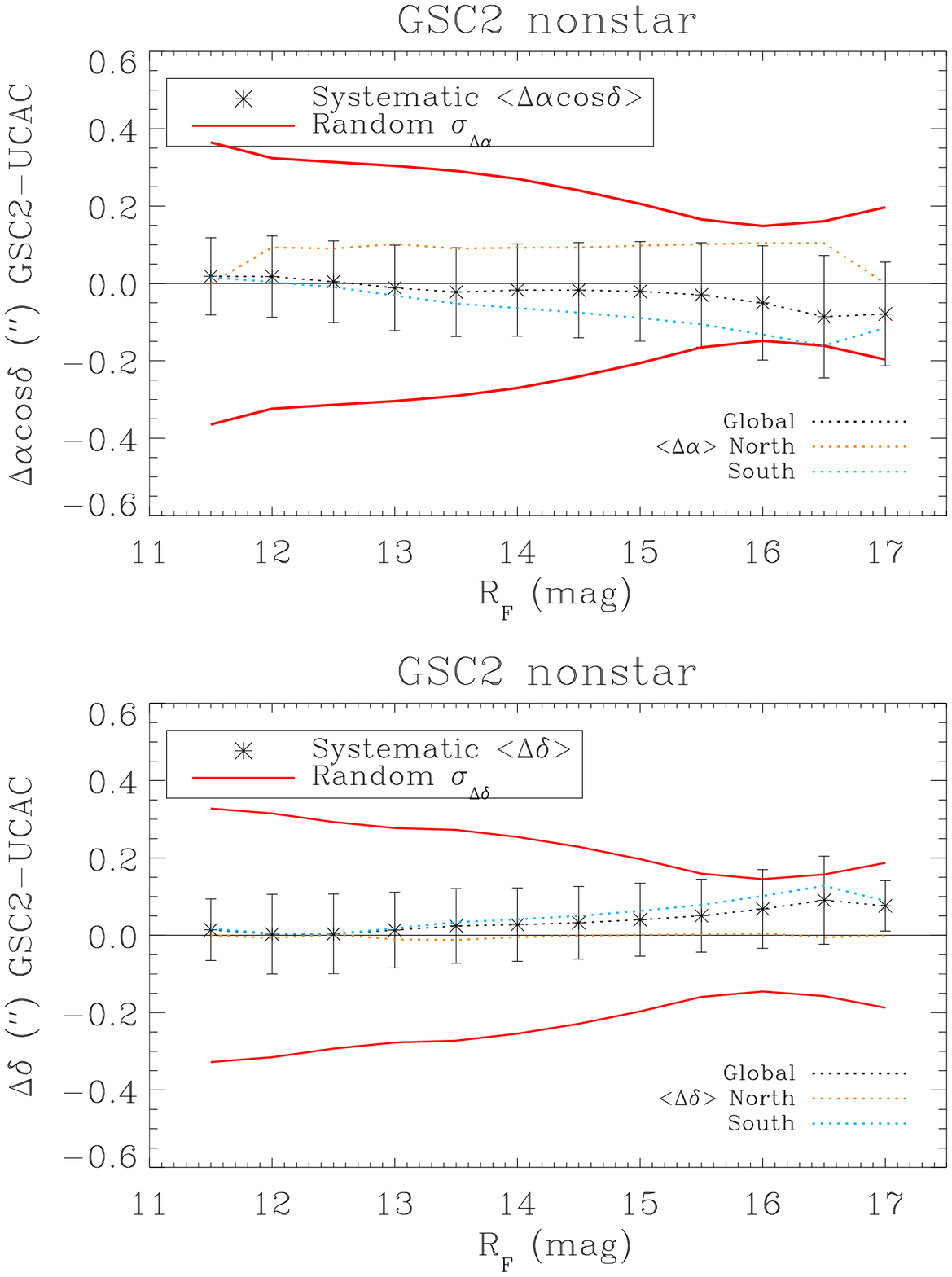}
\caption{GSC~2.3 vs. UCAC~2. Astrometric residuals of GSC~2.3
sources classified as point-like objects ({\it left panels}) and
extended objects ({\it right panels}). The symmetric red solid
lines show the random errors, $\pm \sigma_{\Delta\alpha}$, $\pm
\sigma_{\Delta\delta}$, as a function of the magnitude (see
Equation \ref{eq:random_error}). Asterisks connected by the black
dotted line show the means $\langle\Delta\alpha\rangle$,
$\langle\Delta\delta\rangle$ averaged over all the HTM regions
(Equation \ref{eq:total_systematic_error}), while the error bars
are the standard deviation of the systematic field-to-field
variations (Equation \ref{eq:field_to_field systematic_error}).
The orange and blue dotted lines show the averaged
$\langle\Delta\alpha\rangle$, $\langle\Delta\delta\rangle$ means
over the northern and southern regions, respectively.
\label{fig:GSC2_vs_UCAC2}}
\end{figure*}

\begin{figure*}
\epsscale{2.0} \plottwo{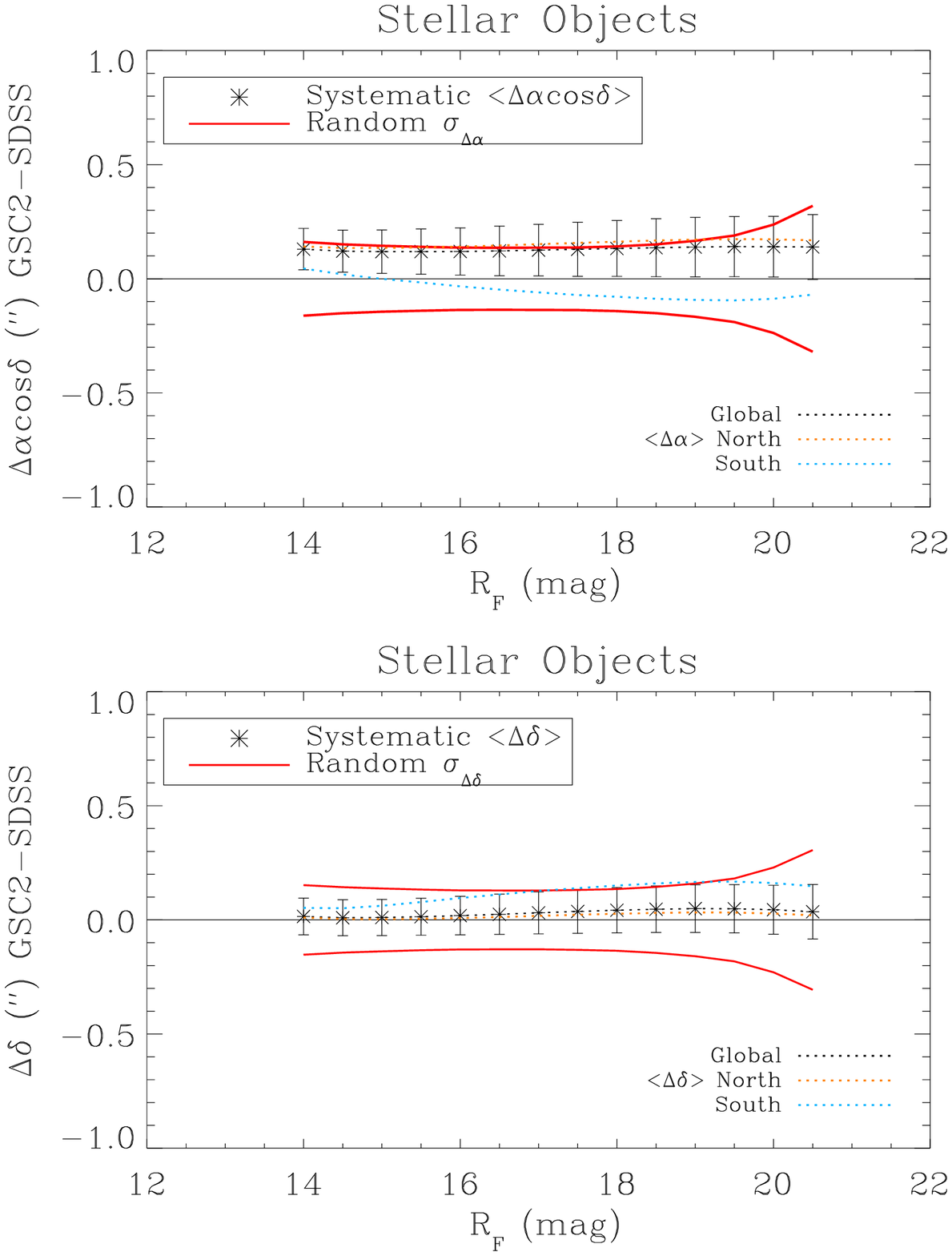}{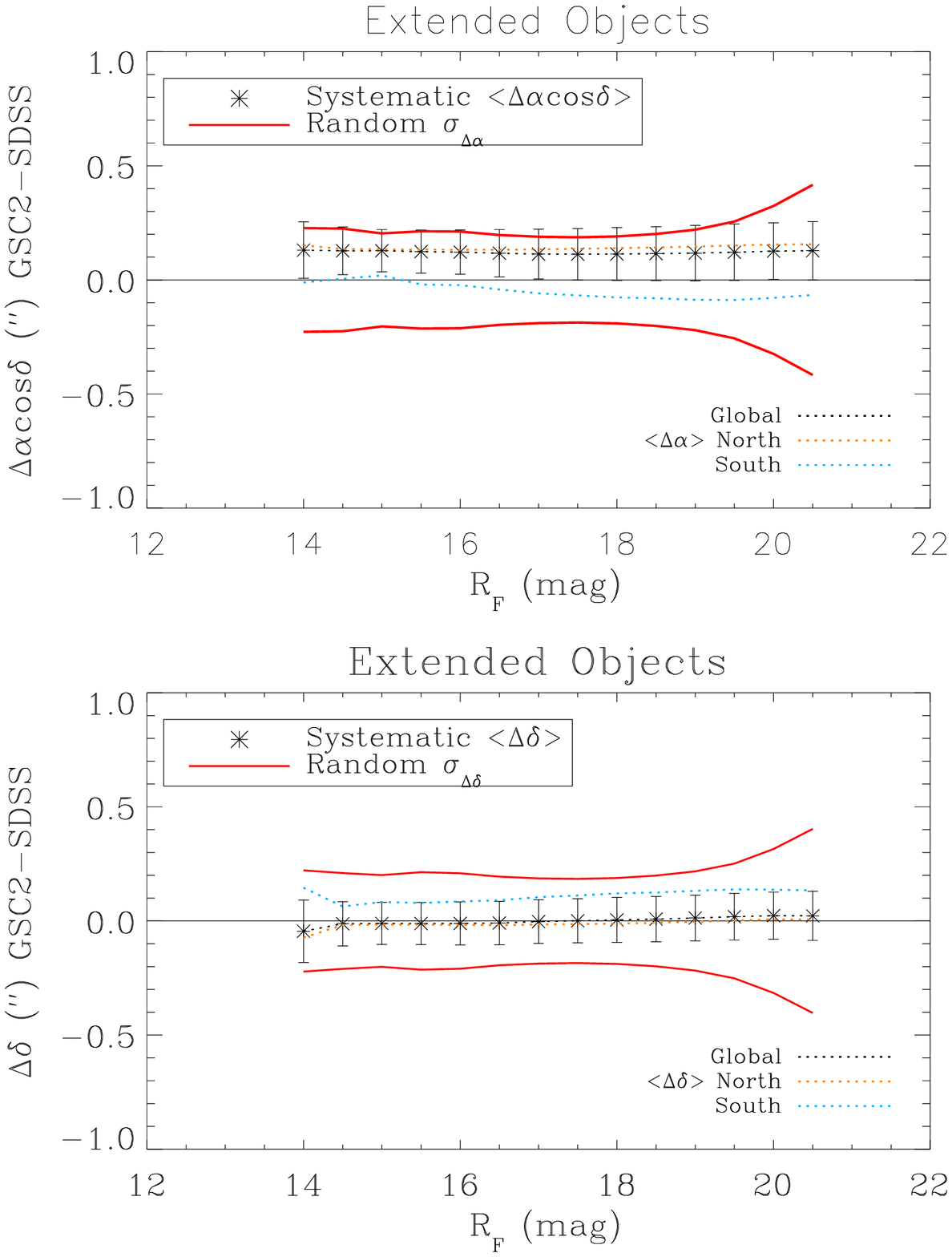}
\caption{Similar to Figure \ref{fig:GSC2_vs_UCAC2} but for the
comparison GSC~2.3 versus SDSS. Astrometric residuals as a function
of the   magnitude of GSC~2.3 sources classified as point-like
objects ({\it left panels}) and extended objects ({\it right
panels}). Same symbols as in Figure \ref{fig:GSC2_vs_UCAC2}.
Random position errors (red solid lines) are almost constant; for
stellar objects
$\sigma_{\Delta\alpha}\simeq\sigma_{\Delta\delta}\simeq 0.14$ down
to $R_F\simeq 18.5$ and comparable with the small scale
field-to-field systematic zero-point variations shown by the error
bars.  The dotted lines confirm the presence of the systematic
offset between the northern/sourthern hemispheres already found at
brighter magnitudes by the comparison against UCAC~2.  Notice that
the global mean residuals (asterisks) are very close to northern
zero-points because of the small fraction of SDSS objects with
negative declinations. \label{fig:GSC2_vs_SDSS}}
\end{figure*}

Both tests adopted a $5\%$ risk level of rejecting the null
hypothesis ($H_0$) of Gaussian distribution of the residuals, and
therefore compared the statistics derived from the sample ($S$)
with $95\%$ quantiles ($Q$) taken from the $\chi^2$ with $k-3$
degrees of freedom and the $K$-$S$ theoretical functions. If the
sample values of the two statistics, $S_{\chi^2}$ or $S_{D_n}$,
were larger than the quantiles, i.e. $S/Q > 1$, the $H_0$
hypothesis was rejected; vice versa, if $S/Q < 1$, the Gaussian
test was accepted.

We have tested the residual differences computed against SDSS and
UCAC~2 inside R.A. strips of 2 h and spanning the entire
decl. range. The results of the tests for different
magnitude bins and for a number of different plates show that,
although not strictly Gaussian, the distributions are well
behaved. As a general trend, the errors in decl. are better
behaved, and the R.A. error increases more toward fainter
magnitudes. As an example, Figure \ref{ucac2_all} shows the
results obtained by comparing GSC~2.3 coordinates against UCAC~2
on a strip of sky between 0 and 2 h and $-90 < \delta < +50$
degrees.

If we do not bin in magnitude, the distribution does not behave
like a Gaussian as would be expected from combining different
Gaussian populations; however, even limiting the magnitude range
to within $R_F=14.5-15$, which gives a total number of matched
(stellar) objects of about 44,000, both tests still fail. In
particular, the distribution of $\Delta\alpha \cos \delta$
diverges from the expected one, and presents a significantly
larger dispersion than the $\Delta\delta$ distribution. This
occurrence is indirect evidence of local systematics errors which
distorts the shape of the estimated distribution function. In
fact, if we restrict the sample to a single plate, and the same
magnitude range (862 stars sample), both tests are better behaved,
and the $H_0$ hypothesis is accepted, as illustrated in Figure
\ref{ucac2_plate}.

\subsubsection{Magnitude-Dependent errors}
\cite{1996AJ....111.1405M,2001AJ....121.1752M} found that
positional differences of GSC-I objects matched on overlapping
plate areas lead to the presence of a {\it radial} magnitude term
with the origin in the plate center. Such an effect was
characterized and removed from GSC-I by means of the Astrographic
Catalog \citep{1998AJ....115.2161U}, whose magnitude limit is
around 14-14.5, hence suitable to probe the faint limit of GSC-I.
However, applying the same precepts to GSC-II data would fail to
detect {\it nonradial} magnitude terms which we know to exist.

\cite{2003AJ....125..984M} were able to use a special catalog
(YS4.0, magnitude limit $V \sim 18$) compiled from the NPM and SPM
plate scans to detect and remove magnitude-dependent residual
errors during the construction of USNO-B by means of the mask
method \citep{1990ApJ...358..359T}.

For our magnitude error analysis, we cross-matched GSC~2.3 against
the adopted external catalogs using a search radius of 3
arcsec to recover the common objects and build the
differences:

$$
\Delta\alpha \equiv (\alpha_{\rm GSC} - \alpha_{\rm cat}) \cos
\delta_{\rm GSC}
\\
\:  \; \Delta\delta \equiv (\delta_{\rm GSC} - \delta_{\rm cat})
$$

Then, we binned the data at 0.5-mag steps and for each of
such bins computed the statistics described below.

We began with estimating the global rms of the residuals as

\begin{equation}
  \epsilon_{\Delta\alpha}^2 = \frac{1}{N} \sum_{i=1}^{N} \left(
  \Delta\alpha_{\rm i} \right)^2
  \label{eq:total_error}
\end{equation}

where $N$ is the number of objects (stellar or nonstellar) inside
a particular magnitude bin. It is evident from this definition
that $\epsilon_{\Delta\alpha}$ represents a catalog {\it average}
of the {\it total} error, as it includes both systematic and
random errors, and it is the quantity closest to the real
astrometric error. Besides, {\it relative} astrometry over small
fields (of the order of half a degree or less) is mainly affected
by {\it random } errors; therefore, next we estimated the
magnitude dependence of the random component of the GSC~2.3 error.
For each of the 32768 HTM regions we measured the {\it local}
systematics, $\langle\Delta\alpha\rangle_{r}$, and its
variance, $\sigma_{\Delta\alpha ,r }^2$ , as

\begin{equation}
    \langle\Delta\alpha\rangle_{r} = \frac{1}{N_{r}}
      \sum_{i=1}^{N_{r}} \Delta\alpha_{i}
   \label{eq:systematic_error}
\end{equation}

and

\begin{equation}
  \sigma_{\Delta\alpha ,r }^2 = \sum_{i=1}^{N_{r}}
  \frac{\left( \Delta\alpha_{i} - \langle\Delta\alpha\rangle_{r} \right)^2}{(N_{r}-1)}
  \label{eq:variance systematic_error}
\end{equation}

respectively, with $N_r$ being the total number of objects per
0.5 mag bin in region $r$. Consequently, we estimated the catalog
{\it random} error as the weighted mean of the standard deviations
in each HTM region, i.e.,

\begin{equation}
  \sigma_{\Delta\alpha}^2 = \frac{1}{N} \sum_{r=1}^{N_R}N_{\rm r}\sigma_{\Delta\alpha ,r }^2
  \label{eq:random_error}
\end{equation}

where $N_R$ is the total number of HTM regions. Outliers due to
mismatches were rejected, and GSC~2.3 entries flagged as {\it Tycho-2}
stars excluded. Also, magnitude bins with poor samples (less than
20 matched objects) or very crowded samples (more that 5000
objects) were not used in this analysis. Analogous quantities were
evaluated for the residuals in decl.

The statistics in Equations \ref{eq:total_error} and
\ref{eq:random_error} are summarized in Table
\ref{tab:GSC23_SDSS_astro}, which reports the results of the
comparisons with UCAC~2 for the bright-to-intermediate magnitude
range and with SDSS for the intermediate-to-faint range. We note
that the SDSS astrometry appears more precise than 2MASS at the
faint limit, and therefore more representative of the actual
GSC~2.3 error, although for only a portion of the sky.

The data confirm the better behavior of the point-like objects
(``stars''), which attain a random error of
$\sigma_{\Delta\alpha}\simeq \sigma_{\Delta\delta} \le 0.14$
arcsec at intermediate magnitude, $14 < R_F < 18.5$,
increasing up to $0.23$ arcsec at the faint end ($R_F \approx
20$). As expected, larger residual errors are obtained for
extended objects (``nonstar''), which may include galaxies and
nebul{\ae}, as well as many blends and unresolved binaries, in
particular at bright magnitudes or in crowded fields toward low
galactic latitudes. In this respect, we note that the higher
errors of the extended objects in the UCAC~2 comparison, as
opposed to the SDSS comparison, are explained by the dominance of
blended images in the former sample, whereas the latter is mostly
made of truly extended objects, being the SDSS catalog confined to
high galactic latitudes.

The total rms errors, $\epsilon_{\Delta\alpha}$ and
$\epsilon_{\Delta\delta}$, which also include  the contribution of
systematic errors \footnote{It is convenient to recall here that
we chose not to remove the errors of the reference catalogs
described in Section \ref{sec:comparison_catalogs}}, are of about
$0.2''$ per component at intermediate magnitudes ($14 < R_F
<18.5$), and increase to $0.3''$ down to $R_F \approx 20$.

An estimate of the systematic field-to-field variations in each
magnitude bin can be obtained by taking the standard deviation of
the $\langle\Delta\alpha\rangle_r$
around their averages computed over all the HTM regions, i.e.

\begin{equation}
  \sigma_{\langle\Delta\alpha\rangle}^2 = \sum_{r=1}^{N_{R}}
  \frac{\left( \langle\Delta\alpha\rangle_r - \langle\Delta\alpha\rangle \right)^2}{(N_R-1)}
  \label{eq:field_to_field systematic_error}
\end{equation}

where $\langle\Delta\alpha\rangle$ is given by

\begin{equation}
    \langle\Delta\alpha\rangle = \frac{1}{N_R}
      \sum_{r=1}^{N_R} \langle\Delta\alpha\rangle_r.
   \label{eq:total_systematic_error}
\end{equation}

Similar expressions hold for the field-to-field variations in
decl. The quantities $\sigma_{\langle\Delta\alpha\rangle}$
and $\sigma_{\langle\Delta\delta\rangle}$ are showed as the error
bars in Figures \ref{fig:GSC2_vs_UCAC2} and \ref{fig:GSC2_vs_SDSS}
and they are close in size to the random errors for stellar
objects in the intermediate magnitude range. The corresponding
averages, i.e. Equation \ref{eq:total_systematic_error} for R.A.,
are indicated by asterisks in the same Figures and represent a
uniform {\it weight} mean. This is a better indicator of the
GSC2.3 astrometric zero-point when compared to the straight mean
taken over all of the residuals of the matched objects. For, the
simple average tends to overrepresent the regions close to the
galactic plane.
The catalog zero point appears quite small for bright objects, as
they are close to the magnitudes of the {\it Tycho-2} stars used to
calibrate the plates, while it increases up to $\sim 150$ mas in
R.A. at 20th magnitude. However, a realistic estimate of the
GSC~2.3 adherence to the ICRF must also account for {\it local}
systematics, which can vary significantly from region to region;
one large-scale example is the northern/southern dicotomy revealed
by the orange and blue lines of Figures \ref{fig:GSC2_vs_UCAC2}
and \ref{fig:GSC2_vs_SDSS}, which depict the magnitude dependence
of the mean astrometric zero point in the northern and southern
regions. This effect will be addressed in the next section.

\begin{figure*}
\epsscale{2.0} \plotone{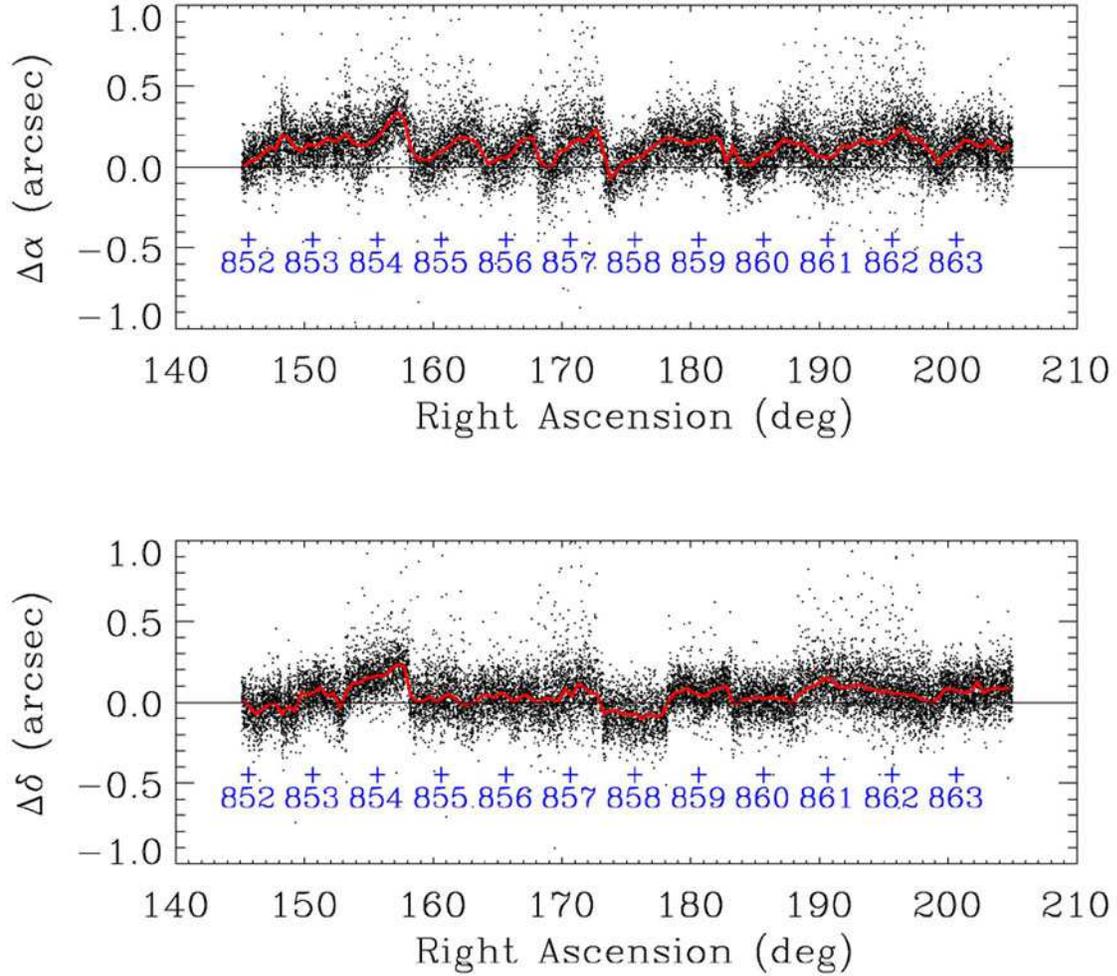}
\caption{GSC~2.3 vs. SDSS.  Astrometric residuals ({\it dots}),
$\Delta\alpha \cos\delta$ and $\Delta\delta$, for 17,507 stars
with $16.5\le R_F\le 17.0$ along an equatorial strip in the range
$145^\circ\le \alpha\le 205^\circ$. The red {\it solid lines} are
the running means of the astrometric residuals which indicate the
geometric systematic errors.  Here, the crosses indicate the R.A.
of the centers of the POSS-II fields along the celestial equator
at $\delta=0$. For this sample, the rms's of the residuals are
$0.15''$ and $0.14''$ for R.A. and decl. respectively, while the
zero-point scatter is $0.07''$ and $0.06''$ as derived by the
standard deviation of the running mean. \label{SDSS_astr}}
\end{figure*}

\begin{figure*}
\centering
\includegraphics[width=0.6\textwidth,angle=+90]{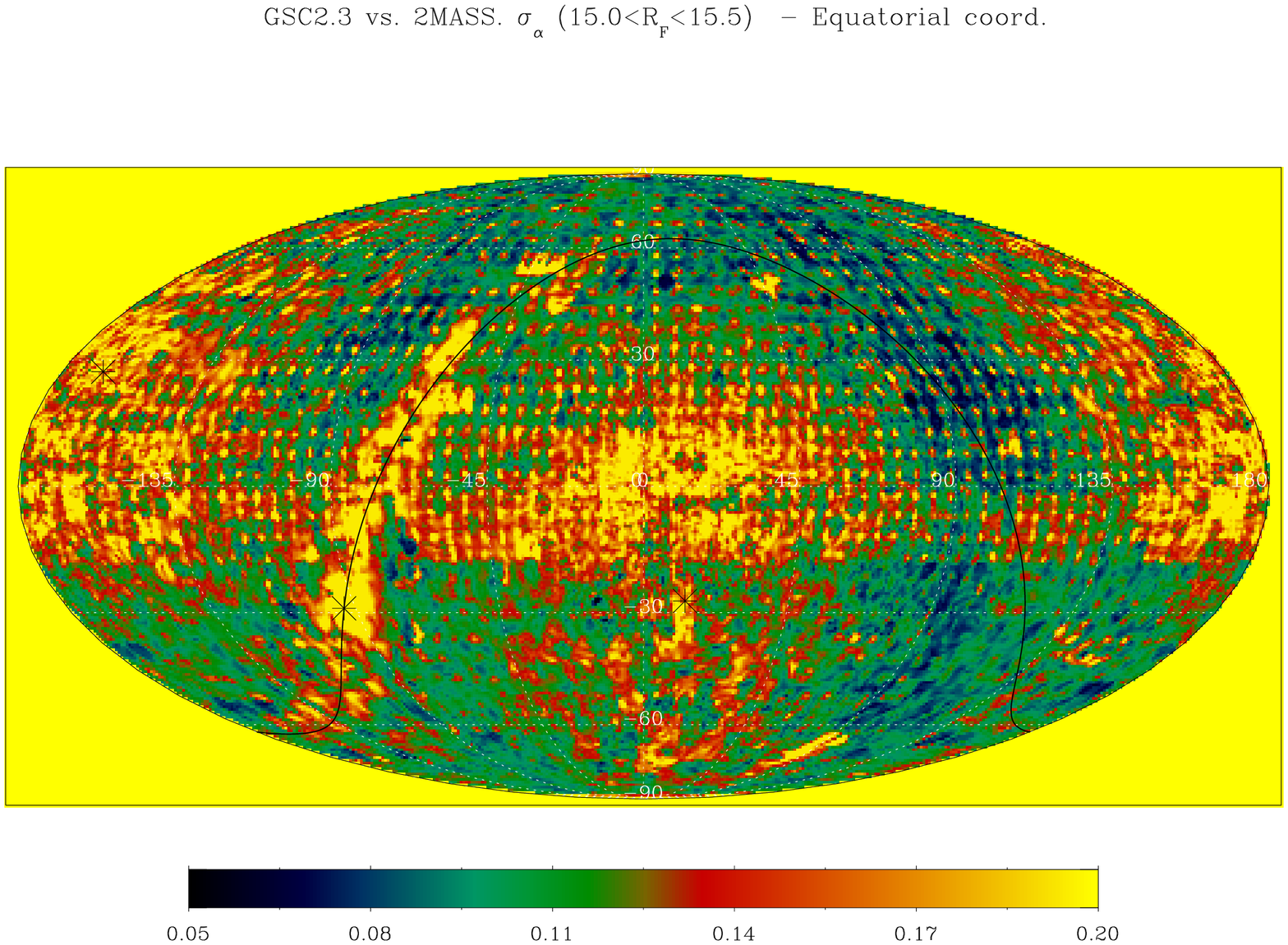}
\includegraphics[width=0.6\textwidth,angle=+90]{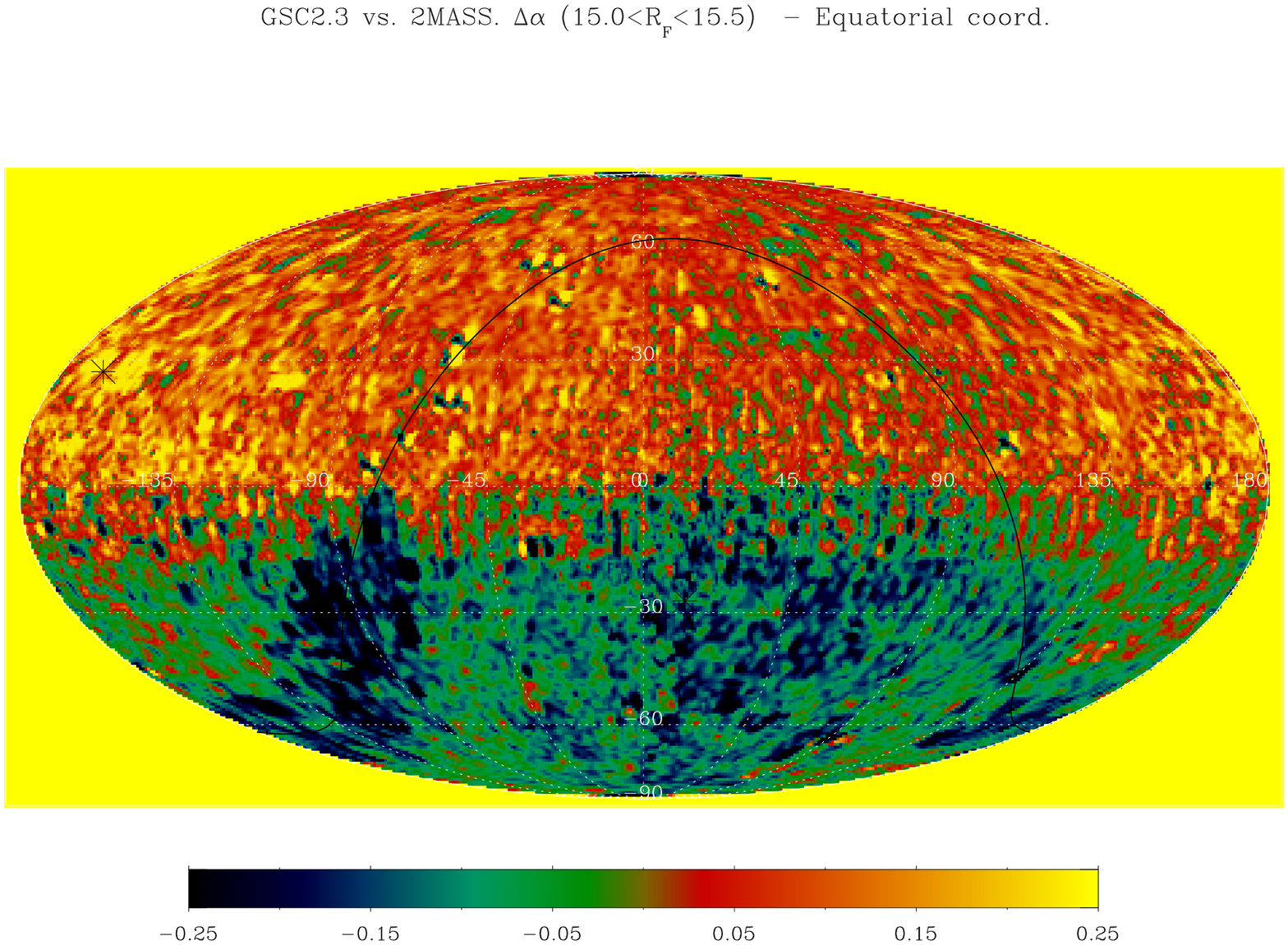}
\caption{Equatorial maps of the GSC~2.3 regional random errors
$\sigma_{\Delta\alpha ,r }$ (top panel) and systematic errors
$\langle\Delta\alpha\rangle$ (bottom panel) versus 2MASS, in the
magnitude range $15<R_F<15.5$. The black solid lines trace the
Galactic equator, while the asterisks represent the Galactic poles
and the Galactic center. Similar maps are found for the
decl. errors. \label{fig:astrometry_maps}}
\end{figure*}

\begin{figure*}[t]
\epsscale{0.9} \plotone{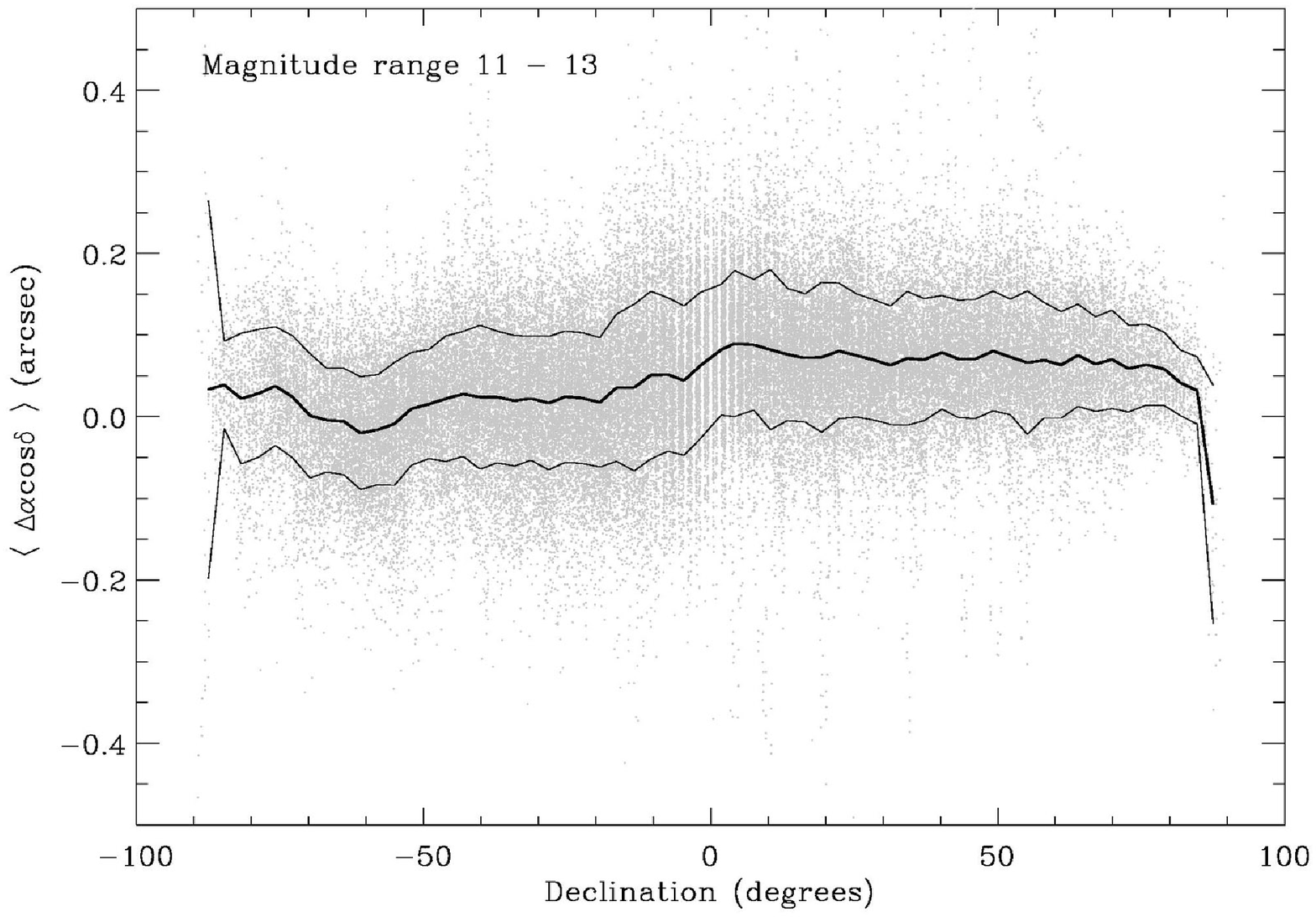}
\plotone{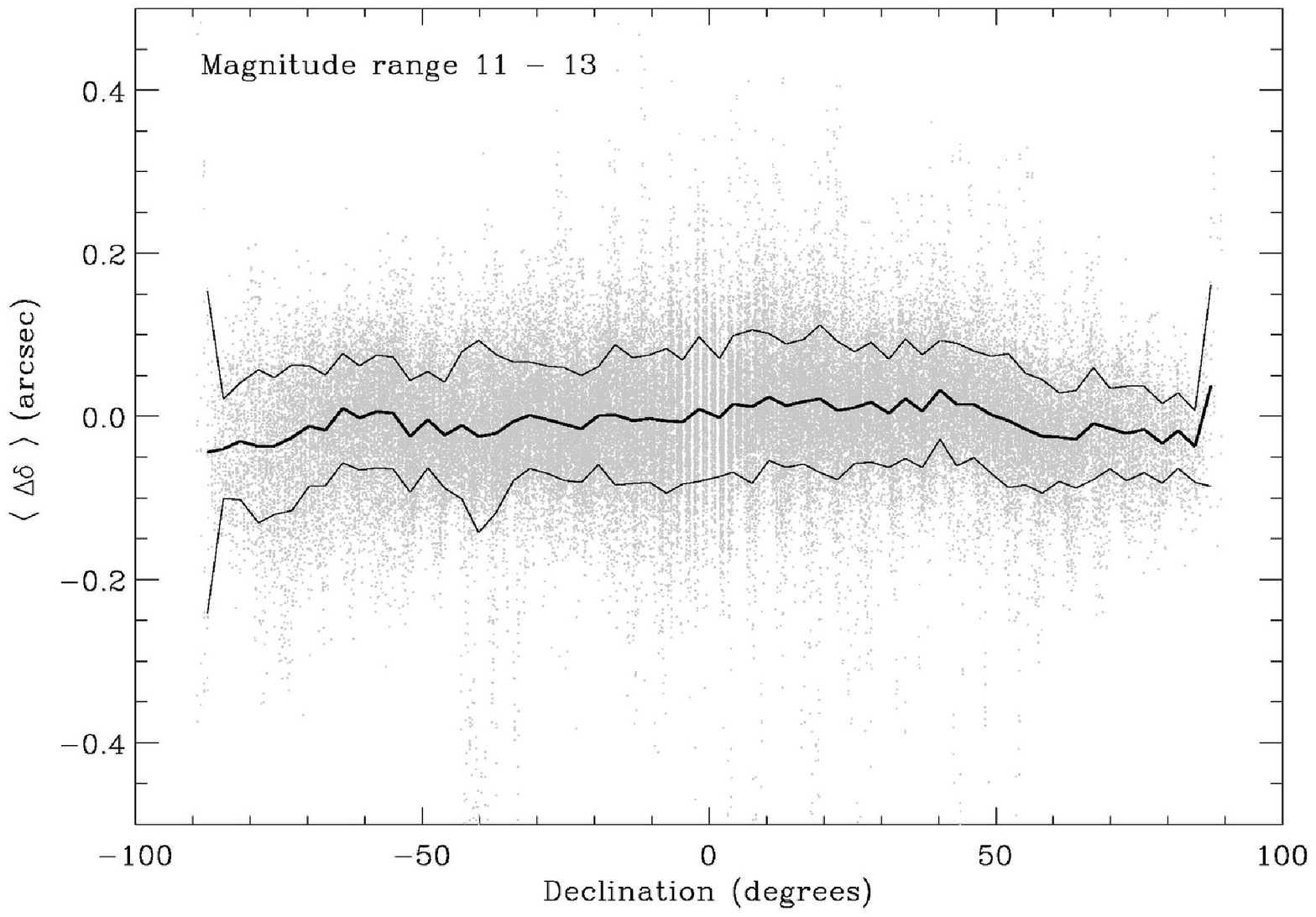}
\plotone{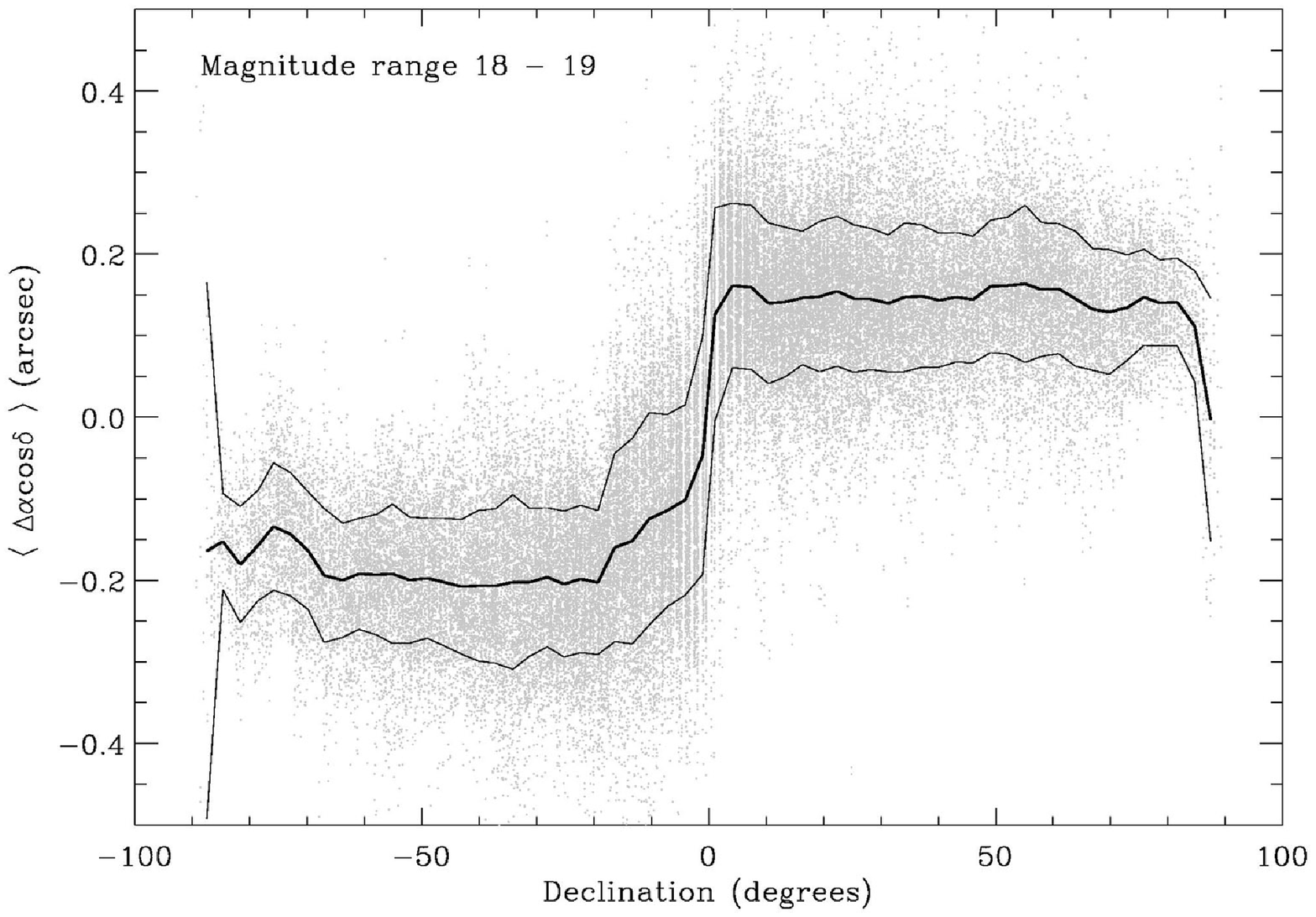}
\plotone{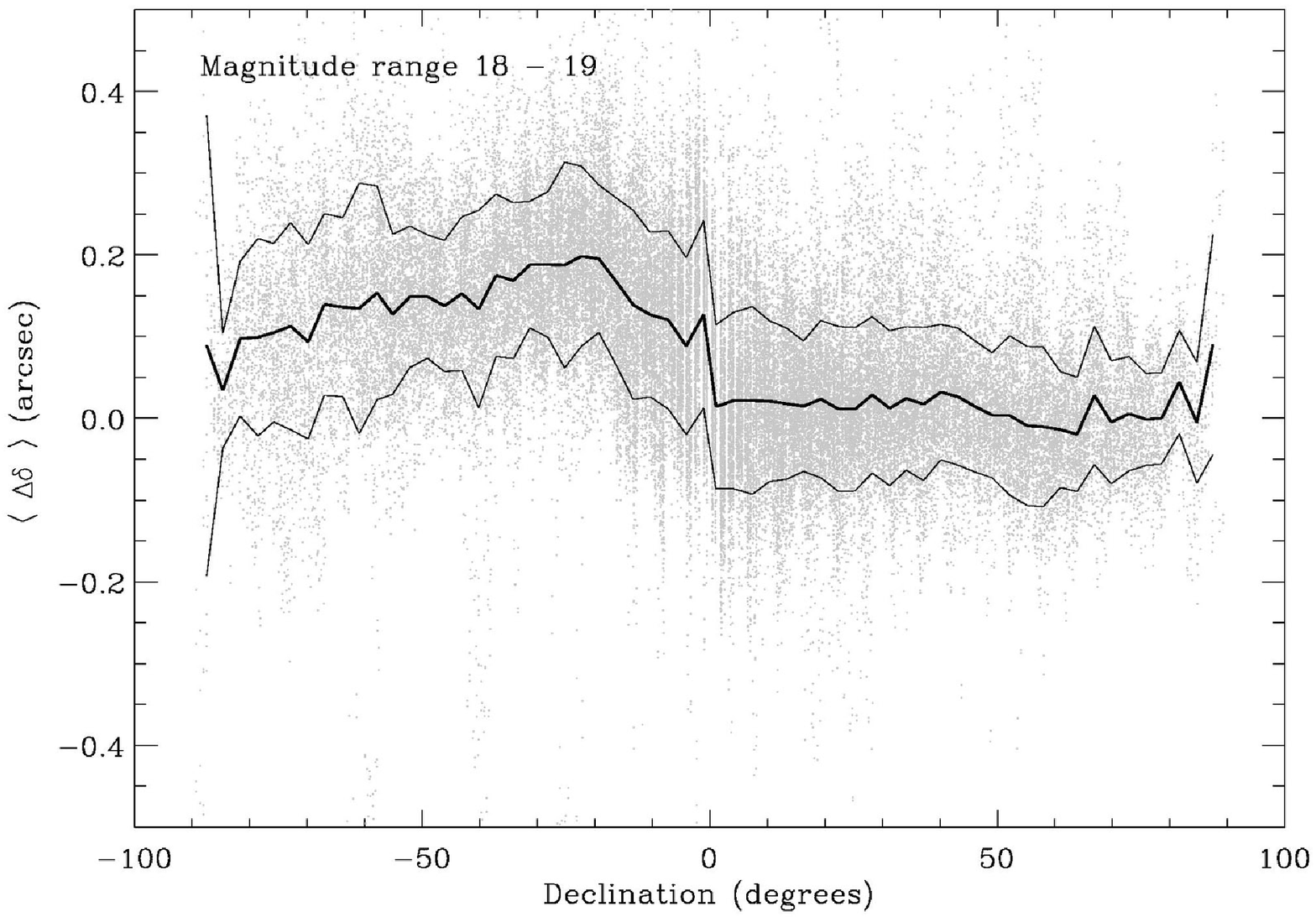}

\caption{GSC~2.3 - 2MASS R.A.\ (left) and decl.
(right) differences as a function of decl. for the $R_F$
magnitude ranges 11-13 (top panels) and 18-19 (bottom panels) .
The central solid line marks the median for the data binned in
decl., while the external ones trace the rms deviations
about each median. \label{fig:astrometry:masks}}
\end{figure*}

\begin{figure*}
\epsscale{2.0} \plottwo{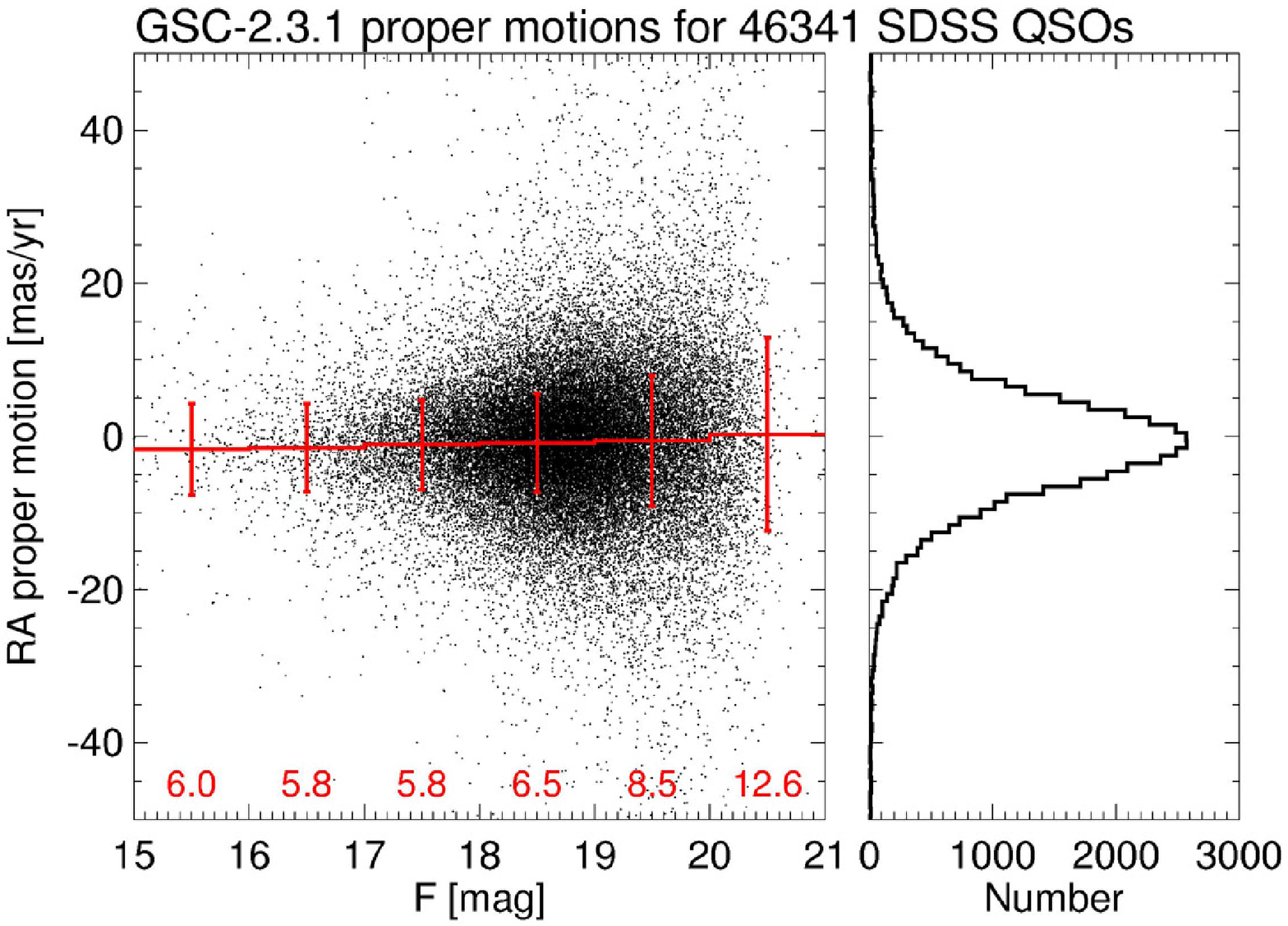}{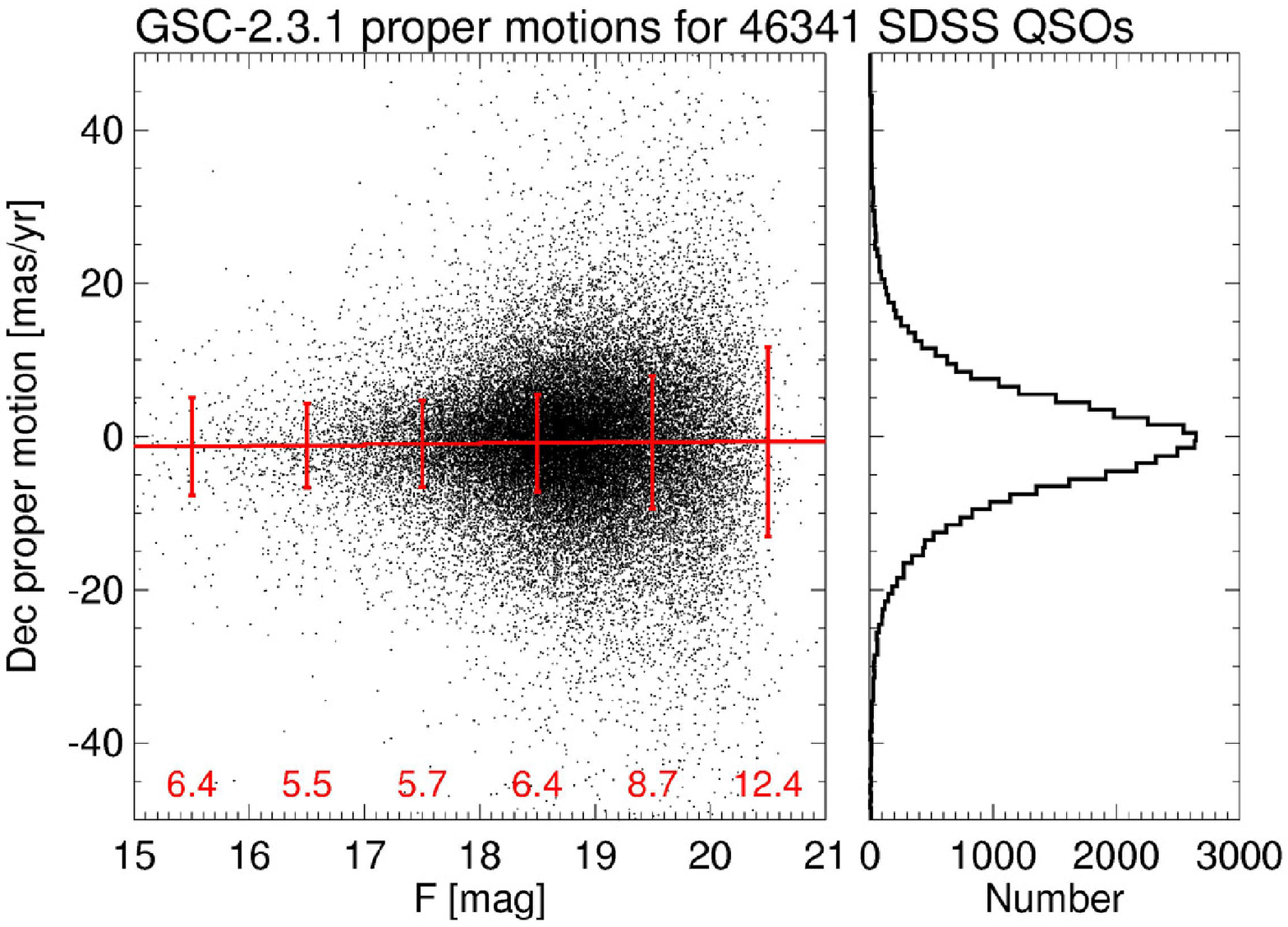}
\caption{Measured proper motions of SDSS quasars in (a) R.A. and
(b) decl. Since QSOs should have zero proper motion, this is a
measure of the proper motion errors (see text). \label{pm_sdss}}
\end{figure*}

\begin{figure}
\epsscale{1.0} \plotone{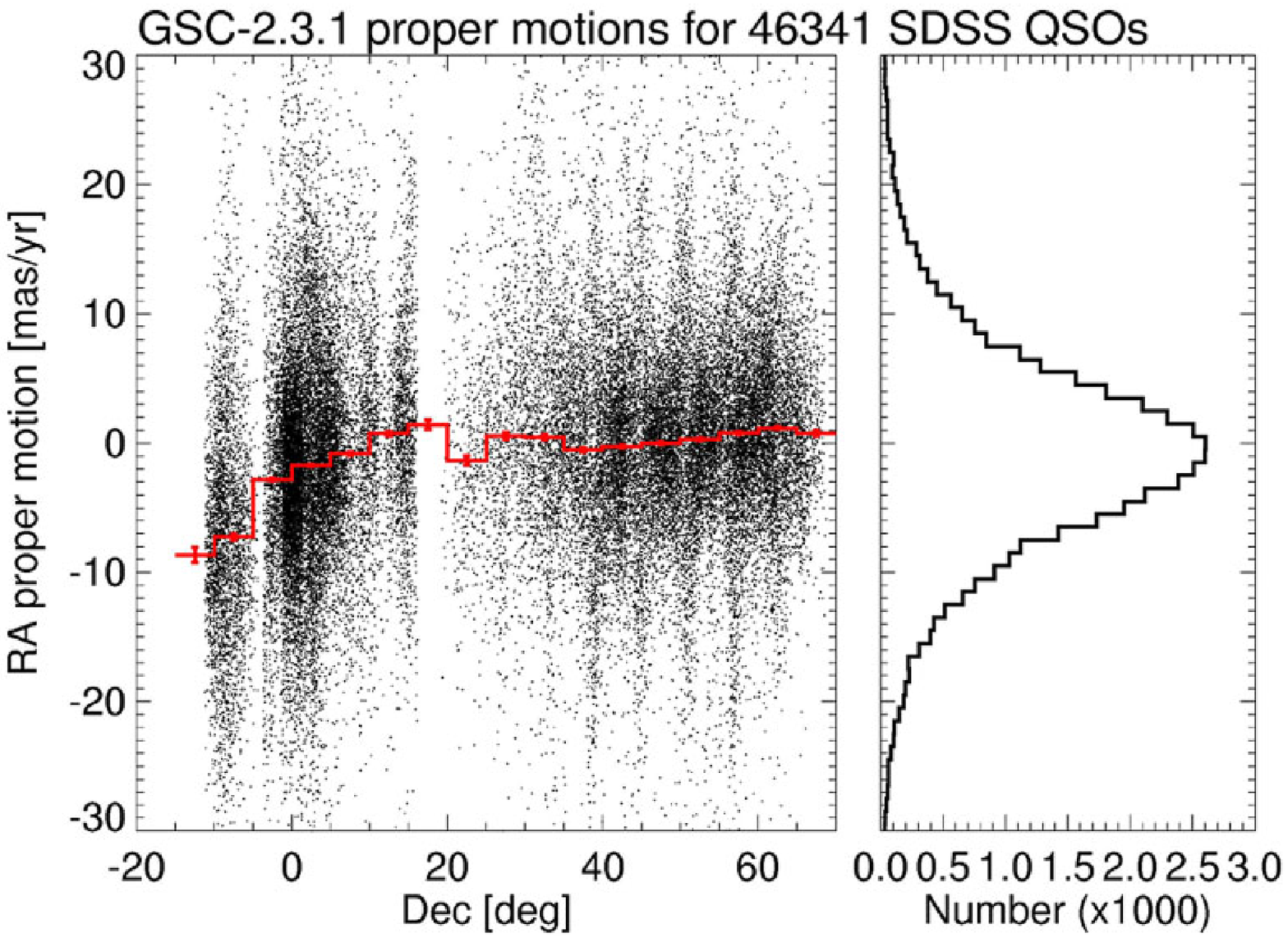}
\caption{Same
measurements as in Figure \ref{pm_sdss} but plotted as a function
of decl.\ to show the systematic proper motion error in the
southern hemisphere \label{pm_sdss_dec}}
\end{figure}

\subsubsection{Spatial Dependent Errors}
To investigate spatial-dependent errors we have checked GSC~2.3 residuals versus SDSS positions
of 17,507 stars along an equatorial strip with $145^\circ\le \alpha\le 205^\circ$
 and for the magnitude range
of $16.5\leq R_F \leq 17.0$. From the plots of Figure \ref{SDSS_astr} one can
immediately note the plate-to-plate discontinuity and the gradients within each plate.
In particular, the systematic zero-point scatter amounts to $0.07''$
and $0.06''$ in R.A. and decl.\ respectively, as derived by the
standard deviation of the running mean of the residuals ({\it red solid
  line}), while the standard deviation of the residuals is $0.15''$ and $0.14''$.

An all-sky view of the GSC~2.3 systematic and random errors
estimated by Equations \ref{eq:systematic_error} and
\ref{eq:variance systematic_error}, and utilizing 2MASS as
comparison catalog, is given by the equatorial maps in Figure
\ref{fig:astrometry_maps} for the magnitude range of $15 < R_F <
15.5$. The random error map shows a nonuniform distribution, with
significant variations with respect to the mean values
($\sigma_{\Delta\alpha}$ and $\sigma_{\Delta\delta}$) reported in
Table \ref{tab:GSC23_SDSS_astro}. In particular, we note the
poorer behavior along the galactic equator (crowding), as well as
the dependence on R.A., which might indicate effects due to poorer
observing conditions (seasonal effects and/or suboptimal hour
angles during exposures). The higher errors at the plate corners
form the peculiar speckled pattern which is most visible along the
equator where the effects of the sphere mapping geometry are
minimized.

The bottom panel shows a definite systematic difference between
north and south, which we further analyzed. To this end, we
generated plate-based vector plots of the R.A. and declination
differences versus 2MASS with a 1-mag step resolution and looked
for possible signatures of different surveys, emulsion/filter
combinations, and decl. These maps present a characteristic
pattern, whose typical amplitude is negligible in the magnitude
range of the reference catalog used for the astrometric solution,
and increases toward fainter stars up to an average of $\sim
0.25$ arcsec (at mag $\simeq$ 20). Moreover, the net residual
is positive or negative according to decl./survey telescope,
as Figure \ref{fig:astrometry:masks} clearly shows. This sign
dichotomy, which is more pronounced in R.A.\ than in
decl., can be explained by noticing that the northeast
are flipped for a northern versus southern telescope mounting, while,
given the similarities of the observing systems utilized
(telescopes, filters, plates, strategies), the systematic
components have similar behavior in plate-based coordinates.
Therefore, the net average astrometric residual as a function of
decl.\ changes in sign around the equator for both
coordinates. As expected, this systematic feature becomes more
pronounced for fainter stars, farthest from the magnitude range of
the reference catalog to which GSC~2.3 has been tied.

The analysis presented in above sections has shown that we are
dealing with a complex systematic pattern of astrometric
residuals, containing magnitude and spatial dependences
ultimately leading down to the local nature of the
observation/reduction process. It is also clear that such
systematics need to be modeled locally in order to be effectively
removed; this task will be addressed in the next GSC-II release.
Notwithstanding, the size of the GSC~2.3 astrometric errors is
such that, although they cannot be neglected for proper motion
determination, they do not impair the operational use of the
catalog.

\begin{figure}[t]
\includegraphics[scale=.35,angle=0]{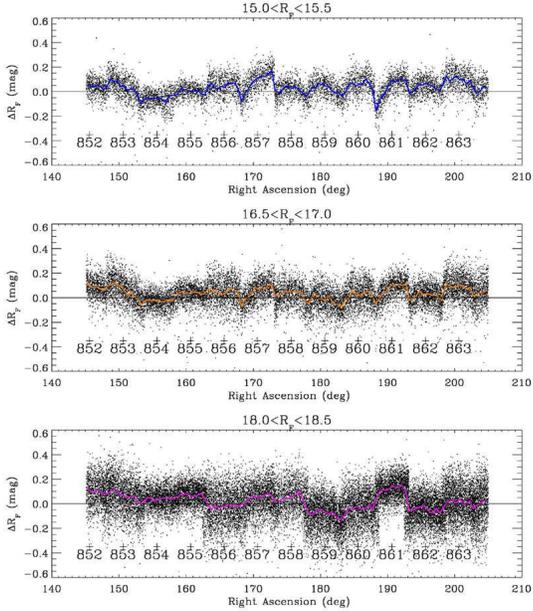}

 \caption{Photometric residuals, $\Delta R_F$, between GSC~2.3 and SDSS for a
$2.5^\circ \times 60^\circ$ strip along the celestial equator. The three panels show the residuals
of 9544, 17,507, and 29,709 stellar objects selected within the magnitude range
$R_F=$15.0-15.5,
16.5-17.0, and 17.5-18.0, respectively.
The thick solid lines show the running means of the residuals.
The crosses at the bottom indicates the position of the centers of the plates covering this
equatorial region.}
 \label{fig:photometry_systematics}
\end{figure}

\subsection{Proper Motion errors}
As mentioned in the previous section, we expected that the complex
astrometric residuals would affect production of absolute proper
motions when derived from simple fitting to the multi-epoch
positions available in the GSC-II database. In order to test this
we took a sample of quasi-stellar objects (QSOs) that had been identified in the SDSS and
computed a proper motion from the GSC~2.3 measurements at all
available epochs using a simple least-squares linear fit. The
QSOs should have zero proper motion since the ICRS reference
frame is tied to the extragalactic dynamical system. The measured
proper motions are therefore an estimate of our errors. These
results are shown in Figure \ref{pm_sdss}. At first glance it
would appear that the proper motions are determined with an rms
error of 6-12 mas yr$^{-1}$ (as a function of magnitude). However, if one
looks at these as a function of decl.\ in Figure
\ref{pm_sdss_dec} we see that, whilst the northern hemisphere
proper motions are determined to 6-8 mas/yr, there is a
significant systematic error for the southern hemisphere which is
increasing the overall error. We believe that whilst the
systematic positional errors tend to cancel out among the
different northern surveys, they do not for the southern ones.
Therefore, although we have exported GSC2.3 proper motions from
the database, these are used internally for telescope operations,
and they will not be released to the scientific community until
the systematic component is dealt with and minimized.

\subsection{Photometric errors}

GSC~2.3 photographic magnitudes result from the photometric plate
calibrations described in Sect. 3.3.2, and  from the selection
criteria which defined the export catalog, as discussed in Sect.
4. Note that the GSC-II pipeline has been tuned for point-like
sources and it is not optimized for galaxy photometry. Similarly
to the astrometry, GSC-II photometry is affected by both position-
and magnitude-dependent systematic errors. In this section, we
assess the GSC~2.3 photometric accuracy via comparisons with the
GSPC-II and the SDSS, once the reference magnitudes have been
transformed into the natural plate bandpasses by means of the
synthetic color transformations discussed in Sect. 3.3.2.

The small-scale test shown in Figure
\ref{fig:photometry_systematics} gives the residuals $\Delta R_F$
with respect to SDSS for a strip of $2.5^\circ \times 60^\circ$
along the celestial equator covered by a dozen plates. The thick
solid lines are running means which evidence a zero-point
systematic error as a function of R.A., clearly
correlated to the plate pattern. Peak-to-peak differences of about
$\pm 0.15$ mag are present, although the standard deviation of the
zero-point variation does not exceed 0.05-0.06 mag. The presence
of a mild magnitude term is also revealed by comparing the three
panels which show the residuals in the magnitude ranges
$R_F=$15.0-15.5, 16.5-17.0, and 17.5-18.0. As expected, the random
scatter around the mean clearly increases as a function of
magnitude, from  $\sigma_R\simeq 0.10$ mag to 0.14 mag of the
first and last panels, respectively.

\begin{figure*}[t]
   \centering
   \includegraphics[scale=0.6,angle=+90]{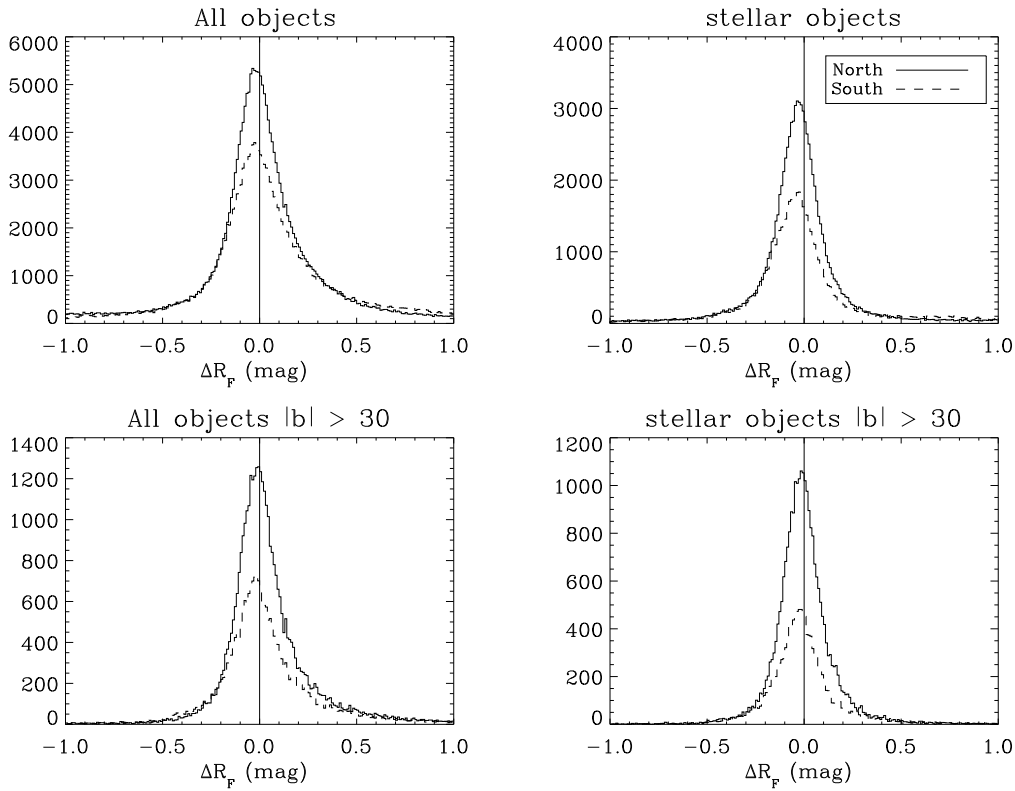}
   \caption{Distributions of the $\Delta R_F$ residuals between GSPC-2 and GSC~2.3 magnitudes for
all the sky (top panels) and high galactic latitudes  (bottom panels). The solid and dashed lines
indicate the statistics computed in the northern and southern hemisphere,
 respectively, for all the objects (left panels) and stellar objects only
(right panels).   }
         \label{fig:gspc2_resid}
   \end{figure*}

An all-sky comparison is the one against GSPC-II, which we call
semi-internal as this catalog was used to calibrate the plates,
even though a  percentage of the matched objects did not
participate in the calibration because many objects were rejected
during the pipeline reduction for various reasons (e.g. color
outside the valid range of the transformations, position close the
plate border, etc.). Figure \ref{fig:gspc2_resid} shows the
distributions of the residuals, $\Delta R_F$, between GSC~2.3 and
GPSC-II magnitudes computed for $204,322$ and  $175,364$ objects
down to the plate limit in the northern and southern hemisphere
respectively.  The distribution is well behaved, with a maximum
close to zero within a few hundredths of magnitudes and a standard
deviation of about 0.3 mag, estimated after outliers rejection at
3$\sigma$. We note an excess of objects in the tail of the
distributions with respect to a pure Gaussian, and also a slight
positive skewness formed  by objects having
$R_F^{\rm GSC~2.3}<R_F^{\rm GSPC-II}$, possibly due to unresolved binaries
which result systematically brighter on the photographic plates.
The residual distribution of the point-like objects is also shown
in the right panel; as expected, it appears similar but with a
smaller dispersion.

The bottom panels show the same statistics for objects at galactic latitudes higher than
$|b|=30^\circ$, where the crowding is less critical. Here the limits of the photometric
accuracy can be directly tested; in the case of stellar objects, dispersions attain 0.13 and
0.16 mag in the northern and southern hemisphere respectively.
The results of the statistics of all the distributions shown Fig.\ref{fig:gspc2_resid} are
listed in Table \ref{tab:GSC23_GPSC2}.

  \begin{figure*}
\epsscale{2.0}
\plottwo{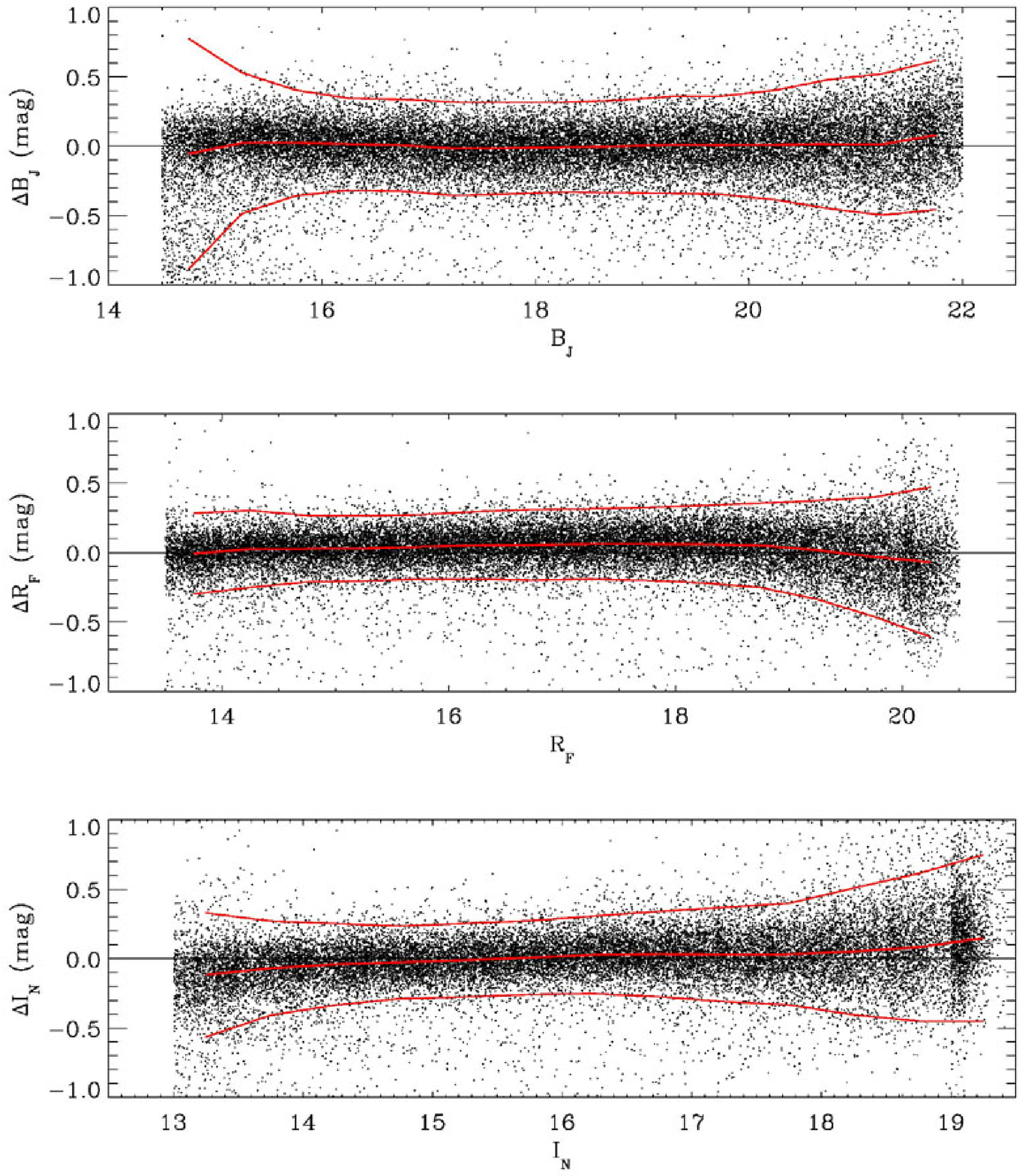}{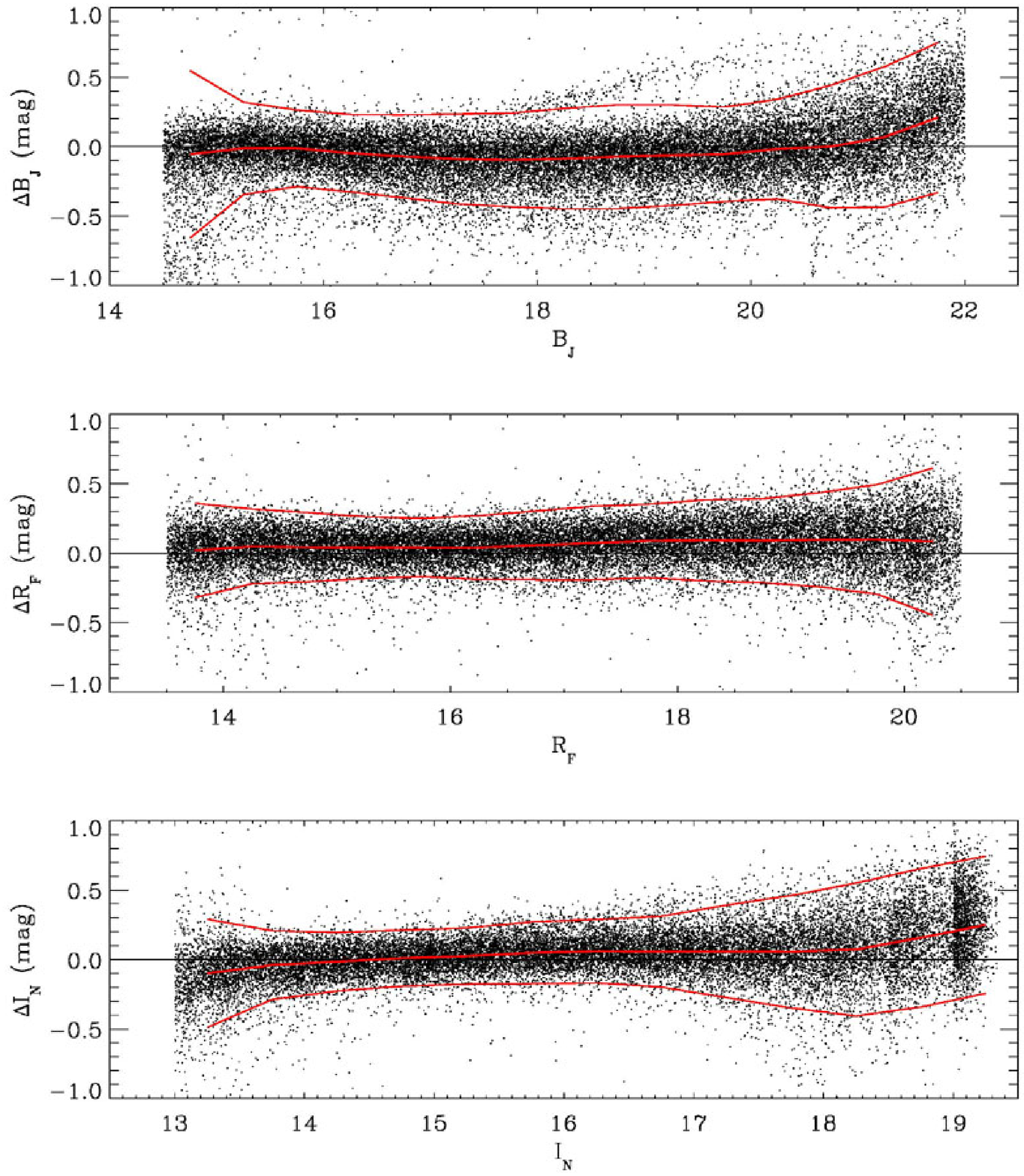}
\caption{Stellar objects in the northern (left panel) and southern
(right panel) hemisphere. Distributions of the $\Delta B_J$
$\Delta R_F$ $\Delta I_N$ residuals between GSC~2.3 and SDSS
magnitudes for stellar objects in the northern hemisphere. Stars
are uniformly distributed in magnitude, with a density of 2000
objects per 0.5 mag bin. The solid lines indicate the mean and the
boundary lines at $\pm 2\sigma$.   }
         \label{fig:sdss_resid_ns}
   \end{figure*}

\begin{figure*}
\epsscale{2.0}
\plottwo{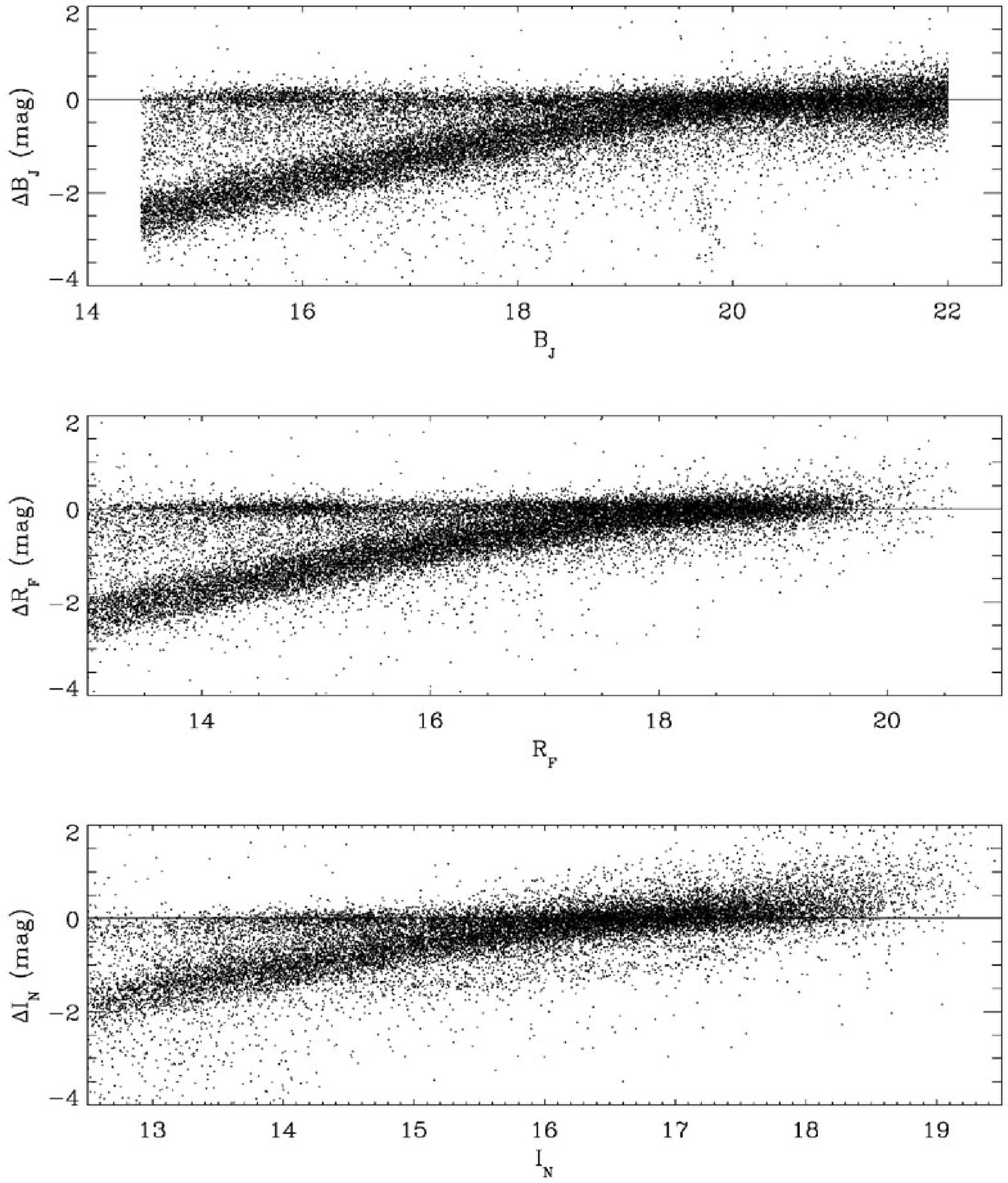}{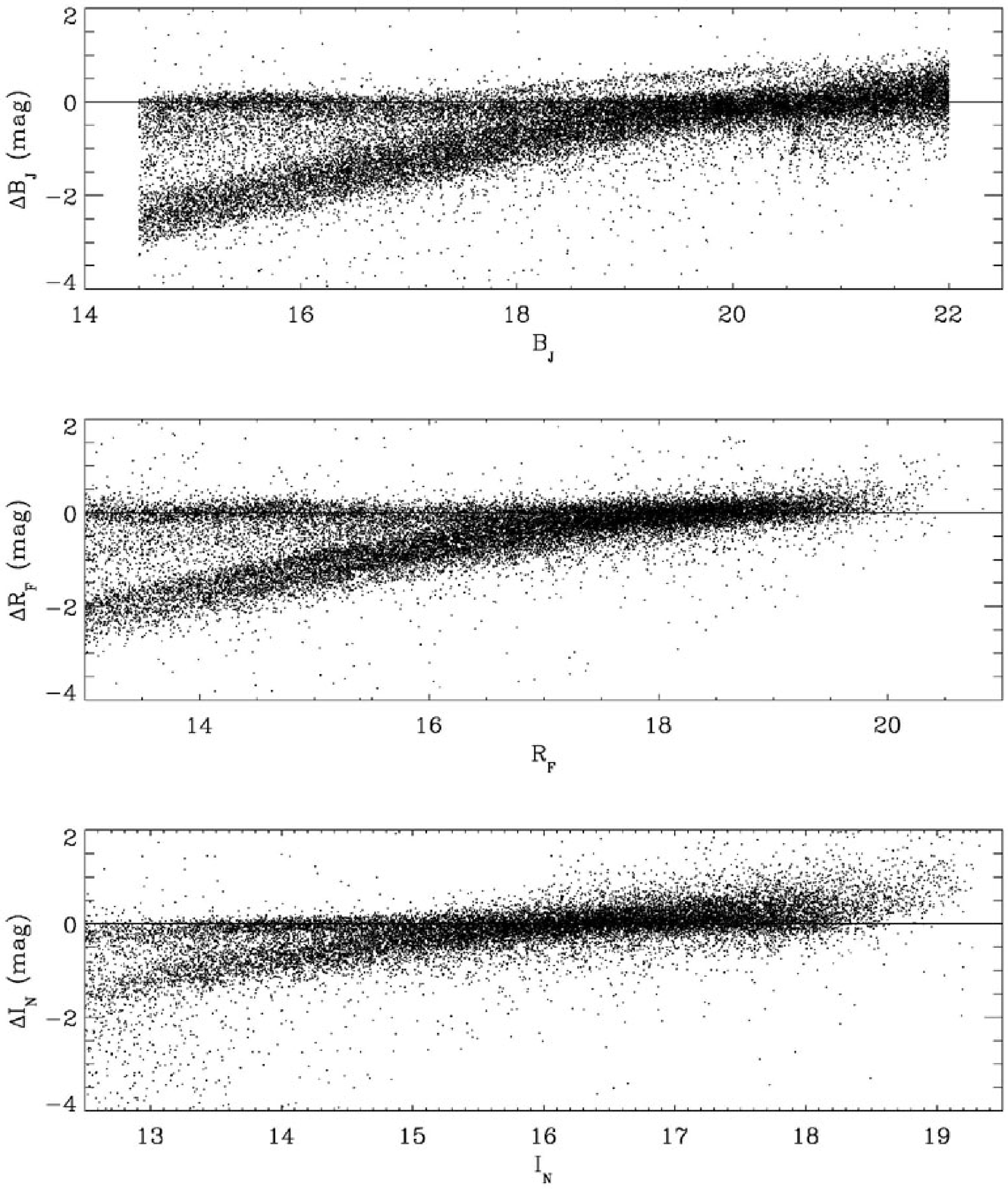}
   \caption{Nonstellar object in the northern (left panels) and southern hemisphere
(right panels). Distributions of the $\Delta B_J$ $\Delta R_F$ $\Delta I_N$ residuals
between GSC~2.3 and SDSS magnitudes for nonstellar objects in the Northern hemisphere.
   Stars are uniformly distributed in magnitude, with a density of 2000 objects per 0.5 mag bin.   }.
         \label{fig:sdss_resid_nonstar}
   \end{figure*}

An independent analysis of GSC~2.3  was done with respect to the SDSS DR5 that provides high
accuracy $ugriz$ photometry which was transformed into the GSC-II photometric system,
$B_JR_FI_N$, and compared with the photographic magnitudes.
In Tables \ref{tab:GSC23_SDSS_J} , \ref{tab:GSC23_SDSS_F}, and \ref{tab:GSC23_SDSS_N} we
report the standard deviation of the residuals of stellar objects as a function
of magnitude, in the northern and southern hemispheres.
In the north, statistics are given for a $20\%$ sample of randomly selected objects, while
all the southern matches were used. The listed values were computed after rejecting
3$\sigma$ outliers. Figure \ref{fig:sdss_resid_ns} shows the same statistics
for a subsample of 2000 objects per 0.5 mag bin, randomly drawn from the whole sample.

At intermediate magnitudes, we attain a precision of $\sigma_{\Delta B_J}\simeq 0.16$,
$\sigma_{\Delta R_F}\simeq 0.12$, and $\sigma_{\Delta I_N}\simeq 0.13$. The errors increase
for brighter and saturated stars, as well as for the faintest object up to 0.20-0.25 mag in
the second-last magnitude bin before the plate limit. Note that these values, which are
averaged on a large sky area, include both random and position-dependent systematic errors
(see Fig.\ \ref{fig:photometry_systematics}).

The global zero points of the stellar objects, $\langle \Delta B_J\rangle$, $\Delta \langle R_F\rangle$,
$\Delta \langle I_N\rangle$, averaged over the entire sky area where GSC~2.3 and SDSS DR5 overlap
(about 8000 square degrees, mostly in the northern hemisphere) are usually very small,
within a few hundredths of magnitude, as expected by local systematic errors which randomize
on large scales.

This is not true in the case of extended objects that, as shown in
Figure \ref{fig:sdss_resid_nonstar}, appear systematically
brighter in GSC~2.3 with respect to SDSS. The magnitude offset,
$\Delta m$, increases monotonically toward brighter magnitudes up
to a few magnitudes, because extended images such as galaxies or
blends are less affected by the saturation of the photographic
emulsion than point-like objects.  In fact,  $\Delta m\rightarrow
0$ for fainter sources which present densities in the linear part
of the characteristic function of the photographic emulsions.
Figure \ref{fig:sdss_resid_nonstar} also shows a certain number of
bright sources having $\Delta m\approx 0$.  Actually, these
correspond to real point-like objects which have been
misclassified by the classification algorithm.

 \begin{figure*}
   \centering
   \includegraphics[scale=0.7,angle=0]{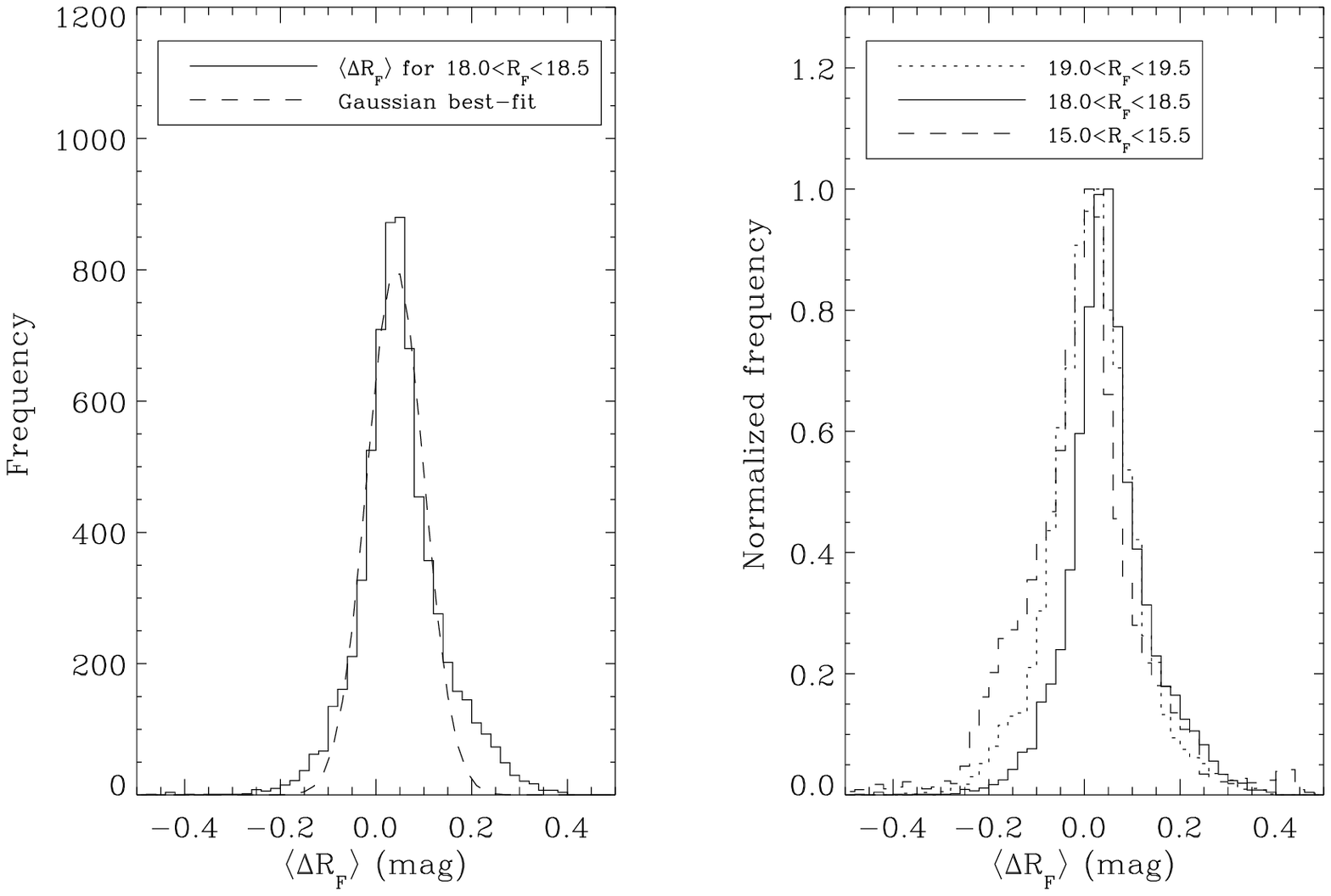}
   \caption{GSC~2.3 -- SDSS residuals.
   Distribution of the mean zero point, $\langle \Delta R_F\rangle$, of stellar objects in the sky
   area $150^\circ<\alpha<220^\circ$ and $20^\circ<\delta<
   50^\circ$.  The mean residuals have been computed within square regions
   of about $30'\times 30'$.}
         \label{fig:sdss_zero_point}
   \end{figure*}

   \begin{figure*}
   \centering
\includegraphics [width=0.44\textwidth] {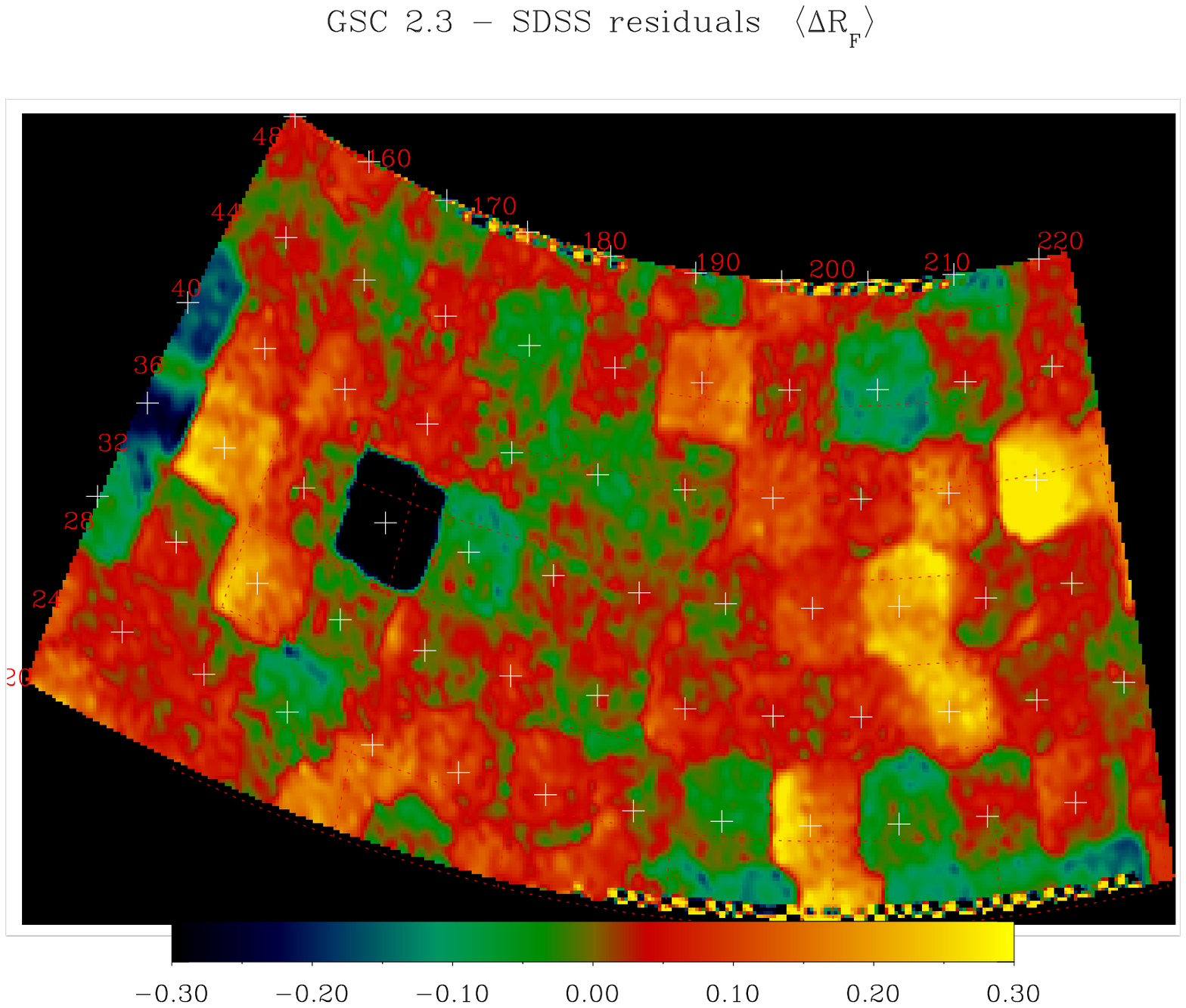}
\hfill
\includegraphics[width=0.43\textwidth] {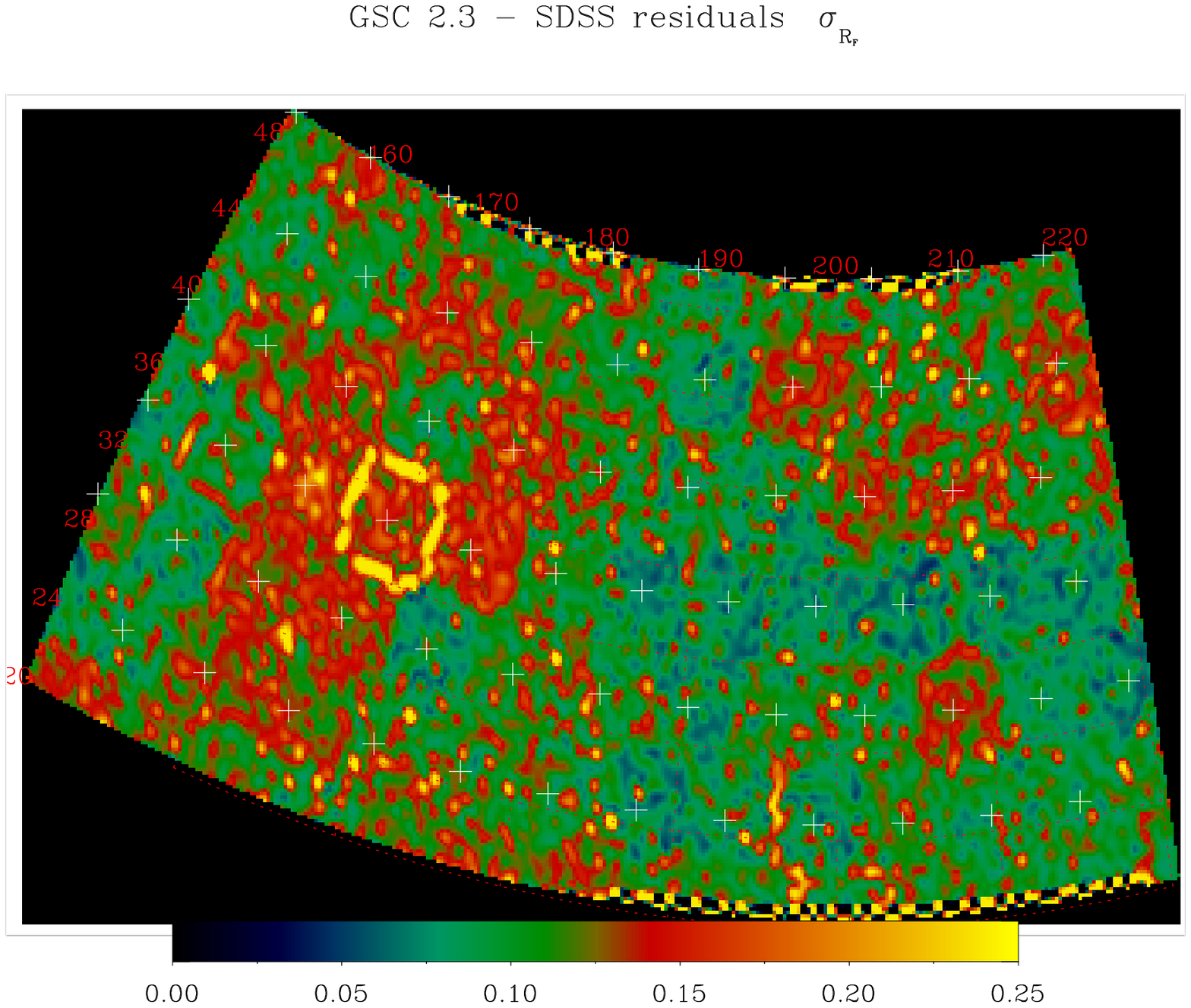}
   \caption{GSC~2.3 -- SDSS residuals. Color map of  $\langle \Delta R_F\rangle$ (left
   panel) and  $\sigma_{\Delta R_F}$ (right panel) of stellar objects measured in the sky area $150^\circ<\alpha<220^\circ$
   and $20^\circ<\delta< 50^\circ$ with a spatial resolution of about $30'\times 30'$.
   The white crosses indicate the centers of the POSS-II plates.
}
         \label{fig:sdss_map}
   \end{figure*}

Local systematic errors have been investigated in details within a
test area of 1700 square degrees selected at ($150^\circ<\alpha<220^\circ$,
$20^\circ<\delta< 50^\circ$). There,  magnitude residuals were
computed in small regions of about $30'\times 30'$ and for
different magnitude bins. The left panel of Figure
\ref{fig:sdss_zero_point} shows the distribution of the mean
residuals, $\langle \Delta R_F\rangle$, of the stellar objects in
the magnitude range, $18.0\le R_F\ < 18.5$, which corresponds to
the typical magnitude limit of the GSPC-II reference stars used for the
photometric calibration of the plates.

 The zero-point distribution  appears well behaved and roughly Gaussian, although with bigger
tails. It presents a mean of +0.05 mag and a
 standard deviation of 0.09 mag whose value represents the typical scatter of the
 local systematic errors that are present in the GSC~2.3 photometry.

The right panel of Figure \ref{fig:sdss_zero_point} compares the
zero-point distribution of three magnitude bins, $15.0\le R_F\ <
15.5$, $18.0\le R_F\ < 18.5$, and $19.5\le R_F\ < 20.0$.
 The two brightest bins show similar distributions, having the same
 standard deviation, $\sigma_{\langle \Delta R_F \rangle}\simeq 0.09$
 mag, but a relative shift of a few hundredths of a magnitude.
 The faintest magnitude bin is close to the plate limit and shows
 a zero-point distribution which appears asymmetric and with a larger dispersion,
 $\sigma_{\langle \Delta R_F \rangle}\simeq 0.12$
 mag, with respect to the previous ones. This depends on the lack
 on some plates of faint photometric reference stars needed to fit the nonlinear
 density to magnitude transformation described in Sect.\ 3.3.2.
Note that $\sim 90\%$ of the sky has been calibrated with photometric sequences down to
$R_F \ga 18$ and $B_J \ga 19.5$, while for $\sim 50\%$ of the sky the photometric sequences
attain $R_F \ga 19.5$ and $B_J \ga 20.5$.
 Careful catalog users can look at the 4th, 5th,
 and 6th digits of the source status flag (\ref{tab:GSC23_fields}) in the export
 catalog in order to check if magnitudes are calibrated by means
 of {\it interpolated} or {\it extrapolated} transformations.

Finally, the scale length and spatial properties of the
photometric systematic errors are graphically represented in
Figure \ref{fig:sdss_map} which show maps of the photometric
zero point, $\langle \Delta R_F \rangle$, and the local random
error, $\sigma_{\langle \Delta R_F \rangle}$, of stellar objects
with $18.0\le R_F\ < 18.5$ over the wide test field of about 1700
square degrees.
 The patchy pattern shown in the left panel of Fig.\ref{fig:sdss_map}
 is clearly correlated to the distribution of the photographic plates, whose centers are
 located on a grid having a step of 5$^\circ$ indicated by the
 crosses in the figure.  The zero-point scatter is generally consistent with
 the 0.09 mag standard deviation derived from the distribution
 shown in Fig. \ref{fig:sdss_zero_point}.  However, here we note
 the presence of a plate located at $\alpha\approx 170^\circ$ and
 $\delta\approx 34^\circ$ severely affected by large zero-point
 error because of the missing of a deep photometric sequence in
 this region.

 The right panel of Figure \ref{fig:sdss_map} shows the maps of
 the local random photometric error, $\sigma_{\langle \Delta R_F
 \rangle}$, which appears quite uniform, thanks to the selection
 criterions described in Sect.\ 4.2.1.  It is also interesting to
 note that a significantly higher error is only present on the
 borders of the critical plate above discussed and which overlap
 with the neighbor plates.

Similar error statistics have been found for the other GSC-II passbands.

\begin{figure*}
\centering
\epsscale{1.9}
\plottwo{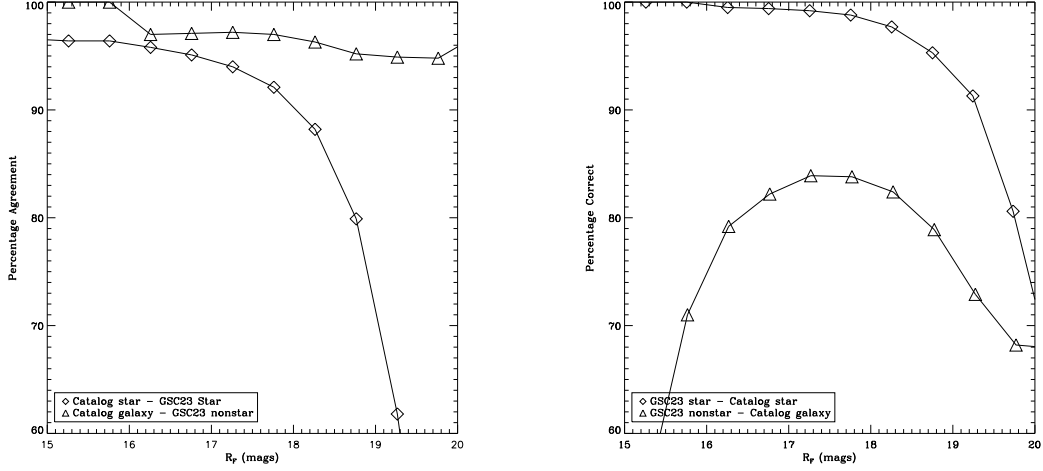}{mp_classification.ps_page_2}
\caption{{\it Left panel:} ratio between the number of correctly classified GSC~2.3 stars/galaxies
 and the total number of matched SDSS stars/galaxies, as a function of $R_F$
magnitude (completeness);
{\it Right panel:} ratio between the number of correctly classified GSC~2.3 stars/galaxies and the
total number of GSC~2.3 classified as stars/galaxies matched with SDSS as a function of  $R_F$ magnitude (sample purity).
\label{class1}}
\end{figure*}

\begin{figure*}
\centering
\epsscale{1.9}
\plottwo {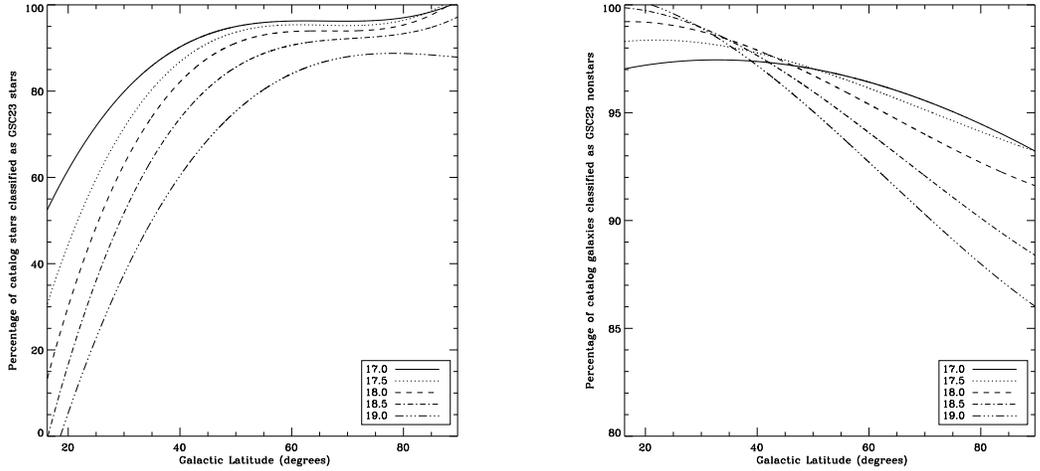}{mp_classification.ps_page_4}
\caption{Probability of a GSC~2.3 classification agreement with the SDSS ({\it left:} stars,
{\it right:} galaxies) as a function of galactic latitude, and for different magnitude ranges.
\label{class2}}
\end{figure*}

\subsection{Classification}
The reliability and efficiency of the classification procedures in the GSC~2.3 are
a complex interaction of many effects: plate-based imaging variations,
external galaxy/star distributions, internal calibrations such as object
magnitude determinations, etc.  A complete performance evaluation  is
beyond the scope of this discussion and we limit ourselves to finding the
reliability of the stellar classification because this is what is crucial
for the operational uses of the GSC~2.3.

As already mentioned, we have tested our classification against various
external catalogs; for this discussion we will limit
ourselves to the SDSS DR5 comparison but note that the results with other
catalogs are similar.

In the SDSS there are eight classifications which, in addition to star and
galaxy, include cosmic rays, defects, star trails, and other
artifacts. We consider here only the objects labeled as galaxy and
star. For each HTM region with more than ten matches we calculated the
stellar classification success probability by comparing the number of
SDSS stars classified as GSC~2.3 stars with the total number of SDSS
stars and similarly for galaxies and nonstars.

We limit our comparison to objects in the magnitude range $R_F=15$--20 as
objects brighter than 15 are dominated by stars, and for objects fainter than 20
we do not have enough pixels to classify. In the left panel of Figure
\ref{class1} we plot the
median probability for the sampled regions as a function of $F$ magnitude. The
percentage of stars correctly classified drops with magnitude from better
than 90\% at 18.0 to less than 50\% at 20.0, while the probability
that a galaxy is classified as nonstar in the GSC~2.3 remains above 90\%
for all the magnitude range.

We expect a variation in galactic latitude but to examine this we restrict the
magnitude range to $R_F=17-19$ where the number of galaxies and stars is
approximately equal and the probability of a correct classification is better
than 80\%.  In Figure \ref{class2} we show the variation of the probability
of a correct classification as a function of galactic latitude and
magnitude. The probability of a correct classification for stars drops
drastically from better than 80\% at latitudes greater than 50 degrees to less
than 40\% for latitudes less than 20 degrees. The galaxy/nonstar agreement
remains better than 80\% but is worse at the poles than at the plane.

As previously stated, the purity of the stellar sample, i.e. that an object classified as a star
is indeed a star, is important for {\it guide stars} use in telescope operations. This leads
to the strategy of adopting a classifier that conservatively classifies
objects as ``stars", but accurately classifies extended objects as
``non-stellar". These properties are reflected in Figure \ref{class2}; the low
``accuracy" of the stellar classifier at faint magnitudes is a result of the
classifier being conservative, which does not affect operations negatively as
the number of stars is ever increasing at fainter magnitudes.

The right panel of figure \ref{class1} plots the reliability or purity of
the stellar
classification as a function of magnitude. This is calculated by comparing the
number of objects classified as stars both in SDSS and GSC~2.3 to the total
number of GSC~2.3 stars. From this we can see that our stellar
classification is reliable at the 90\% level to fainter than 19th magnitude.

\subsection{Completeness}
The magnitude limit of the Schmidt plates on which the GSC-II is
based is nominally 20.5 in $R_F$, so that the catalog is expected to be complete
to at least $R_F = 20$.
This completeness limit is confirmed at high galactic latitudes
using the SDSS (Drimmel et al. 2007, in preparation) along
an equatorial strip covering more the 300 square degrees; more than 99\%
of the SDSS point sources are successfully matched to a GSC~2 object down to
magnitude $R_F
\approx 20$ using a search radius of 5'', while more than 97\% of
the SDSS extended objects were matched at magnitude $R_F = 19$.
However, in very crowded fields near the
Galactic plane the GSC-II may suffer incompleteness at magnitudes
brighter than $R_F=20$ due to crowding effects.
For this reason completeness in the GSC-II is not
uniform over the sky.

To assess the completeness limit for each HTM region we use a
model-independent approach based on the assumption that, for a small
magnitude interval (1--1.5 mag),
the slope of the logarithmic magnitude
distribution, $\partial \log n(m)/\partial m$ ($n(m)$ being the object field density at magnitude $m$, is nearly constant and positive,
consistent with the expectation that the number of objects increases
at ever fainter magnitudes. The constancy of the slope is clearly an
approximation; for galaxies this assumption is very nearly true for
sufficiently large areas of the sky, while for stars the slope changes
slowly. In any case, these assumptions are reasonable for the
magnitudes we are considering, whether we are considering stars or
galaxies or all objects together. Thus, we bin the objects in 0.5 mag bins to magnitude $R_F=20$ for each HTM region and apply the following procedure to estimate the completeness limit of each HTM region.

\begin{figure}[t]
\begin{center}
\includegraphics[width=1\linewidth]{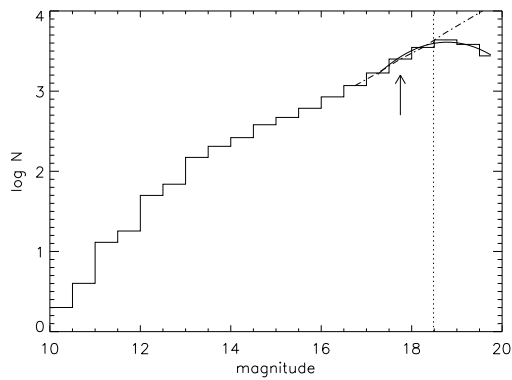}
\end{center}
\caption{Example of determination of the magnitude limit. The histogram shows the
raw GSC~2 object counts from HTM region S2200131, as a function of $R_F$
magnitude. The vertical arrow indicates the last complete bin, the
dash-dotted line shows the expected magnitude distribution, the thick
solid curve represents the actual magnitude distribution, and the
dotted line indicates the determined 90\% completeness limit.
\label{fig:examplereg}}
\end{figure}

First,
the magnitude bin containing the maximum number of stars is
adopted as a first guess of the last (faintest) complete bin. This
guess is checked by performing a linear fit to the log counts of the
previous two bins. If this presumed last complete bin is the last
magnitude bin, or if the fainter bins are empty, then a simple
completeness test is performed: if the count in this bin
agrees to within two times the Poisson error of the count predicted
from the linear fit, then the maximum magnitude of this bin is adopted as
the magnitude limit.  For the large majority of
regions more than 30 degrees from the Galactic plane, the last bin
($R_F =$ 19.5 to 20) meets this criterion and the
limiting magnitude of the counts ($R_F=20$) is adopted as the completeness
limit.

\begin{figure}[h]
\begin{center}
\includegraphics[width=1\linewidth]{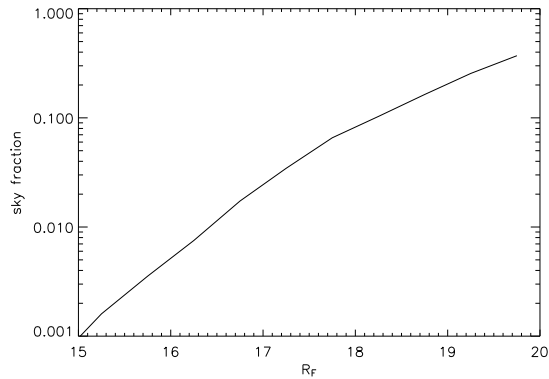}
\end{center}
\caption{Fraction of sky less than 90\% complete, as a function of $R_F$ magnitude.
\label{fig:compplotF}}
\end{figure}

\begin{figure}[h]
\begin{center}
\includegraphics[width=1\linewidth]{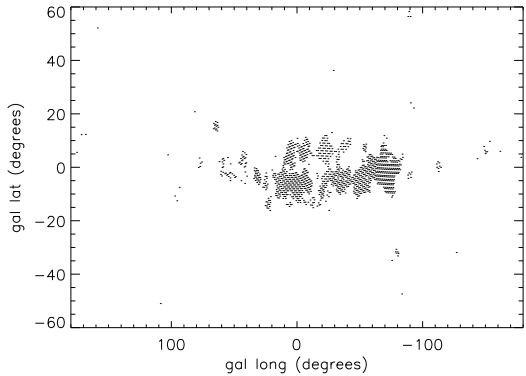}
\end{center}
\caption{Position on the sky of the HTM regions showing less than 90\%
completeness at magnitude $R_F = 18.5$.
\label{fig:compmapF}}
\end{figure}

If the above criterion is not satisfied then a refined estimate
of the last complete bin is made by requiring
that 1) the slope of the log-magnitude distribution decreases less
than 30\% over 1 mag, {\em or} that the count is not less than
2$\sigma$ of the count predicted from a linear fit to the previous two
magnitude bins, and 2) that the local slope of the log-magnitude
distribution is positive. The slope and change in slope of the counts are
evaluated using a second-order polynomial that fits the counts in
the hypothesized last complete bin and the previous two bins. We also
require that the slope of the above-mentioned linear fit be
positive. If any of these conditions are not satisfied, the previous (brighter)
magnitude bin is tested. This procedure is repeated until a
magnitude bin is found that satisfies the above conditions. Note that
if the actual count in the presumed last complete bin {\em exceeds}
the predicted counts it is taken as being the last complete bin.

Once the last complete bin is determined, we estimate the expected
counts after the last complete bin. The expected counts are
based on a local linear fit to the faintest part of the logarithmic
magnitude distribution determined to be complete, though now based on
three magnitude bins, including the last complete bin if the count in
this bin does not differ more than 2$\sigma$ from the estimated count
based on the previous two bins.
If the last complete bin does not satisfy this criteria then the previous three
bins are used in the linear fit. This linear model is our expected log-magnitude distribution, that is $\log n_e$.

A check is then made to see if any fainter bins are within 2 sigma of
our expected magnitude distribution, or exceeds the predicted counts.
If so, the faintest such bin is considered as our faintest complete bin, though
we retain the expected magnitude distribution already found. This
check is necessary only to handle a few regions with particularly
``noisy'' counts. Indeed, the complexity of the procedure described above was necessitated to assure that a last complete bin could be robustly determined for {\em all} HTM regions, including the relatively few pathological cases that can occur in very crowded regions where the photometric calibration was problematic.

We now make a more refined estimate of the completeness limit using a
second-order polynomial to describe the part of the
log-magnitude distribution where incompleteness sets in, from the bin
preceding the last complete bin, and up to two bins beyond the
last complete bin,
using only bins with nonzero counts. Thus, since at this point in the
algorithm the last complete bin does not coincide with the last
nonempty bin, the polynomial is fit to three or more points.
We adopt this polynomial as the measured log-magnitude
distribution, $\log n$. To determine to what limiting magnitude the
counts are, say, 90\% complete, we find the root to the equation
\[
\log n(m) - \log n_e(m) - \log(.9) = 0 ,
\]
that is, the magnitude at which the ratio of the measured and
expected counts, $n/n_e$, is equal to 0.9.

Figure \ref{fig:examplereg} shows an example of the application of this
procedure to the counts of a region showing evidence of incompleteness.

Using the above procedure we estimate the 90\% completeness limit for
each HTM region using the total object counts. As a
statistical measure of completeness we show in figure \ref{fig:compplotF}
the fraction of sky that is
less than 90\% complete as a function of magnitude. For example, 10\% of the
sky shows incompleteness at magnitude $R_F = 18.5$ or brighter, while at 17th
magnitude 3\% of the sky shows evidence of incompleteness.

Incompleteness at magnitudes brighter than about $R_F=19$
is strictly related to
the object field density, as in this case incompleteness arises not
from detector (plate) sensitivity, but from
the photographic plates being unable to distinguish objects from the
``background'' light of fainter objects whose images are blended due
to the finite resolution of the plates.

Figure \ref{fig:compmapF} shows the location on the sky of the HTM regions
showing incompleteness at magnitudes brighter than $R_F$ 18.5.

\section{Conclusions and future work}

In Table \ref{GSCI-IIreqs}, we compare the requirements provided in
the GSC-II {\it implementation plan}---originally foreseen to
reach magnitude 18 in $V$---with the actual performance of the
current version of the catalog down to the same magnitude limit.
In all items we have met or exceeded the specifications except for
the proper motions.

Table \ref{GSCI-IIspecs} summarizes the main characteristics of GSC~2.3, which
essentially superseeds the requirements in that it reaches 2-3 mag depeer.
The astrometric and photometric errors reported in the table are derived from
those of Tables \ref{tab:GSC23_SDSS_astro} and \ref{tab:GSC23_SDSS_F}, and
positional uncertainties are obtained by summing in quadrature the error in each coordinate.
 A consideration of the global statistics shows that even with the inclusion of fainter objects we are
able to meet the original specifications.

Whilst no further improvements are required (or funded) for its use in
{\it HST} operations, we intend to continue development to produce the best possible
catalog for scientific as well as operational uses.
Our future plans include an astrometric recalibration using the UCAC~3 and applying
a magnitude-dependent correction; this should not only reduce the absolute
positional errors but amend the systematic error that is affecting the computed
proper motions. A photometric re-reduction of all plates will also benefit
from the improved overall quality of the available calibrating sequences.
Finally, we plan to reclassify the objects splitting the nonstar
classification into galaxy and blend as well as better determine the
magnitudes of galaxies using an algorithm that does not assume a
stellar profile \citep{2007ApJS..170...33P}.

As of cycle 15, GSC~2.3 is the guide catalog for {\it HST},
and it is a reference catalog for the VLT and Gemini
adaptive optics programs, where real guide stars are still
preferred to laser ``guide stars''; it has also been provided to the
{\it Chandra}, {\it Galex}, {\it XMM-Newton} and {\it Swift} missions.
In the preparation for the
Gaia mission, the current catalog version is used as a snapshot of
what Gaia is expected to observe \citep{2006MmSAI..77.1172D} and
as a base for the Initial Gaia Source List being compiled by OATo
for the Gaia data reduction. Finally, GSC-II future release(s)
will be at the heart of the Astrometric Support System of the very
ambitious Large Sky Area Multi-Object Fiber Spectroscopic
Telescope (LAMOST) undertaken by the Chinese Academy of Science,
and will be part of the guide system for JWST
\citep{Spagna2001,2003AAS...202.0410S}.

In addition to the operational uses of the catalog itself, the GSC-II
database has also been mined for many scientific studies, such as planetary
nebulae \citep{2004A&A...420..207K}, young open clusters \citep{2000A&A...357..460S},
halo white dwarfs \citep{2006A&A...448..579C}, peculiar objects
 \citep{2001Natur.413..139M,2002A&A...393L..45C},  X-ray pulsars \citep{Panzera2003}, and
the structure and kinematics of stellar populations of our Galaxy
\citep{2006A&A...451..125V,2007MNRAS.375.1381K}.

Finally, it is worth going back to the density map presented in
Figure \ref{fig:skymap}, as it does illustrate GSC~2.3: it
provides accurate positions, magnitudes in three bands, and
stellar classification for 4$\pi$ steradians to levels of
completeness commensurate with the underlying stellar density.
Future versions extracted from the GSC-II II database
will only improve on this situation, including proper motions and
increasing the catalog accuracy as a result of improvements in
reference catalogs and reduction procedures.

\section*{Acknowledgements}

The Guide Star Catalog~II is a joint project of the STScI
 and the OATo. Space Telescope Science Institute is operated by the
Association of Universities for Research in Astronomy, for the
National Aeronautics and Space Administration under contract
NAS5-26555. The Osservatorio Astronomico di Torino is operated by
the Italian National Institute for Astrophysics (INAF).

Additional support was provided by the European Southern
Observatory, Space Telescope European Coordinating Facility, the
International GEMINI project, and the European Space Agency
Astrophysics Division.

The DSS was produced at the Space Telescope Science Institute under U.S. Government grant NAG W-2166. Additional support was provided by  Beijing Astronomical Observatory, Canadian Astronomical Data Center, Centre de Donnee Stellaire, European Southern Observatory, and the National Astronomical Observatory Japan. The DSS images are based on photographic data obtained using the Palomar Oschin Schmidt Telescope and the UK Schmidt Telescope. The plates were processed into  compressed digital form with the permission of these institutions.
The National Geographic Society--Palomar Observatory Sky Atlas (POSS-I) was made by the California Institute of Technology with grants from the National Geographic Society. The POSS-II was made by the California Institute of Technology with funds from the National Science Foundation, the National Aeronautics and Space Administration, the National Geographic Society, the Sloan Foundation, the Samuel Oschin Foundation, and the Eastman Kodak Corporation.
The UK Schmidt Telescope was operated by the Royal Observatory Edinburgh, with funding from the UK Science and Engineering Research Council (later the UK Particle Physics and Astronomy Research Council), until June 1988, and thereafter by the Anglo-Australian Observatory. The blue plates of the southern Sky Atlas and its Equatorial Extension (together known as the SERC-J), the near-IR plates (SERC-I), as well as the Equatorial Red (ER), and the Second Epoch [red] Survey (SES) were all taken with the UK Schmidt telescope at the AAO.

Partial support for the preparation of this publication was
provided by INAF through PRIN2005 grant (CRA 1.06.08.02) to OATo.
Also, B.D.\  and M.G.L.\ acknowledge the continuous
support from STScI through the Institute's Visitor Program.

Of the many individuals who gave us their support and assistance
throughout the realization of the GSC-II Project, we are
especially grateful to Piero Benvenuti, Bob Brucato, George
Djorgovski, Rodger Doxsey, Fabio Favata, Attilio Ferrari, Nathalie
Fourniol, Riccardo Giacconi, Fred Gillette, Jim Gray, Helmut
Jenkner, Nigel Hambly, Malcolm Hartley, Harvey MacGillivray,
Giuseppe Massone, David Morgan, Benoit Pirenne, Marc Postman, Phil
Puxley, Mike Read, Neill Reid, Alex Szalay, Sue Tritton, Fred
Watson, Andreas Wicenec, and Peredur Williams.

\clearpage
\begin{deluxetable}{llcccllrc}
\tabletypesize{\scriptsize}
\tablecaption{Plate material utilized for the construction of the GSC-II.
\label{GSCI-IIplates}} \tablewidth{0pt} \tablehead{
\colhead{Source\tablenotemark{a}}  &  \colhead{Survey\tablenotemark{b}} & \colhead{Declination} & \colhead{Epoch} & \colhead{Emulsion} & \colhead{Band} &  \colhead{Depth} & \colhead{Fields} & \colhead{In GSC}  \\
\colhead{Code } &  & \colhead{ Range} & \colhead{ } & \colhead{+Filter } &
\colhead{ } &  \colhead{(mag) } & \colhead{ }  & \colhead{2.2/2.3} }
\startdata
   N\tablenotemark{(1)} & Pal-QV & $\delta \ge 0^\circ$  & 1983-85 & IIaD+W12  & $V_{12}$  & 19.5 & 616 & N/Y \\
   S\tablenotemark{(2)} & SERC J & $\delta < -15^\circ$ & 1975-87 & IIIaJ+GG395 & $B_J$ & 23.0 & 606 & Y/Y \\
   S\tablenotemark{(2)} & SERC EJ & $-15^\circ< \delta \le 0^\circ$ & 1979-88 & IIIaJ+GG395 & B$_J$ & 23.0 & 288  & Y/Y \\
  XE\tablenotemark{(3)} & POSS-I E  & $\delta \ge -30^\circ$ & 1950-58 & 103aE+red plexiglass & $E$  & 20.0 & 935 & N/N \\
  XO & POSS-I O  & $\delta \ge -30^\circ$  & 1950-58 & 103aO unfiltered & $O$  & 21.0 & 935 & N/Y \\
  XJ & POSS-II J & $\delta \ge 0^\circ$   & 1987-00 & IIIaJ+GG385 & $B_J$ & 22.5 & 897 & Y/Y \\
  XP & POSS-II F & $\delta \ge 0^\circ$  & 1987-99 & IIIaF+RG610 & $R_F$  & 20.8 & 897 & Y/Y  \\
  XI & POSS-II N & $\delta \ge 0^\circ$  & 1989-02 & IV-N +RG9   & $I_N$  & 19.5 & 897 & N/Y  \\
  XS & AAO-SES   & $\delta < -15^\circ$ & 1990-00 & IIIaF+OG590 & $R_F$  & 22.0 & 606 & Y/Y   \\
  ER & SERC ER   & $-15^\circ< \delta \le 0^\circ$ & 1990-98 & IIIaF+OG590 & $R_F$  & 22.0 & 288 & Y/Y  \\
  IS & SERC I    & $\delta \le 0^\circ$ & 1990-02 & IV-N +RG715 & $I_N$  & 19.5 & 731 & N/Y \\
  IS & MW Atlas & $\delta \le 0^\circ$ & 1978-85 & IV-N +RG715 & $I_N$  & 19   & 173 & N/Y \\
  XV\tablenotemark{(4)} & SERC-QV   & $-70^\circ < \delta \le 0^\circ$  & 1987-88 & IIaD+GG495 & $V_{495}$  & 14 &  94 & N/Y \\
  GR\tablenotemark{(5)} & AAO-SR    & $-70^\circ < \delta \le 0^\circ$ & 1996-99 & IIIaF+OG590 & $R_F$  & 20   & 118 & Y/Y \\
\enddata
\tablenotetext{a} {The source code used by the GSC-II database to identify individual plates. For example, XJ442,  XP442, and XI442 identity the blue, red, and near-infrared plates of the 442th POSS-II field.}
\tablenotetext{b} {Survey-filter abbreviations: see NOTE below.}
 \tablenotetext{1} {The Pal-QV survey was taken specifically for GSC-I.  Schott filter GG495
replaced W12 starting 3 June 1984. Scanned with 25 micron sampling (1.7 arcsec/pixel).}
 \tablenotetext{2} {S plates originally scanned with 25 micron sampling, 440 plates at lower latitudes rescanned with 15 micron sampling (1.0 arcsec/pixel).}
  \tablenotetext{3} {POSS-I E filter was red plexiglass no.~2444; scanned at
    25 micron sampling except 123 plates scanned at 15 micron, mostly south of $-18^\circ$.}
 \tablenotetext{4} {SERC ''Quick V'' survey taken specifically for GSC-I with
4-min exposures to cover crowded southern Milky Way fields ($|b|<15^\circ$, $-112^\circ<l<34^\circ$ plus two plates on the LMC).
(Three similar ''XX'' plates were centered on M31 and each of the
Magellanic Clouds.)}
 \tablenotetext{5} {AAO-SR survey taken specifically for GSC-II with 5-min exposures to
 cover crowded southern Milky Way fields ($|b|<15^\circ$, $-112^\circ<l<34^\circ$ plus two plates on the LMC). }
\tablecomments{\rm Here, Pal-QV refers to the Palomar "Quick V" survey (\cite{1990AJ.....99.2019L}), SERC refers to the Science \& Engineering Research Council surveys, POSS I and II are the Palomar Observatory Sky Surveys I and II, AAO-SES and -SR refer to the Anglo-Australian Observatory Second Epoch Survey and Short-Red survey, and the MW Atlas is the SERC I/SR Atlas of the Milky Way and Magellanic Clouds (ref?). POSS-I and Pal-QV surveys utilized a $6^\circ$ grid of telescope pointings, with minimum plate overlap, while the other surveys used a $5^\circ$ degree grid. Unless stated otherwise, all plates scanned at 15 micron sampling (1 arcsec/pixel). Scans performed on glass copies of the S, XE and XO plates, rather than on the original plates.}
\end{deluxetable}

\begin{deluxetable}{cc}
\tablecaption{Classification parameters.\label{features} }
\tablewidth{0pt}
\tablehead{
 \colhead{Parameter}   &  \colhead{Definition}      }
\startdata
1         & Integrated density \\
 2         & Peak density\\
 3         & Semimajor axis \\
 4         & Semiminor axis  \\
 5         & Ellipticity   \\
 6,7,8     & $\sigma^2_x, \sigma^2_y, \sigma^2_{xy}$ \\
 9-12      & Texture features (\cite{1983sma..conf...69M})  \\
 13-14     & Spike features (\cite{1990AJ.....99.2019L})  \\
 15-30     & Area at 16 detection thresholds \\
\enddata
\end{deluxetable}

\begin{deluxetable}{rlrcr}
 \tablecaption{GSC~2.3 (July 2006).  Export Binary Table Fields and Types
   \label{tab:GSC23_fields}}
 \tablewidth{0pt}
 \tablehead{  \colhead{Number} & \colhead{Field name} &
\colhead{Format} &   \colhead{Unit} & \colhead{Notes}   }
 \startdata
1 & gscID2  & Integer*4 (J)  & id  & GSC~2.3 object id \\
2 & gsc1ID  & Character*11 (11A) & id & GSC 1.1 object id\\
3 & hstID   & Character*11 (11A)  & id & {\it HST} object id \\
4 & RightAsc & Real*8 (D) & rad & Right ascension ICRF \\
5 & Declination & Real*8 (D) & rad & Declination ICRF \\
6 & PositionEpoch & Real*4 (E) & year & Position epoch\tablenotemark{1} \\
7 & raEpsilon  &  Real*4 (E) & arcsec & Reference error on R.A.\tablenotemark{2} \\
8 & decEpsilon  &  Real*4 (E) & arcsec & Reference error on DEC\tablenotemark{2} \\
14 & FpgMag    &  Real*4 (E) & mag     & $R_F$ photographic magnitude\\
15 & FpgMagErr    &  Real*4 (E) & mag     & Reference error on $R_F$\tablenotemark{2}\\
16 & FpgMagCode    &  Integer*2 (I) & --     & Filter code of $R_F$\\
17 & JpgMag    &  Real*4 (E) & mag     & $B_J$ Photographic magnitude\\
18 & JpgMagErr    &  Real*4 (E) & mag     & Reference error on $B_J$\\
19 & JpgMagCode    &  Integer*2 (I) & --     & Filter code of $B_J$\\
20 & VMag    &  Real*4 (E) & mag     & $V$ [photographic] magnitude\tablenotemark{3}\\
21 & VMagErr    &  Real*4 (E) & mag     & Reference error on $V$\tablenotemark{2}\\
22 & VMagCode    &  Integer*2 (I) & --     & Filter code of $V$\\
23 & NpgMag    &  Real*4 (E) & mag     & $I_N$ photographic magnitude\\
24 & NpgMagErr    &  Real*4 (E) & mag     & Reference error on $I_N$\tablenotemark{2}\\
25 & NpgMagCode    &  Integer*2 (I) & --     & Filter code of $I_N$\\
29 & BMag    &  Real*4 (E) & mag     & $B$  Magnitude\tablenotemark{4}  \\
30 & BMagErr    &  Real*4 (E) & mag     & Reference error on $B$\\
31 & BMagCode    &  Integer*2 (I) & --     & Filter code of $B$\\
47 & classification & Integer*4 (J) & -- & Morphological classification\tablenotemark{5} \\
48 & semiMajorAxis  & Real*4 (E) &  pixels &  Image semimajor axis\tablenotemark{6} \\
49 & eccentricity  &  Real*4 (E) & --   &   Image eccentricity\tablenotemark{6} \\
50 & positionangle &  Real*4 (E) & degrees & Image orientation\tablenotemark{6}\\
51 & sourceStatus  & Integer*4 (J) & -- & Object processing status flag \\
 \enddata
\tablenotetext{1} {\small Plate epoch for GSC-II objects. For
Tycho 2 objects, for which $T_\alpha\neq T_\delta$,  the R.A.
epoch is given. } \tablenotetext{2} {\small These astrometric and
photometric errors are not formal statistical uncertainties but a
raw and conservative estimates to be used for telescope
operations.} \tablenotetext{3}{\small This field may include: (a)
photographic $V_{12}$ or $V_{495}$ from IIaD plates,  (b) $V_T$ of
{\it Tycho-2} stars, or  (c) Johnson $V$ from SKY2000.}
 \tablenotetext{4} {\small $B_T$ of {\it Tycho-2} stars or Johnson B from  SKY2000 or photographic O
from POSS-I}
\tablenotetext{5} {\small Image classification: 0 =  ``star''
(i.e. point-like object) and 3 = ``nonstar'' (i.e. extended
object).}
 \tablenotetext{6}{\small Morphological parameters of the {\it
 same} image used for the position.}
\end{deluxetable}

\begin{deluxetable}{rllr}
 \tablecaption{GSC~2.3 bandpass codes.   \label{tab:GSC23_pht_codes}}
 \tablewidth{0pt}
 \tablehead{  \colhead{Code} & \colhead{Band} &
\colhead{Emulsion+Filter} &   \colhead{Survey/Catalog}  }
\startdata
   0  & $B_J$ & IIIaJ + GG395   &    SERC-J, SERC-EJ\\
   1  & $V_{12}$ &   IIaD  + W12 & Pal-QV\\
   3 & B  &  Johnson B &  SKY2000  \\
   4  & V  & Johnson V  & SKY2000 \\
   5  & $R_F$  &     IIIaF + RG630  &     ESO-R \\
   6  & $V_{495}$ &  IIaD + GG495   & Pal-QV, SERC-SV\\
   7  &   O       &  103aO unfiltered  &     POSS-I O \\
    18  &   $B_J$   & IIIaJ + GG385 &      POSS-II J\\
   35  &   $R_F$              & IIIaF + RG610   &   POSS-II F\\
  36  &   $R_F$   & IIIaF + OG590 &  SERC-ER, SERC-SR, \\
      &                  &               &  AAO-R,  AAO-GR\\
  37  &   $I_N$    &  IV-N + RG9 &   POSS-II IR \\
  38  &   $I_N$    &  IV-N + RG715 &   SERC-IR  \\
  41  &   $B_T$    &            &  Tycho B  \\
  42  &   $V_T$    &            &  Tycho V  \\
\enddata
\end{deluxetable}

\begin{deluxetable}{ll}
\tablecaption{Source Status Flag Codes.\label{tab:status}}
\tablewidth{0pt}
\tablehead{
\colhead{Digits(*)}   & \colhead{Flag Meaning}}
\startdata
1 and 2   & Number of observations, some observations maybe excluded from GSC~2.3. \\
3         & Centroider type used:  0=barycenter,  1=circular \tablenotemark{a}, 2=elliptical,\\
          &  ~~~~~~~~~~~~~~   3=FPA + applied barycenter \tablenotemark{b}, 4=multicircular, \\
          &  ~~~~~~~~~~~~~~   5=multielliptical, 6=FPA + circular, 7 FPA + elliptical  \\
4         & Quality of exported $R_F$ magnitude: 0=not present, 1=fit, 2=extrapolated \\
5         & Quality of exported $B_J$ magnitude: 0=not present, 1=fit, 2=extrapolated \\
6         & Quality of exported $V$ magnitude: 0=not present, 1=fit, 2=extrapolated \\
7     & Classification unanimity: 0=mixed vote, 1=unanimous vote, 3=unanimous artifact  \\
8     & Classification voters: 0=multiple 15um scans,
                1=One 15um scan, \\
      &   ~~~~~~~~~~~~~~   2=multiple 25um scans, 3=One 25um scan  \\
9     & Processing status: 0=completed processing, \\
          &   ~~~~~~~~~~~~~~   1=object too big to be cut out on  at least one plate  \\
10        & Deblended object: 0=single object on all plates, \\
          &   ~~~~~~~~~~~~~~   1=child (deblended) object on at least one plate \\
\enddata
\tablenotetext{*}{counting from right to left; flags for Tycho and SKYMAP
  objects are 99999900 and 88888800 respectively}
\tablenotetext{a}{All non barycentric centroiding using Gaussian fits}
\tablenotetext{b} {FPA = Fractional Pixel Allocation}
\end{deluxetable}

\begin{deluxetable}{rrrrr}
 \tablecaption{GSC~2.3 global statistics \label{tab:GSC23_stats}}
 \tablewidth{0pt}
 \tablehead{ \colhead{ } & \colhead{Northern} & \colhead{Southern} &
\colhead{All} &  \\
 \colhead{ } & \colhead{hemisphere} & \colhead{hemisphere } & \colhead{sky}  }
 \startdata
Objects              & $396\,700\,598$  & $548\,892\,085$ & {$\mathbf{  945\: 592\: 683}$} \\
Point-like objects   & $88\:481\:909$ & $118\:206\:935$  & { $206\:688\:844$ }\\
Extended objects     & $308\:218\:689$  & $430\:685\:150$  & { $738\:903\:839$ }\\[5pt]
\tableline  \\
 $R_F$ magnitudes     & $354\:300\:890$ &  $452\:765\:492$  & {$807\:066\:382$} \\
 $B_J$ magnitudes    & $307\:949\:064$  &  $403\:822\:005$  & {$711\:771\:069$} \\
 $V_{\rm pg}$ magnitudes & $154\:786\:490$  &  $67\:947\:515$  & {$222\:734\:005$} \\
 $I_N$ magnitudes    & $245\:102\:118$  &  $329\:725\:734$  & { $574\:827\:852$} \\
 $B_J$ and $R_F$ magnitudes & $275\:035\:930$ & $330\:759\:735$ & $605\:795\:665$ \\
 $B_J$, $R_F$, and  $I_N$ magnitudes & $210\:601\:441$ & $236\:097\:358$ & $446\:668\:799$ \\[5pt]
 \tableline\\
 GSC 1.1 objects     &  $9\:071\:325$     & $9\:598\:246$  &  $18\:669\:571$ \\
 Tycho 2 stars$^1$       &  $1\:201\:897$     & $1\:321\:635$       &  $2\:523\:552$   \\
 Sky2000/Skymap$^1$      & $348\:458$       & $190\:364$    &  $473\:943$ \\
\enddata
\tablenotetext{1}{\small Without double/multiple entries (see explanation in
section 4.2): 2,519,152 {\it Tycho-2} stars and 175,632 SKY2000 stars}
\end{deluxetable}

\begin{deluxetable}{rrrrr}
 \tablecaption{GSC~2.3 (July 2006).  All-sky cumulative counts\tablenotemark{1}. \label{tab:GSC23_counts}}
 \tablewidth{0pt}
 \tablehead{ \colhead{mag} & \colhead{N$_{\rm obj}(B_J)$} &
\colhead{N$_{\rm obj}(R_F)$} &  \colhead{N$_{\rm obj}(I_N)$} & \colhead{N$_{\rm obj}(B_J+R_F+I_N)$}\tablenotemark{2}\\
 }
 \startdata
      10.0 &      $20\:715$  &     $35\:949$  &    $192\:984$ & $50\:794$ \\
      10.5 &      $34\:491$  &     $82\:556$  &    $443\:489$ & $175\:119$\\
      11.0 &      $54\:584$  &    $239\:047$  &   $1\:010\:629$ & $564\:997$ \\
      11.5 &      $86\:600$  &    $706\:719$  &   $2\:190\:201$ &  $1\:519\:395$ \\
      12.0 &     $154\:228$  &   $1\:829\:779$  &   $4\:415\:724$ & $3\:367\:609$\\
      12.5 &     $364\:579$  &   $3\:994\:288$  &   $8\:089\:061$ &   $6\:305\:430$ \\
      13.0 &    $1\:060\:240$  &   $7\:458\:931$  &  $13\:612\:182$ & $10\:657\:925$\\
      13.5 &    $2\:603\:441$  &  $12\:638\:946$  &  $21\:619\:677$ & $16\:943\:348$ \\
      14.0 &    $5\:056\:922$  &  $20\:196\:571$  &  $32\:990\:022$ & $25\:837\:448$ \\
      14.5 &    $8\:668\:396$  &  $31\:041\:175$  &  $48\:953\:930$ & $38\:151\:152$ \\
      15.0 &   $13\:898\:051$  &  $46\:366\:494$  &  $71\:017\:843$ & $54\:914\:864$ \\
      15.5 &   $21\:361\:566$  &  $67\:781\:378$  & $101\:286\:676$ & $77\:348\:928$ \\
      16.0 &   $31\:792\:178$  &  $97\:002\:324$  & $142\:016\:145$ & $107\:002\:624$ \\
      16.5 &   $46\:052\:408$  & $135\:717\:387$  & $196\:139\:738$ & $145\:725\:200$\\
      17.0 &   $65\:119\:781$  & $186\:270\:427$  & $267\:399\:004$ & $195\:532\:864$ \\
      17.5 &   $90\:046\:889$  & $250\:652\:663$  & $349\:775\:619$ & $254\:950\:176$ \\
      18.0 &  $12\:1721\:348$  & $327\:112\:112$  & $434\:137\:985$ & $319\:880\:384$ \\
      18.5 &  $161\:745\:847$  & $405\:027\:755$  & $510\:834\:052$ & $383\:714\:272$\\
      19.0 &  $210\:802\:304$  & $489\:823\:020$  & $562\:417\:267$ & $427\:113\:344$\\
      19.5 &  $270\:784\:037$  & $583\:790\:690$  & $574\:291\:825$ & $443\:187\:232$\\
      20.0 &  $340\:876\:792$  & $685\:566\:332$  & $574\:700\:787$ & $446\:393\:632$ \\
      20.5 &  $417\:992\:984$  & $783\:784\:283$  & $574\:823\:790$ & $446\:588\:640$\\
      21.0 &  $497\:409\:326$  & $805\:951\:305$  & $574\:827\:840$ & $446\:645\:408$\\
      21.5 &  $573\:255\:946$  & $806\:469\:272$  & $574\:827\:852$ & $446\:675\:616$\\
      22.0 &  $649\:675\:534$  & $806\:698\:629$  & $574\:827\:852$ & $446\:689\:312$ \\
      22.5 &  $707\:695\:255$  & $806\:828\:565$  & $574\:827\:852$ & $446\:696\:416$\\
      23.0 &  $711\:708\:557$  & $806\:937\:362$  & $574\:827\:852$ & $446\:698\:848$ \\
      23.5 &  $711\:722\:103$  & $807\:017\:618$  & $574\:827\:852$ & $446\:699\:680$ \\
      24.0 &  $711\:736\:633$  & $807\:061\:276$  & $574\:827\:852$ & $446\:699\:680$ \\
      $all$  & {$\mathbf {711\:771\:069}$} & {$\mathbf{ 807\:066\:382}$} &
      { $\mathbf{ 574\:827\:852}$} & { $\mathbf{ 466\:699\:774}$} \\
\enddata
\tablenotetext{1} {\small Cumulative counts from 0 magnitude to $m$ given
in the first column}
\tablenotetext{2} {\small The number of objects with computed photometry in
the 3 passbands are accumulated according to magnitude steps in $R_F$}
\end{deluxetable}

\begin{deluxetable}{crrrrrrrrrrrr}
\tablecaption{GSC~2.3 - UCAC~2/SDSS: Astrometric residuals.
\label{tab:GSC23_SDSS_astro} }
\tablewidth{0pt}
\tablehead{ & \multicolumn{5}{c}{stellar objects}&\multicolumn{5}{c}{extended objects}\\
 \colhead{$R_F$} & $N_{\rm star}$  & \colhead{$\sigma_{\Delta\alpha}$} &\colhead{$\epsilon_{\Delta\alpha}$} &
 \colhead{$\sigma_{\Delta\delta}$} &  \colhead{$\epsilon_{\Delta\delta}$}&$N_{\rm ext}$  &
\colhead{$\sigma_{\Delta\alpha}$} &
\colhead{$\epsilon_{\Delta\alpha}$} &
\colhead{$\sigma_{\Delta\delta}$}
&  \colhead{$\epsilon_{\Delta\delta}$}\\
 \colhead{(mag) } &   & \colhead{($''$)} & \colhead{($''$)}& \colhead{($''$)}& \colhead{($''$)}
&  & \colhead{($''$)}& \colhead{($''$)}& \colhead{($''$)}& \colhead{($''$)}
}
\startdata
{UCAC~2}&          &          &       &         & & &&&&\\
  11.5-12.0 &  $   786\,859 $ &     0.16 &    0.20  & 0.17  &    0.20 &$    81\,641 $ &    0.63   &   0.67  &0.59   & 0.63\\
  12.5-13.0 &  $2\,688\,254 $ &     0.16 &    0.19  & 0.16  &    0.19 &$    438\,091$  &    0.57   &   0.62  &0.53   & 0.58\\
  13.5-14.0 &  $5\,467\,031 $ &     0.14 &    0.19  & 0.14  &    0.18 &$   1\,559\,410$ &     0.55  &   0.60  & 0.52 &  0.57\\
  14.5-15.0 &  $9\,741\,299 $ &     0.13 &    0.19  & 0.13  &    0.17 &$   4\,559\,916$ &     0.48  &   0.54  & 0.46 &  0.51\\
  15.5-16.0 &  $16\,289\,992$  &     0.12 &    0.20  & 0.13  &    0.18 & $  9\,553\,274$ &     0.42  &   0.48  & 0.40 &  0.45\\
{SDSS}&          &          &       &         &   &&&&&\\
  14.5- 15.0&  $  566\,319   $  &     0.14 &    0.21  & 0.14  &    0.16 &$    37\,169 $ &     0.21  &  0.27 &  0.21  &  0.24\\
  15.5-16.0 &  $  878\,746   $  &     0.13 &    0.21  & 0.13  &    0.16 &$   115\,608 $ &     0.22  &  0.28 &  0.22  &  0.25\\
  16.5-17.0 &  $ 1\,194\,037 $ &     0.13 &    0.22  & 0.13  &    0.16 & $  394\,648  $&     0.18  &  0.25 &  0.18  &  0.21\\
  17.5-18.0 &  $ 1\,528\,006 $ &     0.14 &    0.23  & 0.14  &    0.17 & $ 1\,011\,007$  &     0.18  &  0.25 &  0.18  &  0.21\\
  18.5-19.0 &  $ 1\,896\,983 $ &     0.16 &    0.26  & 0.16  &    0.20 & $ 2\,759\,341$  &     0.21  &  0.28 &  0.21  &  0.24\\
  19.5-20.0 &  $ 1\,608\,896 $ &     0.23 &    0.31  & 0.23  &    0.25 & $ 7\,965\,088$  &     0.32  &  0.37 &  0.31  &  0.34\\
\enddata
\end{deluxetable}

\begin{deluxetable}{lrrrrrr}
 \tablecaption{GPSC-2  - GSC~2.3 photometric residuals \label{tab:GSC23_GPSC2}}

 \tablewidth{0pt}
 \tablehead{ \colhead{Objects} & \multicolumn{3}{c}{Northern Hemisphere} & \multicolumn{3}{c}{Southern Hemisphere }   \\
 \colhead{ } &
\colhead{Nbr} & \colhead{$\langle \Delta R_F \rangle $  } & \colhead{$\sigma_{\Delta R_F} $ } &
\colhead{Nbr} & \colhead{$\langle \Delta R_F \rangle $  } & \colhead{$\sigma_{\Delta R_F} $ }
 }

\startdata

All objects                  & $204\,322$ & 0.01 & 0.28 & $175\,364$ & 0.04 & 0.33  \\
Stellar objects              & $84\,996$ & $-$0.03 & 0.18 &  $60\,613$ & $-$0.02 & 0.25  \\
All objects (high latitude)      & $36\,714$ & 0.03 & 0.18 & $23\,977$ &  0.02 & 0.22  \\
Stellar objects (high latutude)  & $25\,508$  & $-$0.01 &  0.13 & $12\,846$ & $-$0.02 & 0.16  \\

\enddata

\end{deluxetable}

\begin{deluxetable}{lrrrr}
\tablecaption{GSC~2.3 - SDSS photometric residuals $\Delta B_{\rm J}$
\label{tab:GSC23_SDSS_J}}

\tablewidth{0pt}
\tablehead{ \colhead{ } & \multicolumn{2}{c}{North} &
\multicolumn{2}{c}{South} \\
\colhead{$B_{\rm J}$} &
\colhead{n} & \colhead{$\sigma_{\Delta B_{\rm J}} $ } &
\colhead{n} & \colhead{$\sigma_{\Delta B_{\rm J}} $ }
}
\startdata
14.5-15.0 & 44333 & 0.42 & 23713 & 0.31\\
15.5-16.0 & 83030 & 0.19 & 47614 & 0.14\\
16.5-17.0 & 129263 & 0.17 & 76672 & 0.14\\
17.5-18.9 & 167603 & 0.17 & 102888 & 0.17\\
18.5-19.0 & 203192 & 0.16 & 114972 & 0.19\\
19.5-20.0 & 246944 & 0.18 & 118327 & 0.18\\
20.5-21.0 & 263612 & 0.22 & 109137 & 0.22\\
21.5.22.0 & 150603 & 0.27 & 58421 & 0.27\\
\enddata

\end{deluxetable}

\begin{deluxetable}{lrrrr}
\tablecaption{GSC~2.3 - SDSS photometric residuals $\Delta R_{\rm F}$
\label{tab:GSC23_SDSS_F}}

\tablewidth{0pt}
\tablehead{ \colhead{ } & \multicolumn{2}{c}{North} &
\multicolumn{2}{c}{South } \\
\colhead{$R_{\rm F}$} &
\colhead{N} & \colhead{$\sigma_{\Delta R_{\rm F}} $ } &
\colhead{N} & \colhead{$\sigma_{\Delta R_{\rm F}} $ }
}
\startdata
13.5-14.0 & 49732 & 0.16 & 26838 & 0.17\\
14.5-15.0 & 91999 & 0.13 & 52713 & 0.12\\
15.5-16.0 & 146229 & 0.12 & 84616 & 0.11\\
16.5-17.0 & 198843 & 0.12 & 117889 & 0.13\\
17.5-18.0 & 254016 & 0.13 & 140616 & 0.13\\
18.5-19.0 & 306114 & 0.15 & 143052 & 0.15\\
19.5-20.0 & 222294 & 0.22 & 88951 & 0.21\\
\enddata
\end{deluxetable}

\begin{deluxetable}{lrrrr}
\tablecaption{GSC~2.3 - SDSS photometric residuals $\Delta I_{\rm N}$
\label{tab:GSC23_SDSS_N}}

\tablewidth{0pt}

\tablehead{ \colhead{ } & \multicolumn{2}{c}{North} &
\multicolumn{2}{c}{South } \\
\colhead{$I_{\rm N}$} &
\colhead{N} & \colhead{$\sigma_{\Delta I_{\rm N}} $ } &
\colhead{N} & \colhead{$\sigma_{\Delta I_{\rm N}} $ }
}
\startdata
13.5-14.0 & 56684 & 0.17 & 34795 & 0.12\\
14.5-15.0 & 120778 & 0.13 & 66884 & 0.10\\
15.5-16.0 & 186497 & 0.14 & 106972 & 0.11 \\
16.5-17.0 & 258430 & 0.15 & 150240 & 0.13\\
17.5-18.0 & 308639 & 0.19 & 164387 & 0.21 \\
18.5-19.0 & 143655 & 0.26 & 59368 & 0.25 \\
\enddata
\end{deluxetable}

\clearpage
\begin{deluxetable}{lll}
\tablecaption{GSC~II specifications\tablenotemark{a} and performance ($V_{lim} = 18$)\label{GSCI-IIreqs}}
\tablewidth{0pt}
\tablehead{ \colhead{} & \colhead{Specifications} & \colhead{Performance} }
\startdata
Astrometry                                       &
&                 \\
~~~~~Reference frame                              & ICRF
&   ICRF         \\
~~~~~Absolute position error                 & $<$ 0.``5
& $<$ 0.``3 \tablenotemark{b}              \\
~~~~~Relative position error  over a 0.$^\circ$5 field & $\le$ 0.``2
& $<$ 0.``2 \tablenotemark{b} \\
~~~~~Adherence to reference frame             & $<$ 0.``15
& $<$ 0.``15 \tablenotemark{c}  \\
~~~~~Proper motion error (total)                  & $<$ 0.004 ``
year$^{-1}$& Not released   \\
Photometry                                       &
&                \\
~~~~~Passbands                                    & At least 2 (1 color)
& $B_J$, $R_F$, $I_N$, Some $V$ and $O$ \\
~~~~~Magnitude error                              & 0.1-0.2 mag
& $<$ 0.2               \\
~~~~~Completeness to magnitude limit              & Yes
& For 90\% of the sky \tablenotemark{d}    \\
Stellar classification:                                   & 95\%
& 98\% \tablenotemark{e}\\
\enddata
\tablenotetext{a}{source: GSC-II Implementation Plan
\citep{BMLMGL1994}. Performance statistics are given for objects
to magnitude limit V=18, as in the original implementation plan}
\tablenotetext{b}{From a global comparison to DR5 of the SDSS, for
objects classified as stellar. The random contribution for
extended sources is approximately 20\% worse}
\tablenotetext{c}{Averaged over all plates, all magnitudes}
\tablenotetext{d}{Completeness is reached outside of the galactic
plane; for details see Section 5.6} \tablenotetext{e}{98\% of
objects classified as stars are confirmed as stellar by a
comparison to SDSS; see Section 5.5 for details}
\end{deluxetable}

\clearpage
\begin{deluxetable}{lcc}
\tablecaption{GSC~2.3 Global properties \label{GSCI-IIspecs}}
\tablewidth{0pt}
\tablehead{
\colhead{} & \colhead{} & \colhead{}
}
\startdata
Total objects & 945,592,683 & \\
Magnitude limit &  $B_J = 22.5$,  $R_F = 20.5$, and $I_N = 19.5$ & \\
Mean epoch of positions & 1992.5 &\\
Reference frame & ICRF & \\
Astrometric reference catalogs & ACT~+~{\it Tycho-2}  &\\
Average positional accuracy \tablenotemark{(a)} & Stellar objects & Extended objects \\
    $R_F$ $<$ 18.5 & 0.''28 &  0.''35 \\
    18.5 $\le$ $R_F$ $\le$ 19.5 & 0.''30 & 0.''37 \\
    $R_F$ $>$ 19.5 & 0.''40  &  0.''50\\
Average positional precision\tablenotemark{(b)}& Stellar objects & Extended objects \\
    $R_F$ $<$ 18.5 &  0.''20 &  0.''25 \\
    18.5 $\le$ $R_F$ $\le$ 19.5 & 0.''22 & 0.''30 \\
    $R_F$ $>$ 19.5 & 0.''32 &   0.''44 \\
Photometric reference catalogs & GSPC2~+~{\it Tycho} &\\
Average photometric accuracy(stellar sources)\tablenotemark{(c)}: & & \\
    $R_F$ $<$ 18.5  & 0.13 mag & \\
    18.5 $\le$ $R_F$ $\le$ 19.5 & 0.15 mag & \\
    $R_F$ $>$ 19.5 & 0.22 mag & \\
Completeness  & $>$ 98\% & Up to $R_F$=20, $l > 30^{\circ}$\tablenotemark{(d)} \\
\enddata
\tablenotetext{(a)}{Accuracy referes to the combined contribution
of random and systematic errors. The values are global averages
for the SDSS DR5 sample. Note that plate-to-plate as well as
north-south discrepancies can reach a few tenths of an arcsecond.
This is mainly due to residual systematic effects, as described in
Section 5.2, which will be addressed in the next catalog release.}
\tablenotetext{(b)}{These values, estimated using formula
\ref{eq:random_error}, only show the random part contribution to
the positional error.} \tablenotetext{(c)} {Based on the SDSS DR5
sample. Systematic offsets of the order few hundredths of a
magnitude; photometry of non-stellar sources suffers from
systematic errors which, for very bright objects, can be as high
as $\approx$ 2 mag, see Fig. \ref{fig:sdss_resid_nonstar}.}
\tablenotetext{(d)} {For the Galactic plane, the fraction of sky
with less than 90\% completeness limit is shown in Fig.
\ref{fig:compplotF} as function of magnitude.}
\end{deluxetable}


\begin{thebibliography}{}

\bibitem[{Adelman-McCarthy} et~al., 2007]{2007ApJS..172..634A}
{Adelman-McCarthy}, J.~K., {Ag{\"u}eros}, M.~A., {Allam}, S.~S.,
{Anderson},
  K.~S.~J., {Anderson}, S.~F., {Annis}, J., {Bahcall}, N.~A., {Bailer-Jones},
  C.~A.~L., {Baldry}, I.~K., {Barentine}, J.~C., and 144 coauthors
  2007,
\newblock { \apjs}, 172, 634

\bibitem[{Beard} et~al., 1990]{1990MNRAS.247..311B}
{Beard}, S.~M., {MacGillivray}, H.~T., and {Thanisch}, P.~F.
 1990,
\newblock { \mnras}, 247, 311

\bibitem[{Bucciarelli} et~al., 2001]{2001A&A...368..335B}
{Bucciarelli}, B., {Garc{\'{\i}}a Yus}, J., {Casalegno}, R.,
{Postman}, M.,
  {Lasker}, B.~M., {Sturch}, C., {Lattanzi}, M.~G., {McLean}, B.~J., {Costa},
  E., {Falasca}, A., {Le Poole}, R., {Massone}, G., {Potter}, M., {Rosenberg},
  A., {Borgman}, T., {Doggett}, J., {Morrison}, J., {Pizzuti}, A., {Pompei},
  E., {Rehner}, D., {Siciliano}, L., and {Wolfe}, D. 2001,
\newblock {\aap}, 368, 335

\bibitem[{Bucciarelli~et~al.}, 2006]{2006yCat.2272....0B}
{Bucciarelli}, B. and {et al.} 2006,
\newblock {Guide Star Photometric Catalog V2.4},
\newblock {VizieR Online Data Catalog II 272}

\bibitem[{Bushouse} and {Simon}, 1994]{1994ASPC...61..339B}
{Bushouse}, H. and {Simon}, B. 1994,
 in ASP Conf. Ser. 61, Astronomical Data Analysis Software and Systems III,
ed.  D.R. Crabtree, R.J. Hanisch,  and J. Barnes (San Francisco, CA: ASP),
  339

\bibitem[{Carollo} et~al., 2006]{2006A&A...448..579C}
{Carollo}, D., {Bucciarelli}, B., {Hodgkin}, S.~T., {Lattanzi},
M.~G.,
  {McLean}, B., {Morbidelli}, R., {Smart}, R.~L., {Spagna}, A., and
  {Terranegra}, L. 2006,
 \aap, 448, 579

\bibitem[{Carollo} et~al., 2002]{2002A&A...393L..45C}
{Carollo}, D., {Hodgkin}, S.~T., {Spagna}, A., {Smart}, R.~L.,
{Lattanzi},
  M.~G., {McLean}, B.~J., and {Pinfield}, D.~J. 2002,
 \aap, 393, L45

\bibitem[{Cornuelle} et~al., 1997]{1997ASPC..127...55C}
{Cornuelle}, C.~S., {Aldering}, G., {Humphreys}, R.~M., {Larsen},
J., and
  {Cabanela}, J. 1997,
The APS Catalog of the POSS I, Image Database, and Luyten Proper
  Motion Catalog, in.ASP Conf. Ser. 127, Proper Motions and Galactic
  Astronomy, ed. R.~M. Humphreys (San Francisco, CA: ASP),  55

\bibitem[{Djorgovski} et~al., 2003]{2003BASBr..23Q.197D}
{Djorgovski}, S.~G., {Carvalho}, R.~R., {Gal}, R.~R., {Odewahn},
S.~C., {Mahabal}, A.~A., {Brunner}, R., {Lopes}, P.~A.~A., and
{Kohl Moreira}, J.~L. 2003,
\newblock {The digital Palomar observatory sky survey (DPOSS): general
  description and the public data release},
 Bull. Astron. Soc. Braz., 23, 197

\bibitem[{Drimmel} et~al., 2005]{2005ESASP.576..163D}
{Drimmel}, R., {Bucciarelli}, B., {Lattanzi}, M.~G., {Spagna}, A.,
{Jordi}, C.,  {Robin}, A.~C., {Reyl{\'e}}, C., and {Luri}, X. 2005,
What Gaia Will See: All-Sky Source Counts from the GSC2,
in The Three-Dimensional Universe with Gaia, ESA Special Publication 576,
ed. C. Turon, K.~S. O'Flaherty,  and M.~A.~C. {Perryman},
 (Noordwijk: ESA), 163

\bibitem[{Drimmel} et~al., 2006]{2006MmSAI..77.1172D}
{Drimmel}, R., {Spagna}, A., {Bucciarelli}, B., {Lattanzi}, M.,
and {Smart}, R.
  2006,
Mem. Soc. Astron. Ital., 77, 1172




\bibitem[Graham et al.(2007)]{2007MNRAS.378..198G} Graham, A.~W., Driver,
S.~P., Allen, P.~D., \& Liske, J.\ 2007, \mnras, 378, 198


\bibitem[{Gunn} and {Stryker}, 1983]{1983ApJS...52..121G}
{Gunn}, J.~E. and {Stryker}, L.~L. 1983,
\newblock {Stellar spectrophotometric atlas, wavelengths from 3130 to 10800 \AA{}},
\newblock { \apjs}, 52, 121

\bibitem[{Hambly} et~al., 2001]{2001MNRAS.326.1279H}
{Hambly}, N.~C., {MacGillivray}, H.~T., {Read}, M.~A., {Tritton},
S.~B.,
  {Thomson}, E.~B., {Kelly}, B.~D., {Morgan}, D.~H., {Smith}, R.~E., {Driver},
  S.~P., {Williamson}, J., {Parker}, Q.~A., {Hawkins}, M.~R.~S., {Williams},
  P.~M., and {Lawrence}, A. 2001,
\newblock {\mnras}, 326, 1279

\bibitem[{H{\o}g} et~al., 2000]{2000A&A...355L..27H}
{H{\o}g}, E., {Fabricius}, C., {Makarov}, V.~V., {Urban}, S.,
{Corbin}, T.,
  {Wycoff}, G., {Bastian}, U., {Schwekendiek}, P., and {Wicenec}, A. 2000,
\newblock {\em \aap}, 355, L27

\bibitem[{Infante}, 1993]{1993ASPC...51..304I}
{Infante}, L. 1993,
\newblock {Counts and Colours from the NGP CFHT Faint Galaxy Survey},
in ASP Conf. Ser. 51,  Observational Cosmology,
ed. G.~L. {Chincarini}, A. {Iovino}, T. {Maccacaro},  and
  D. {Maccagni} (San Francisco, CA: ASP),  304

\bibitem[{Irwin} and {McMahon}, 1992]{1992IAUIn...2...31I}
{Irwin}, M. and {McMahon}, R. 1992,
\newblock {APM Northern Sky Catalogue},
 in IAU Commission on Instruments, vol.~2, 31

\bibitem[{Ivezi{\'c}} et~al., 2004]{2004AN....325..583I}
{Ivezi{\'c}}, {\v Z}., {Lupton}, R.~H., {Schlegel}, D., {Boroski},
B.,
  {Adelman-McCarthy}, J., {Yanny}, B., {Kent}, S., {Stoughton}, C.,
  {Finkbeiner}, D., {Padmanabhan}, N., {Rockosi}, C.~M., {Gunn}, J.~E.,
  {Knapp}, G.~R., {Strauss}, M.~A., {Richards}, G.~T., {Eisenstein}, D.,
  {Nicinski}, T., {Kleinman}, S.~J., {Krzesinski}, J., {Newman}, P.~R.,
  {Snedden}, S., {Thakar}, A.~R., {Szalay}, A., {Munn}, J.~A., {Smith}, J.~A.,
  {Tucker}, D., and {Lee}, B.~C. 2004,
\newblock {Astron. Nachr.}, 325, 583

\bibitem[{Jenkner} et~al., 1990]{1990AJ.....99.2082J}
{Jenkner}, H., {Lasker}, B.~M., {Sturch}, C.~R., {McLean}, B.~J.,
{Shara},
  M.~M., and {Russel}, J.~L. 1990,
 \aj, 99, 2082

\bibitem[{Kerber} et~al., 2004]{2004A&A...420..207K}
{Kerber}, F., {Mignani}, R.~P., {Pauli}, E.-M., {Wicenec}, A., and
  {Guglielmetti}, F. 2004,
\aap, 420, 207


\bibitem[Kinman et al.(2007)]{2007MNRAS.375.1381K} Kinman, T.~D., Cacciari,
C., Bragaglia, A., Buzzoni, A., \& Spagna, A.\ 2007, \mnras,
375, 1381

\bibitem[{Klemola} et~al., 1994]{1994gsso.conf...20K}
{Klemola}, A.~R., {Hanson}, R.~B., and {Jones}, B.~F. 1994,
Lick NPM program: NPM1 Catalog and its applications,
in Galactic and  Solar System Optical Astrometry,
ed. L.~V. {Morrison}  and G.~F. {Gilmore} (Cambridge: Cambridge
University Press),   20

\bibitem[Konig, 1962]{kon}
Konig, A. 1962,
\newblock Astronomical Techniques,
\newblock (Chicago, IL: Chicago University Press)

\bibitem[{Kunszt} et~al., 2000]{2000ASPC..216..141K}
{Kunszt}, P.~Z., {Szalay}, A.~S., {Csabai}, I., and {Thakar},
A.~R. 2000,
\newblock {The Indexing of the SDSS Science Archive},
\newblock in ASP Conf. Ser. 216,
Astronomical Data Analysis Software and Systems IX,
ed. N. {Manset}, C. {Veillet},  and D. {Crabtree}
(San Francisco, CA: ASP),  141

\bibitem[{Kunszt} et~al., 2001]{2001misk.conf..631K}
{Kunszt}, P.~Z., {Szalay}, A.~S., and {Thakar}, A.~R. 2001,
\newblock {The Hierarchical Triangular Mesh},
\newblock in Mining the Sky, ed.  A.~J.
{Banday},  S. {Zaroubi}, and M. {Bartelmann}
(Berlin: Springer-Verlag),   631

\bibitem[{Laidler} et~al., 1994]{1994AAS...184.2701L}
{Laidler}, V.~G., {Greene}, G.~R., {Ray}, K., {Evzerov}, A., and
{Lasker},
  B.~M. 1994,
\newblock {GAMMA---A New-High-speed Microdensitometer Built on a PDS
  Substrate},
\newblock  BAAS, 26, 897

\bibitem[{Lasker} et~al., 1998]{1998asal.confE...3L}
{Lasker}, B.~M., {Greene}, G.~R., {Lattanzi}, M.~J., {McLean},
B.~J., and
  {Volpicelli}, A. 1998,
\newblock {DSS-II and GSC-II: STScI All-Sky Image and Catalog Databases},
\newblock in {Astrophysics and Algorithms; a DIMACS Workshop
on Massive Astronomical Data Sets}

\bibitem[{Lasker} and {Lattanzi}, 1994]{BMLMGL1994}
{Lasker}, B.~M. and {Lattanzi}, M.~G. 1994,
\newblock The Guide Star Catalog II Construction Project:
Implementation Plan,  Technical Note

\bibitem[{Lasker} et~al., 1988]{1988ApJS...68....1L}
{Lasker}, B.~M., {Sturch}, C.~R., {Lopez}, C., {Mallamas}, A.~D.,
{McLaughlin},
  S.~F., {Russell}, J.~L., {Wisniewski}, W.~Z., {Gillespie}, B.~A., {Jenkner},
  H., {Siciliano}, E.~D., {Kenny}, D., {Baumert}, J.~H., {Goldberg}, A.~M.,
  {Henry}, G.~W., {Kemper}, E., and {Siegel}, M.~J. 1988,
\newblock {\apjs}, 68, 1

\bibitem[{Lasker} et~al., 1990]{1990AJ.....99.2019L}
{Lasker}, B.~M., {Sturch}, C.~R., {McLean}, B.~J., {Russell},
J.~L., {Jenkner},
  H., and {Shara}, M.~M. 1990,
\newblock {\aj}, 99, 2019

\bibitem[{Loveday}, 1996]{1996MNRAS.278.1025L}
{Loveday}, J. 1996,
\newblock {\mnras}, 278, 1025

\bibitem[Lutz, 1980]{1980ComputJ.....23.262}
Lutz, R.~K. 1980,
\newblock {Comput. J.}, 23(3), 262

\bibitem[{MacGillivray} and {Stobie}, 1984]{1984VA.....27..433M}
{MacGillivray}, H.~T. and {Stobie}, R.~S. 1984,
\newblock {New results with the COSMOS machine},
\newblock {Vistas Astron.}, 27, 433

\bibitem[{Malagnini}, 1983]{1983sma..conf...69M}
{Malagnini}, M.~L. 1983,
\newblock {A classification algorithm for star-galaxy counts},
\newblock in ESA SP-201: Statistical Methods in
  Astronomy, ed. E.~J. {Rolfe} (Noordwijk: ESA),  69

\bibitem[{Mirabel} et~al., 2001]{2001Natur.413..139M}
{Mirabel}, I.~F., {Dhawan}, V., {Mignani}, R.~P., {Rodrigues}, I.,
and
  {Guglielmetti}, F. 2001,
\newblock {\nat}, 413, 139

\bibitem[{Monet} et~al., 2003]{2003AJ....125..984M}
{Monet}, D.~G., {Levine}, S.~E., {Canzian}, B., {Ables}, H.~D.,
{Bird}, A.~R.,
  {Dahn}, C.~C., {Guetter}, H.~H., {Harris}, H.~C., {Henden}, A.~A., {Leggett},
  S.~K., {Levison}, H.~F., {Luginbuhl}, C.~B., {Martini}, J., {Monet},
  A.~K.~B., {Munn}, J.~A., {Pier}, J.~R., {Rhodes}, A.~R., {Riepe}, B., {Sell},
  S., {Stone}, R.~C., {Vrba}, F.~J., {Walker}, R.~L., {Westerhout}, G.,
  {Brucato}, R.~J., {Reid}, I.~N., {Schoening}, W., {Hartley}, M., {Read},
  M.~A., and {Tritton}, S.~B. 2003,
\newblock {\aj}, 125, 984

\bibitem[{Morgan}, 1995]{1995ASPC...84..137M}
{Morgan}, D.~H. 1995,
\newblock {Sky Surveys and Atlases from the Large Schmidt Telescopes},
\newblock in  ASP Conf. Ser. 84, IAU Colloq. 148, The Future Utilisation of
  Schmidt Telescopes,
Ed. J. {Chapman}, R. {Cannon}, S. {Harrison},  and B. {Hidayat}
(San Francisco, CA: ASP),  137

\bibitem[{Morrison} et~al., 1996]{1996AJ....111.1405M}
{Morrison}, J.~E., {R{\"o}ser}, S., {Lasker}, B.~M., {Smart}, R.~L.,
and {Taff},
  L.~G. 1996,
\newblock {\aj}, 111, 1405

\bibitem[{Morrison} et~al., 2001]{2001AJ....121.1752M}
{Morrison}, J.~E., {R{\"o}ser}, S., {McLean}, B., {Bucciarelli},
B., and
  {Lasker}, B. 2001,
\newblock {\aj}, 121, 1752


\bibitem[Murthy et al.(1994)]{1994cs........8103M} Murthy, S.~K., Kasif,
S., \& Salzberg, S.\ 1994,
arXiv cs/9408103

\bibitem[{Myers} et~al., 2002]{Myers2002}
{Myers}, J.~R., {Sande}, C.~B., {Miller}, A.~C., {Warren},
J.~W.~H., and
  {Tracewell}, D.~A. 2002,
\newblock SKY2000 Master Catalog, Version 4,
\newblock ftp://cdsarc.u-strasbg.fr/pub/cats/V/109

\bibitem[{Panzera} et~al., 2003]{Panzera2003}
{Panzera}, M.~R., {Campana}, S., {Covino}, S., {Lazzati}, D.,
{Mignani}, R.~P.,
  {Moretti}, A., and {Tagliaferri}, G. 2003,
\newblock {\aap}, 399, 351

\bibitem[{Perryman} and {ESA}, 1997]{1997hity.book.....P}
{Perryman}, M.~A.~C. and {ESA} 1997,
\newblock The {\it Hipparcos} and {\it Tycho} catalogues. Astrometric and photometric
  star catalogues derived from the ESA {\it Hipparcos} Space Astrometry Mission.
(ESA SP Series Vol. 1200 ISBN: 9290923997 (set)), (Noordwijk: ESA Publications Division)

\bibitem[{Petrosian} et~al., 2007]{2007ApJS..170...33P}
{Petrosian}, A., {McLean}, B., {Allen}, R.~J., and {MacKenty},
J.~W. 2007,
\newblock {\apjs}, 170, 33

\bibitem[{Pier} et~al., 2003]{2003AJ....125.1559P}
{Pier}, J.~R., {Munn}, J.~A., {Hindsley}, R.~B., {Hennessy},
G.~S., {Kent},
  S.~M., {Lupton}, R.~H., and {Ivezi{\'c}}, {\v Z}. 2003,
\newblock {\aj}, 125, 1559

\bibitem[{Platais} et~al., 2002]{2002yCat.1277....0P}
{Platais}, I., {Girard}, T.~M., {Kozhurina-Platais}, V., {van
Altena}, W.~F.,
  {Lopez}, C.~E., {Mendez}, R.~A., {Ma}, W.-Z., {Yang}, T.-G., {MacGillivray},
  H.~T., and {Yentis}, D.~J. 2002,
\newblock {SPM Catalog 2.0},
\newblock {VizieR Online Data Catalog} I/277

\bibitem[{Postman} et~al., 1996]{1996AJ....111..615P}
{Postman}, M., {Lubin}, L.~M., {Gunn}, J.~E., {Oke}, J.~B.,
{Hoessel}, J.~G.,
  {Schneider}, D.~P., and {Christensen}, J.~A. 1996,
\newblock {\aj}, 111, 615

\bibitem[{Russell} et~al., 1990]{1990AJ.....99.2059R}
{Russell}, J.~L., {Lasker}, B.~M., {McLean}, B.~J., {Sturch},
C.~R., and
  {Jenkner}, H. 1990,
\newblock {\aj}, 99, 2059

\bibitem[{Sciortino} et~al., 2000]{2000A&A...357..460S}
{Sciortino}, S., {Micela}, G., {Favata}, F., {Spagna}, A., and
{Lattanzi},
  M.~G. 2000,
\newblock {\aap}, 357, 460

\bibitem[{Skrutskie} et~al., 2006]{2006AJ....131.1163S}
{Skrutskie}, M.~F., {Cutri}, R.~M., {Stiening}, R., {Weinberg},
M.~D.,
  {Schneider}, S., {Carpenter}, J.~M., {Beichman}, C., {Capps}, R., {Chester},
  T., {Elias}, J., {Huchra}, J., {Liebert}, J., {Lonsdale}, C., {Monet}, D.~G.,
  {Price}, S., {Seitzer}, P., {Jarrett}, T., {Kirkpatrick}, J.~D., {Gizis},
  J.~E., {Howard}, E., {Evans}, T., {Fowler}, J., {Fullmer}, L., {Hurt}, R.,
  {Light}, R., {Kopan}, E.~L., {Marsh}, K.~A., {McCallon}, H.~L., {Tam}, R.,
  {Van Dyk}, S., and {Wheelock}, S. 2006,
\newblock {\aj}, 131, 1163

\bibitem[{Spagna}, 2001]{Spagna2001}
{Spagna}, A. 2001,
\newblock {Guide Star Requirements for NGST: Deep NIR Starcounts and Guide Star
  Catalogs},
\newblock STScI-NGST-r-0013b Technical Report

\bibitem[{Stys} and {Kriss}, 2003]{2003AAS...202.0410S}
{Stys}, J. and {Kriss}, G. 2003,
\newblock {The Suitability of Guide Star Catalog 2 (GSC-2) as a Source for JWST
  Guide Stars},
\newblock BAAS 202, 410

\bibitem[{Taff} et~al., 1990]{1990ApJ...358..359T}
{Taff}, L.~G., {Lattanzi}, M.~G., and {Bucciarelli}, B. 1990,
\newblock {\apj}, 358, 359

\bibitem[{Thanisch} and {Robin}, 1984]{Thanisch}
{Thanisch}, P.~F., M. B.~V. and {Robin}, A. 1984,
\newblock {Image Vis. Comput.}, 2, 191

\bibitem[{Urban} et~al., 1998]{1998AJ....115.2161U}
{Urban}, S.~E., {Corbin}, T.~E., and {Wycoff}, G.~L. 1998,
\newblock {\aj}, 115, 2161

\bibitem[{Vallenari} et~al., 2006]{2006A&A...451..125V}
{Vallenari}, A., {Pasetto}, S., {Bertelli}, G., {Chiosi}, C.,
{Spagna}, A., and
  {Lattanzi}, M. 2006,
\newblock {\aap}, 451, 125

\bibitem[{White}, 1997]{1997scma.conf..135W}
{White}, R.~L. 1997,
\newblock {Object Classification in Astronomical Images},
\newblock in {Statistical  Challenges in Modern Astronomy II}
ed. G.~J. {Babu}  and E.~D. {Feigelson} (Berlin: Springer-Verlag), 135

\bibitem[{White} and {Percival}, 1994]{1994SPIE.2199..703W}
{White}, R.~L. and {Percival}, J.~W. 1994,
\newblock {Compression and progressive transmission of astronomical images},
\newblock in Proc. SPIE: Advanced Technology Optical Telescopes V,
Vol. 2199, ed. L.~M. {Stepp},  703

\bibitem[{White} et~al., 1992]{1992doss.conf..167W}
{White}, R.~L., {Postman}, M., and {Lattanzi}, M.~G. 1992,
\newblock {Compression of the Guide Star Digitised Schmidt Plates},
\newblock in ASSL: Digitised Optical Sky Surveys, Vol. 174,
ed. H.~T. {MacGillivray}, and E.~B. {Thomson},  167

\bibitem[{Zacharias} et~al., 2004]{2004AJ....127.3043Z}
{Zacharias}, N., {Urban}, S.~E., {Zacharias}, M.~I., {Wycoff},
G.~L., {Hall},
  D.~M., {Monet}, D.~G., and {Rafferty}, T.~J. 2004,
\newblock {\aj}, 127, 3043

\end{thebibliography}
\end{document}